\documentclass[12pt]{iopart}

\usepackage{iopams}

\usepackage{textcomp} % Provides math symbols that can be used in text mode
\usepackage{amsopn}
\usepackage{mathrsfs,amssymb} %\mathscr{ABCD} \mathfrak{ABCD,abcd}
\usepackage{setstack} %\overset \underset ...

\usepackage{bm}       % Provides bold-faced math symbols
\usepackage{booktabs} % Provides improved table formatting
\usepackage{dcolumn}  % Provides table columns aligned at decimal points
\usepackage{graphicx} % Standard package to incorporate graphics
\usepackage[printonlyused]{acronym} % Provides a method for incorporating

\usepackage[normalem]{ulem} % to use \sout
\usepackage{verbatim} %block comment

\usepackage{capt-of}

\usepackage{relsize}

\usepackage{xcolor}

\usepackage{multirow}
\usepackage{tabularx}

\usepackage{enumitem}
\usepackage{slashed}

\usepackage[colorlinks,breaklinks]{hyperref}

\usepackage{etoolbox}
\makeatletter
\def\@mkboth#1#2{}
\newlength\appendixwidth
\preto\appendix{\addtocontents{toc}{\protect\patchl@section}}
\newcommand{\patchl@section}{%
  \settowidth{\appendixwidth}{\textbf{Appendix }}%
  \addtolength{\appendixwidth}{1.5em}%
  \patchcmd{\l@section}{1.5em}{\appendixwidth}{}{\ddt}%
}
\makeatother

\newcommand{\bra}{\langle}
\newcommand{\ket}{\rangle}
\newcommand{\threej}[6]{
     \left( \begin{array}{ccc}
              #1 & #2 & #3 \\
              #4 & #5 & #6 
            \end{array}  \right) } 
\newcommand{\sixj}[6]{
     \left\{ \begin{array}{ccc}
              #1 & #2 & #3 \\
              #4 & #5 & #6 
            \end{array}  \right\} }

\renewcommand{\bs}[1]{\ensuremath{\boldsymbol{#1}}}

\newcommand{\be}{\begin{equation}}
\newcommand{\ee}{\end{equation}}
\newcommand{\bea}{\begin{eqnarray}}
\newcommand{\eea}{\end{eqnarray}}
\newcommand{\beq}{\begin{equation}}
\newcommand{\eeq}{\end{equation}}
\newcommand{\beqa}{\begin{eqnarray}}
\newcommand{\eeqa}{\end{eqnarray}}

\newcommand{\nn}{\nonumber\\}

\newcommand{\eqref}[1]{(\ref{#1})}
\newcommand{\iint}{\int\int}

\renewcommand{\textrm}[1]{\rm{#1}}
\renewcommand{\text}[1]{\rm{#1}}

\newenvironment{align}{\begin{eqnarray}}{\end{eqnarray}} %%to use align environment

\DeclareMathOperator*{\SumInt}{%
\mathchoice%
  {\ooalign{$\displaystyle\sum$\cr\hidewidth$\displaystyle\int$\hidewidth\cr}}
  {\ooalign{\raisebox{.14\height}{\scalebox{.7}{$\textstyle\sum$}}\cr\hidewidth$\textstyle\int$\hidewidth\cr}}   
  {\ooalign{\raisebox{.2\height}{\scalebox{.6}{$\scriptstyle\sum$}}\cr$\scriptstyle\int$\cr}}
  {\ooalign{\raisebox{.2\height}{\scalebox{.6}{$\scriptstyle\sum$}}\cr$\scriptstyle\int$\cr}}  
}

\DeclareMathOperator\arctanh{arctanh}

\begin{document}

\title[]{Ab initio calculation of nuclear structure corrections in muonic atoms}

\author{C.~Ji$^{1}$, S.~Bacca$^{2,3,4}$, N.~Barnea$^{5}$,  O.~J.~Hernandez$^{2,3,6}$, N.~Nevo-Dinur$^{3}$}

\address{$^1$ Key Laboratory of Quark and Lepton Physics (MOE) and Institute of Particle Physics,  Central China Normal University, Wuhan 430079, China}
\address{$^2$ Institut f\"ur Kernphysik and PRISMA Cluster of Excellence, Johannes Gutenberg-Universit\"at Mainz, 55128 Mainz, Germany}
\address{$^3$ TRIUMF, 4004 Wesbrook Mall, Vancouver, BC V6T 2A3, Canada}
\address{$^4$ Department of Physics and Astronomy, University of Manitoba, Winnipeg, MB R3T 2N2, Canada}
\address{$^5$ Racah Institute of Physics, The Hebrew University, Jerusalem 91904, Israel}
\address{$^6$ Department of Physics and Astronomy, University of British Columbia, Vancouver, BC, V6T 1Z4, Canada}

\ead{jichen@mail.ccnu.edu.cn, s.bacca@uni-mainz.de, nir@phys.huji.ac.il, javierh@phas.ubc.ca, nnevodinur@triumf.ca}

\date{\today}

\begin{abstract}
   The measurement of the Lamb shift in muonic hydrogen and the subsequent emergence of the 
  proton-radius puzzle have motivated an experimental campaign devoted to measuring the 
  Lamb shift in other light muonic atoms, such as muonic deuterium and helium. 
  For these systems it has been shown that two-photon exchange
  nuclear structure corrections are the largest 
  source of uncertainty and consequently the bottleneck for exploiting
  the experimental precision to extract the nuclear charge radius. 
  Utilizing techniques and methods developed to study electromagnetic reactions in light nuclei,
  recent calculations of 
  nuclear structure corrections to the muonic Lamb shift have reached
  unprecedented precision, reducing the uncertainty with respect to previous estimates by a factor of 5 in 
  certain cases. 
  These results will be useful for shedding light on the nature of the 
  proton-radius puzzle and other open questions pertaining to it. 
  Here, we review and update calculations for muonic 
  deuterium and  tritium atoms, and for muonic helium-3 and helium-4 ions.  
  We present a thorough derivation of the  
  formalism and discuss the results in relation to other approaches where available. We also describe how to  assess  theoretical uncertainties, for which the 
  language of chiral effective field theory furnishes a systematic approach 
  that could be further exploited in the future.
\end{abstract}

\vspace{2pc}
\maketitle
%

%\maketitle

\newpage
\setcounter{tocdepth}{2} %%%%%hide subsubsection

{
\hypersetup{linkcolor=black}
\tableofcontents
}

\section{Introduction}

 In 2010, a disagreement between the determination of the proton charge radius ${r_p}$ from experiments involving muonic hydrogen and those based on electron-proton systems was discovered~\cite{Pohl:2010zza}. This gave rise to the so called ``proton radius puzzle'', which has received significant attention since its inception: our understanding of a simple quantity, the size of the proton, was in fact put into question.
 Earlier measurements of the proton charge radius depended solely on electronic hydrogen spectroscopy and electron scattering data. The CODATA 2010 evaluation, based on the compilation of the above two types of experimental data provided ${r_p}=0.8775(51)$ fm~\cite{Mohr:2012tt}. In contrast, the CREMA (Charge Radius Experiment with Muonic Atoms) collaboration determined the  proton radius  via laser spectroscopy measurements of the Lamb shift~\cite{Lamb47} -- the 2$S$--2$P$ atomic transition -- in an experiment with muonic hydrogen atoms ($\mu {\rm H}$) performed  at the Paul Scherrer Institute (PSI) in Switzerland.  The first results were published in Ref.~\cite{Pohl:2010zza} and later confirmed in Ref.~\cite{Antognini13}. The charge radius ${r_p}$ was found to be $0.84087(39)$ fm~\cite{Antognini13}, an order of magnitude more precise  and $4\%$ smaller than the CODATA 2010 value~\cite{Mohr:2012tt}, leading to a difference of about 7 combined standard deviations ($7 \sigma$). This disagreement has now been updated to a still significant $5.6 \sigma$ after the CODATA 2014 compilation (${r_p}$=0.8751(61) fm~\cite{CODATA_2014}).

 The high accuracy of the muonic hydrogen experiment is due to the fact that the muon's mass $m_{\mu}$ is 207 times larger than that of an electron $m_e$. This results in a seven orders of magnitude larger corrections $\sim \left(m_{\mu}/m_e\right)^3$ to the atomic spectrum due to finite size effects proportional to $r_p^2$.
 Compared to the various electronic data, the muonic hydrogen result  deviates by $4 \sigma$ from the global average of electronic hydrogen ($e$H) spectroscopy~\cite{CODATA_2014} and by $3\sigma\sim 5\sigma$ from the world-average electron scattering data~\cite{Sick:2003gm,Blunden:2005jv,Sick:2018}, among which the most recent measurements are from the Mainz Microtron (MAMI)~\cite{Bernauer:2010hp} and the Jefferson Laboratory (JLab)~\cite{Zhan:2011hp}.

Based on lepton flavor universality, the proton is expected to interact identically with the muon and electron. Therefore, this large discrepancy pushed for re-examining  the consistency among the different types of experiments and re-investigating their systematic uncertainties. Other interpretations of the discrepancy have been sought; most notable are novel aspects of hadronic structure\cite{Miller:2012ne,Hill:2016bjv} and beyond-the-standard-model theories, leading to lepton universality violations (see \cite{Pohl_review} and references therein).

\begin{figure}[htb]
\centering
\includegraphics[width=13cm]{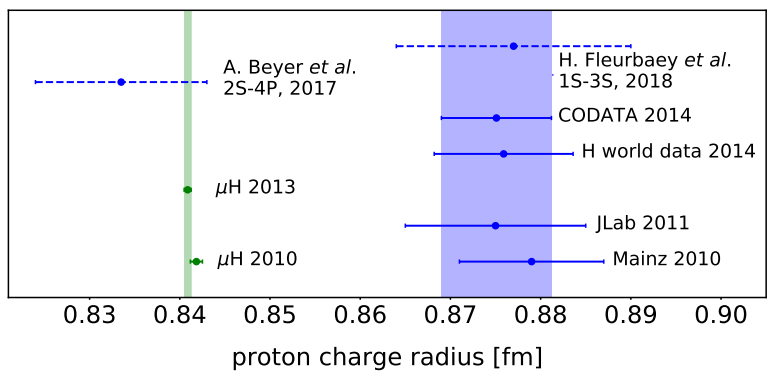} 
 \caption{The present status of the proton radius puzzle:  proton charge radius determinations from $\mu {\rm H}$~\cite{Pohl:2010zza,Antognini13} in comparison to the 
   CODATA-2014 evaluation~\cite{CODATA_2014}, data taken in Mainz~\cite{Bernauer:2010hp} and at JLab~\cite{Zhan:2011hp} for electron scattering off the proton and results from world average ordinary hydrogen spectroscopy~\cite{CODATA_2014}. The latest hydrogen spectroscopy measurements by Beyer {\it et al.}~\cite{Beyer79} and Fleurbaey {\it et al.}~\cite{Paris} are also shown.}
 \label{ppuzzle}
\end{figure}

To date, no commonly accepted explanation  exists. Very recently, two new measurements were performed based on spectroscopy of ordinary hydrogen, leading yet again to two contradicting results: the Garching experiment measured the $2S$--$4P$ transition frequency in $e$H yielding a small radius $r_p$ = 0.8335(95) fm~\cite{Beyer79} compatible with muonic hydrogen, while the Paris experiment examined $1S$--$3S$  transition frequency in $e$H obtaining  $r_p$=0.877(13) fm~\cite{Paris}, in very good agreement with the current CODATA-recommended value. The present situation with all the above mentioned results is depicted in Fig.~\ref{ppuzzle}.
 While this picture may suggest that systematic uncertainties in the various experiments need to be revisited, it is fair to say that the proton radius puzzle is yet to be solved and further investigations are needed.

 To understand this discrepancy, new experiments have been proposed to measure precisely the electron-proton scattering at low  momentum transfer down to $Q^2 \sim 10^{-4}~(\textrm{GeV/c})^2$~\cite{JLab-E1211106,PRad2013,MAMI2013} and to investigate the low-$Q^2$ muon-proton elastic scattering in the MUSE experiment~\cite{Gilman:2013eiv,MUSE2013}.
 An alternative approach is to study the mean-square charge radii $r_{nucl}$ of other light nuclei by measuring Lamb shifts in muonic atoms with different nuclear charges or mass numbers, such as muonic hydrogen isotopes ($\mu^2{\rm H}$ and $\mu^3{\rm H}$) and muonic helium ions ($\mu^3{\rm He}^+$ and $\mu^4{\rm He}^+$). Through a systematic comparison between $r_{nucl}$ extracted from experiments involving, respectively, electron-nucleus and muon-nucleus systems, one can test whether the discrepancy persists or is enhanced in systems with different number of protons $Z$,  number of neutrons $N$, or different mass number $A=Z+N$. The CREMA collaboration at PSI has started to perform a series of Lamb shifts experiments in light muonic atoms~\cite{Antognini11}. Results on $\mu^2{\rm H}$ lead to the discovery of a deuteron-radius puzzle~\cite{Pohl669}. Results on helium isotopes will be released in the near future.

\begin{figure}[htb]
\centerline{
\includegraphics[angle=0,scale=0.35]{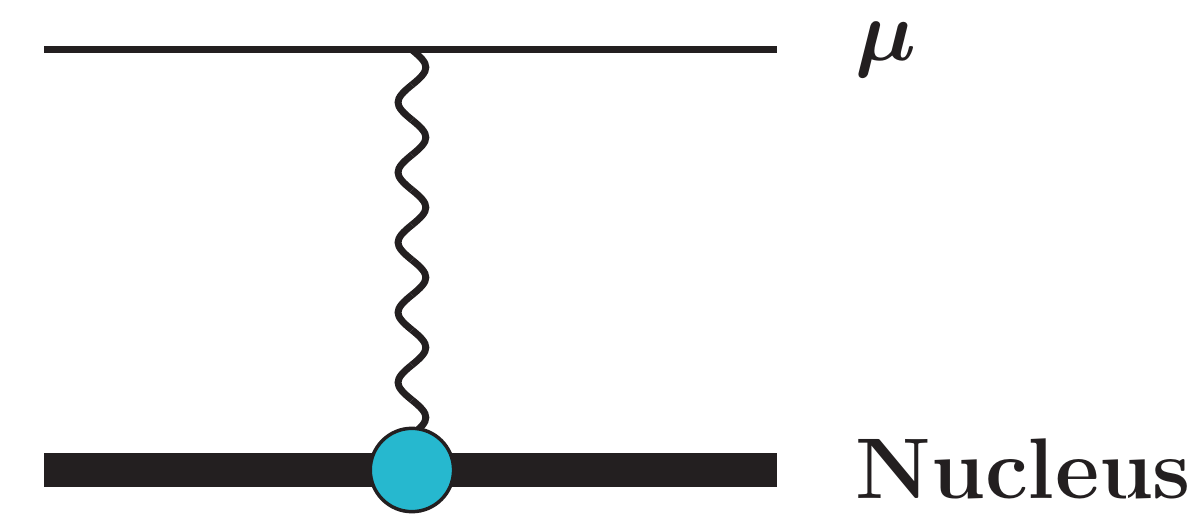}}
\caption{
The muon-nucleus one-photon exchange.
} 
\label{fig:ope} 
\end{figure}

 In the Lamb shift measurements, the accuracy in determining $r_{nucl}$ relies not only on the experimental precision, but also on how accurately one can calculate quantum electro-dynamics (QED) and nuclear-structure corrections.
   In light muonic atoms, unlike their electronic counter parts, QED corrections to the Lamb shifts are dominated by vacuum polarization rather than vertex corrections and the level ordering of the 2$S$ and 2$P$ states is reversed. Owing to the heavier mass, the muon orbits much closer to the nucleus than does the electron, thus nuclear-structure corrections are considerably larger than in electronic atoms~\cite{Borie:1982ax, Borie:2012zz}.
  The Lamb shift $\delta_{\rm LS}$ in a muonic atom/ion with nuclear charge $Z$ can be generally related to the charge radius of a nucleus $r_{nucl}$ (in units of $\hbar=c=1$) by   
\begin{equation} \label{eq:E2s2p}
\delta_{\rm LS} = \delta_{\rm QED} +  
                \mathcal{A}_{\rm OPE}\, r^2_{nucl}  +\delta_{\rm TPE}, 
\end{equation}
where the $\delta_{\rm QED}$ term is composed mainly of QED photon vacuum polarization, muon self energy, and  relativistic recoil corrections, whose dominant effect, i.e., the Uehling term, is of order $\alpha(Z\alpha)^2$~\cite{Eides2001} with $\alpha$ denoting the fine-structure constant. 
Beyond the leading contribution, various QED corrections of higher orders (e.g., up to $(Z\alpha)^6$, $\alpha^2(Z\alpha)^4$, $\cdots$) have been calculated by many groups to very good accuracy (see Refs.~\cite{Borie:2012zz, Eides2001} for reviews).
The other two terms in Eq.~\eqref{eq:E2s2p} are  nuclear-structure corrections.
The term proportional to $r^2_{nucl}$ is dominated by the exchange of one photon between the muon and the nucleus (Fig.~\ref{fig:ope}), where the nuclear electric form factor is inserted into the photon-nucleus vertex. Such dominant effect determines the coefficient $\mathcal{A}_{\rm OPE} \approx m_r^3 (Z\alpha)^4/12$,
where $m_r=m_{\mu} M_A /(m_{\mu}+M_A)$ is the reduced mass in the muon-nucleus center of mass system, with the nuclear mass denoted by $M_A$. Higher-order corrections to $\mathcal{A}_{\rm OPE}$ from relativistic, QED and nuclear finite-size effects have been calculated to great accuracy (see Refs.~\cite{Borie:2012zz,Friar:1978wv} for reviews).

\begin{figure}[htb]
\centerline{
\includegraphics[angle=0,scale=0.35]{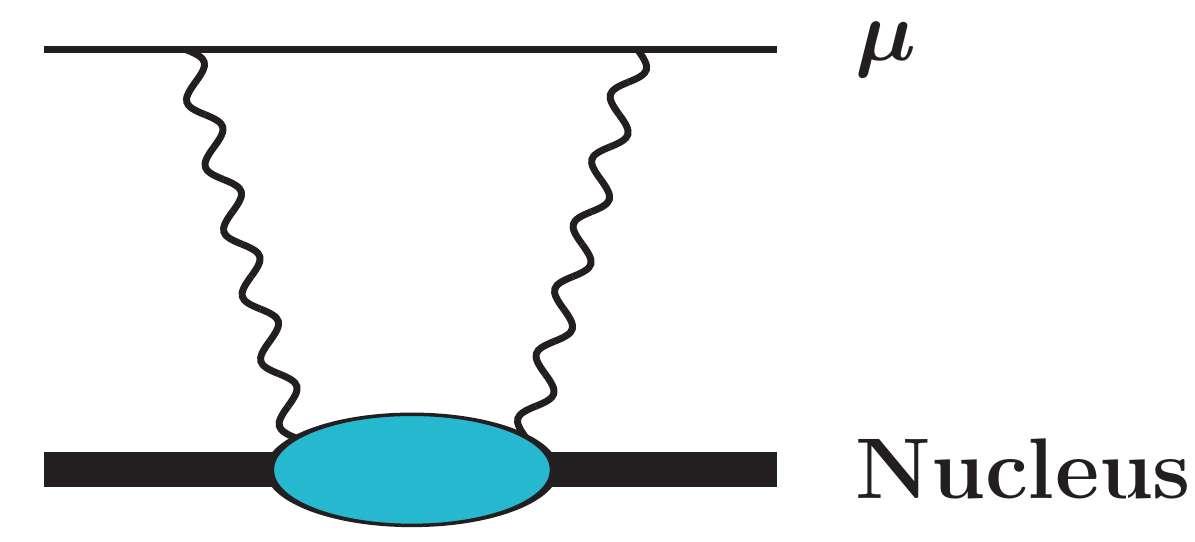}}
\caption{
The muon-nucleus two-photon exchange.
} 
\label{fig:tpe} 
\end{figure}

The $\delta_{\rm TPE}$, which is of order $(Z\alpha)^5$, originates from the two-photon exchange (TPE) contribution (Fig.~\ref{fig:tpe}) and can be separated into elastic and inelastic parts, $\delta_{\rm TPE}=\delta_{\rm Zem}+\delta_{\rm pol}$. The elastic part $\delta_{\rm Zem}$ was derived by Friar as the dominant nuclear finite-size effect~\cite{Friar:1978wv}. $\delta_{\rm Zem}$ is proportional to the third electric Zemach moment~\cite{Zemach:1956zz}, also called Friar moment, which is expressed as an integral of the nuclear charge density $\rho_E(\bs{R})$: 
\begin{equation}
\label{eq:zemach3}
\delta_{\rm Zem} = - \frac{m_r^4}{24}(Z\alpha)^5 \iint d^3{R} d^3{R}' \left|\bs{R}-\bs{R}'\right|^3 \rho_E(\bs{R})\rho_E(\bs{R}').
\end{equation}
The inelastic part $\delta_{\rm pol}$ is called the nuclear polarizability and reflects the excitation and deexcitation of the nucleus/nucleon through two-photon-exchange interaction with the muon shown in Fig.~\ref{fig:tpe}.
Due to the energy-scale separation between the nuclear and the nucleon excitation energies, $\delta_{\rm pol}$ can be further separated into a nuclear contribution $\delta^{A}_{\rm pol}$ related to the few-nucleon dynamics and a hadronic part $\delta^{N}_{\rm pol}$, related to the intrinsic nucleon dynamics. Their effects can be studied independently using effective theories at different scales.

To  understand the physical meaning of $\delta^A_{\rm pol}$, one can naively imagine that the protons are pulled away from the nuclear center of mass due to the Coulomb attractions to the lepton, thus generating mostly nuclear  dipole-excited states. Such a distorted charge distribution then tries to follow the orbiting lepton, similar to Earth's equipotential tidal bulges lagging behind  the Moon~\cite{Friar:2013rha}.

The spectroscopic measurements of Lamb shift $\delta_{\rm LS}$ can reach very high
accuracy, and so can the calculation of $\delta_{\rm QED}$. Therefore, a key ingredient for extracting $r_{nucl}$ from Eq.~\eqref{eq:E2s2p} is the accurate determination of $\delta_{\rm TPE}$.
Ab initio nuclear-structure calculations of $\delta_{\rm TPE}$ have already  impacted  this field, as we shall present in this review. A precision of the order of a few percent can be reached, which is presently  better than any other method based on phenomenology or experimental extractions of the $\delta_{\rm TPE}$ contribution.
\begin{table}[htb]
\centering
\caption{Experimental uncertainty in the measured Lamb-shift energy of muonic atoms  
  compared to the theoretical uncertainty in  ab initio calculations of $\delta_{\rm TPE}$. Data taken from Refs.~\cite{Antognini13,Pohl669,Krauth:2015nja,Franke:2017tpc,Hernandez2018,Nevo_Dinur_2016,Ji13,Diepold:2016cxv}.
}
\label{tab:1}
\footnotesize
\renewcommand{\tabcolsep}{1.0mm}
\begin{tabular}{l l l}
\hline \hline \noalign{\smallskip}
&Experiment &  Theory\\
\noalign{\smallskip}\hline\noalign{\smallskip}
$\mu{\rm H}$                 & 2.3 $\mu$eV  & 2 $\mu$eV  \\
$\mu^2{\rm H}$                  & 0.034 meV  &  0.05 meV \\ 
$\mu ^{3} \text{He}^{+}$ & 0.08 meV     &   0.4 meV\\ 
$\mu ^{4} \text{He}^{+}$ & 0.06 meV     &  0.4 meV\\ 

\noalign{\smallskip}\hline\hline
\end{tabular}
\end{table}
To appreciate the importance of determining nuclear-structure corrections and reducing their uncertainties,  the experimental uncertainty in the Lamb shift energy measurements is compared in Table~\ref{tab:1} to the theoretical uncertainties in  $\delta_{\rm TPE}$. One can see that for the $\mu{\rm H}$ case both uncertainties are of the same order of magnitude. However, for $\mu^2{\rm H}$, $\mu ^3$He$^+$ and  $\mu ^4$He$^+$ the ratio between them is dramatically increased. This indicates, that for light muonic atoms TPE corrections constitute the real bottleneck to exploit the experimental precision in the extraction of the charge radius. It is important to note that ab initio nuclear-structure calculations performed for muonic atoms from $\mu^2{\rm H}$ to $\mu ^4$He$^+$ have so far provided the most precise determination of $\delta_{\rm TPE}$, substantially reducing the uncertainties with respect to other methods and approaches. Moreover, regardless of the source of the proton radius discrepancy, the TPE correction is a necessary theoretical input that determines the attainable precision of nuclear charge radii extracted from spectroscopic measurements of muonic atoms.

The purpose of this review is to present a thorough derivation of the formalism used to calculate $\delta_{\rm TPE}$ with ab initio methods
and to compare our recent results to other approaches, emphasizing the  reduction in uncertainty  obtained by using first principle nuclear physics techniques.

The  review is structured as follows.  Section~\ref{theory} will be dedicated to the theoretical formalism.  In Section~\ref{sec:numerics} we briefly outline the few-body methods used in our computations and in Section~\ref{sec:uncert} we explain how we estimate theoretical uncertainties. Finally, in Section~\ref{sec:results} we discuss our results in the context of other approaches and of the newly risen experimental questions, before drawing conclusions in Section~\ref{sec:conclude}.

\section{Theoretical formulation}
\label{theory}

\subsection{Summary of formulas}
For readers interested only in the final expressions of the formulas, we present here a prescription for computing nuclear-structure corrections to the $2S$ state energy of a hydrogen-like muonic atom (or ion), in which a single muon orbits a nucleus ${}^A_Z{\rm X}$ with charge number $Z$ and mass number $A$. The $2P$ state is less influenced by the nucleus, due to the fact that the muon $2P$ wave function overlaps much less with the nucleus.

The entire two-photon exchange contribution in a muonic atom, $\delta_{\rm TPE}$, contains corrections from the nucleus structure $\delta_{\rm TPE}^{A}$ and the intrinsic nucleon structure $\delta_{\rm TPE}^N$, each of which is further separated into elastic component (Zemach contribution) and inelastic one (polarizability). Therefore, these four contributing terms are categorized in two ways:
\numparts
\begin{eqnarray}
\label{eq:TPE-AN}
\delta_{\rm TPE} &=& \delta_{\rm TPE}^{A} + \delta_{\rm TPE}^N
= \left[\delta_{\rm pol}^A + \delta_{\rm Zem}^A \right] + \left[\delta_{\rm pol}^N + \delta_{\rm Zem}^N \right] {,}
\\
\label{eq:TPE-FP}
\delta_{\rm TPE} &=& \delta_{\rm Zem}\, + \delta_{\rm pol} \;\;
= \left[\delta_{\rm Zem}^A + \delta_{\rm Zem}^N \right] + \left[\delta_{\rm pol}^A + \delta_{\rm pol}^N \right] {.}
\end{eqnarray}
\endnumparts
The nuclear polarizability, $\delta_{\rm pol}^{A}$,  consists of four major contributions: non-relativistic $\delta_{\rm pol}^{\rm NR}$ (Section~\ref{sec:polar-nonrel}), Coulomb distortion $\delta_{\rm pol}^{\rm C}$ (Section~\ref{sec:Coul-corr}), relativistic $\delta_{\rm pol}^{\rm R}$ (Section~\ref{sec:relativstic}), and nucleon-size $\delta_{\rm pol}^{\rm NS}$ (Section~\ref{sec:NS-correct}) corrections. The four parts of $\delta_{\rm pol}^{A}$, together with  $\delta_{\rm Zem}^{A}$, are further divided into smaller fragments, which are shown in the square brackets of Eqs.~(\ref{eq:Pol-A}, \ref{eq:Zem-A}).
\numparts
\begin{eqnarray}
\label{eq:Pol-A}
\delta_{\rm pol}^A &=& \delta_{\rm pol}^{\rm NR} + \delta_{\rm pol}^{\rm C} + \delta_{\rm pol}^{\rm R} + \delta_{\rm pol}^{\rm NS}
\nn
&=&\left[ \delta^{(0)}_{D1} + (\delta^{(1)}_{R3} +\delta^{(1)}_{Z3}) +(\delta^{(2)}_{R^2} +\delta^{(2)}_{Q} +\delta^{(2)}_{D1D3}) \right]
+\left[\delta^{(0)}_C\right]
\nn
&&+\left[ \delta^{(0)}_{L}+\delta^{(0)}_{T}+\delta^{(0)}_{M} \right]
+\left[ \delta^{(1)}_{R1}+\delta^{(1)}_{Z1} + \delta^{(2)}_{NS} \right] {,}
\\
\label{eq:Zem-A}
\delta_{\rm Zem}^A &=&  -\left[\delta^{(1)}_{Z3}+\delta^{(1)}_{Z1}\right] {.}
\end{eqnarray}
\endnumparts

Each term in Eqs.~\eqref{eq:Pol-A} and \eqref{eq:Zem-A} will be explained in the following sub-sections. Here, we list the expressions for calculating each term: 
\numparts
\begin{eqnarray}
\label{eq:sum-d0}\fl
\delta^{(0)}_{D1} 
&=& -\frac{16\pi^2}{9}(Z\alpha)^2 \phi^2(0) \int^\infty_{0} d\omega 
    \sqrt{\frac{2m_r}{\omega}} S_{D_1}(\omega) {,}
\\
\label{eq:sum-R3}\fl
\delta^{(1)}_{R3} 
&=& -\frac{\pi}{3} m_r (Z\alpha)^2\phi^2(0) 
    \iint d^3 R d^3 R' |\bs{R}-\bs{R}'|^3 \rho_0^{pp}(\bs{R},\bs{R}) {,}
\\
\label{eq:sum-Z3}\fl
\delta^{(1)}_{Z3} 
&=& \hspace{3ex} \frac{\pi}{3} m_r (Z\alpha)^2\phi^2(0) 
    \iint d^3 R d^3 R' |\bs{R}-\bs{R}'|^3 \rho_0^{p}(\bs{R})\rho_0^{p}(\bs{R}') {,}
\\
\label{eq:sum-R2}\fl
\delta^{(2)}_{R^2} 
&=& \hspace{3ex} \frac{4\pi}{9}  m_r^2 (Z\alpha)^2\phi^2(0)  \int^\infty_{0} d\omega 
    \sqrt{\frac{\omega}{2m_r}} S_{R^2}(\omega)  {,}
\\
\label{eq:sum-Q2}\fl
\delta^{(2)}_{Q} 
&=& \hspace{3ex} \frac{64}{225} \pi^2 m_r^2 (Z\alpha)^2\phi^2(0)  \int^\infty_{0} d\omega
    \sqrt{\frac{\omega}{2m_r}} S_{Q}(\omega)  {,}
\\
\label{eq:sum-D1D3}\fl
\delta^{(2)}_{D1D3} 
&=& -\frac{64}{45} \pi^2 m_r^2 (Z\alpha)^2\phi^2(0)  \int^\infty_{0} d\omega
    \sqrt{\frac{\omega}{2m_r}}  S_{D_1 D_3}(\omega) {,}
\\
\label{eq:sum-C0}\fl
\delta_C^{(0)}
&=& -\frac{16\pi^2}{9} (Z\alpha)^3\phi^2(0)  \int_{0}^{\infty} d\omega
    \frac{m_r}{\omega} \ln\frac{2(Z\alpha)^2 m_r}{\omega}\, S_{D_1}(\omega) {,}
\\
\label{eq:sum-L0}\fl
\delta_{L}^{(0)} 
&=& \hspace{3ex} \frac{32\pi}{9} (Z\alpha)^2 \phi^2(0) \int^{\infty}_{0} d\omega\,
    \mathcal{F}_L(\omega/m_r) S_{D_1} (\omega){,}
\\
\label{eq:sum-T0}\fl
\delta_{T}^{(0)} 
&=& \hspace{3ex} \frac{16\pi}{9} (Z\alpha)^2 \phi^2(0) \int^{\infty}_{0} d\omega\,
    \mathcal{F}_T (\omega/m_r) S_{D_1}(\omega)  {,}
\\
\label{eq:sum-M0}\fl
\delta_{M}^{(0)} 
&=& \hspace{3ex} \frac{1}{3 m_p^2} (Z\alpha)^2  \phi^2(0) \int^{\infty}_{0} d\omega\,
    \mathcal{F}_M (\omega/m_r ) S_{M_1} (\omega)  {,}
\\
\label{eq:sum-R1}\fl
\delta^{(1)}_{R1} 
&=& -8\pi m_r (Z\alpha)^2 \phi^2(0)        
    \int\int d^3 R d^3 R' |\bs{R}-\bs{R}'| 
    \left[\frac{2}{\beta^2}\rho_0^{pp}(\bs{R},\bs{R}')-\lambda\rho_0^{np}(\bs{R},\bs{R}')\right] {,}
\\
\label{eq:sum-Z1}\fl
\delta^{(1)}_{Z1} 
&=& \hspace{3ex} 8\pi m_r (Z\alpha)^2 \phi^2(0)        
    \iint d^3 R d^3 R' |\bs{R}-\bs{R}'| \rho_0^p(\bs{R})
    \left[\frac{2}{\beta^2} \rho_0^p(\bs{R}')-\lambda\rho_0^n(\bs{R}')\right] {,}
\\
\label{eq:sum-NS2}\fl
\delta^{(2)}_{NS} 
&=& -\frac{128}{9}\pi^2 m_r^2 (Z\alpha)^2 \phi^2(0) \left[\frac{2}{\beta^2}+\lambda\right]
    \int^\infty_{0} d\omega \sqrt{\frac{\omega}{2m_r}} S_{D_1}(\omega) {.}
\end{eqnarray}
\endnumparts
Here, $\phi^2(0)=(m_r Z\alpha)^{3}/8\pi$ is the norm of the muonic $2S$-state wave function. The parameters $\beta$ and $\lambda$ are defined as $\beta=\sqrt{12/r_p^2}$ and $\lambda = -r_n^2/6$, with $r_p$ and $r_n$ denoting the proton and neutron charge radius. The expressions above are in general energy-weighted integrations of nuclear electromagnetic response functions, called sum rules. Besides power-law and logarithmic energy weights, expressions of more complicated ones, {i.e.}, $\mathcal{F}_L$, $\mathcal{F}_T$, and $\mathcal{F}_M$, are given respectively in Eqs.~(\ref{eq:mathFL}, \ref{eq:mathFT}, \ref{eq:mathFM}). A response function $S_{O}$ is defined as
\begin{equation}
\label{eq:response-O}
S_{O}(\omega) = \frac{1}{2J_0+1} \SumInt \limits_{N\neq N_0,J} |\bra N_0 J_0 ||\hat O ||N J\ket|^2 \delta(\omega-\omega_N) {,}
\end{equation}
where $\SumInt$ indicates the sum of nuclear excited states (both discrete and continuum), $\hat O$ is an electromagnetic operator, and $\bra N_0 J_0 ||\hat O || N J\ket$ denotes a reduced matrix element. $\omega_N$ is the excitation energy between nuclear states $|NJ\ket$ and $|N_0 J_0\ket$. Readers can find the specific response functions $S_{D_1}$ in Eq.~\eqref{eq:d1-response}, $S_{R^2}$ in Eq.~(\ref{eq:S-R2}), $S_{Q}$ in Eq.~\eqref{eq:S-Q}, $S_{D_1 D_3}$ in Eq.~\eqref{eq:S-D1D3}, and $S_{M_1}$ in Eq.~\eqref{eq:S-MSL}.

The one- and two-body point-nucleon densities are defined by
\numparts
\begin{eqnarray}\label{eq:rhoc_0}
\rho_0^{c}(\bs{R}) &=&
      \bra N_0 |\frac{1}{Z}
                \sum_{a}^A\delta(\bs{R}-\bs{R}_a)
                \hat{e}_{c,a}
      | N_0 \ket {,}
      \\
 \label{eq:rhocc_0}
 \rho_0^{cc'}(\bs{R},\bs{R}') &=&
      \bra N_0 |\frac{1}{Z^2}
                \sum_{ab}^A\delta(\bs{R}-\bs{R}_a)\delta(\bs{R}'-\bs{R}_b)
                \hat{e}_{c,a}\hat{e}_{c',b}
      | N_0 \ket {,}
\end{eqnarray}
\endnumparts
where $c,c'$ equals $p$ or $n$. $\hat e_{p,a}$ and $\hat e_{n,a}$ are the proton and neutron projection isospin operators, defined as
\begin{equation}
\label{eq:e_pn}
\hat e_{p,a} = (1+\hat{\tau}_{z,a})/2,\quad \quad
\hat e_{n,a} = (1-\hat{\tau}_{z,a})/2~,
\end{equation}
where $\hat{\tau}_{z,a}=\pm 1$ with the sign determined by the $a$th nucleon being a proton or neutron.

Eq.~\eqref{eq:sum-d0}, calculated in Section~\ref{sec:NR0}, represents the leading contribution to the non-relativistic polarizability effect $\delta_{\rm pol}^{\rm NR}$. Eqs.~(\ref{eq:sum-R3}, \ref{eq:sum-Z3}) are the sub-leading corrections to $\delta_{\rm pol}^{\rm NR}$, given in Section~\ref{sec:NR1}. Eqs.~(\ref{eq:sum-R2}, \ref{eq:sum-Q2}, \ref{eq:sum-D1D3}) form the sub-sub-leading contributions to $\delta_{\rm pol}^{\rm NR}$, provided in Section~\ref{sec:NR2}. Eq.~\eqref{eq:sum-C0} represents the Coulomb-distortion correction, given in Section~\ref{sec:Coul-corr}. Eqs.~(\ref{eq:sum-L0}, \ref{eq:sum-T0}, \ref{eq:sum-M0}) are the relativistic corrections, which are derived in Section~\ref{sec:relativstic}. Eqs.~(\ref{eq:sum-R1}, \ref{eq:sum-Z1}) and Eq.~\eqref{eq:sum-NS2} show the leading and subleading nucleon-size effects, which can be found in Section~\ref{sec:NS-correct}.

%================================================================
%================================================================

\subsection{Non-relativistic calculations}\label{sec:polar-nonrel}

The muonic atom (or ion) is a hydrogen-like system consisting of a muon and a nucleus.
The non-relativistic Hamiltonian of the muonic atom has three components, {i.e.}, the nuclear Hamiltonian $H_{\rm nucl}$, the muon Hamiltonian $H_{\mu}$, and the nuclear-structure correction $\Delta H$:
\begin{equation}
\label{eq:H-muA}
H = H_{\rm nucl} +H_\mu + \Delta H {.}
\end{equation}
$H_{\rm nucl}$ describes the internal structure of a nucleus ${}_Z^A{\rm X}$. It is written in terms of nucleon degrees of freedom, and the nuclear potential is represented by two- and three-nucleon interactions. We use a shorthand notation $|N\ket$ to denote the $N$th nuclear eigenstate:
\begin{equation}\label{eq:H-nucl}
H_{\rm nucl}|N\ket = E_N|N\ket {,}
\end{equation}
with $E_N$ indicating the corresponding eigenenergy. $|N\ket$ refers to both discrete and continuum states, whose quantum numbers, such as the total angular momentum $J$ and its $z$-component $M$, are omitted for simplicity. In later cases, we also refer to $|N\ket$ as $|NJ\ket$ or $|NJM\ket$, when specific quantum numbers are required. For the ground state, whose energy, total angular momentum, and $z$-component are respectively $E_{N_0}$, $J_0$ and $M_0$, we refer to it as $|N_0\ket$, $|N_0 J_0\ket$ or $|N_0 J_0 M_0\ket$.

$H_\mu$ is the muon Hamiltonian. The muon is bound to a point-like nucleus by an attractive Coulomb interaction. In the non-relativistic limit, $H_\mu$ is written as
\begin{equation}
\label{eq:H-mu}
H_{\mu} = \frac{q^2}{2m_r} - \frac{Z\alpha}{r} {,}
\end{equation}
where $q$ and $r$ are the relative momentum and distance between the muon and the nucleus.
The eigenenergy $\epsilon_{\mu}$ of the Hamiltonian $ H_{\mu}$, leads to the unperturbed (not necessarily ground-state) atomic spectrum; while $|\mu\ket$ indicates the corresponding eigenstate. When atomic quantum numbers need to be specified, we use a full notation $|{\mu}_{n\ell m}\ket$ (or $|{\mu}_{n\ell}\ket$) to specify the principle ($n$), orbital ($\ell$) and magnetic ($m$) quantum numbers of an atomic state, whose coordinate-space representation is, 
\begin{equation}
\label{eq:mu-nlm}
\bra \bs r| {\mu}_{n\ell m}\ket = \phi_n(0) \left(\frac{4\pi}{2\ell+1}\right)^{1/2} R_{n\ell}(r)\; Y_{\ell m}(\hat r) {,}
\end{equation}
where $\phi_n(0)$ is the wave-function normalization constant. Since we focus on the Lamb shift in muonic atoms, we take only $n=2$ in this article, and drop the subscript in $\phi_2(0)$ for simplicity. So we have $\phi(0)=\nu^{3/2}/\sqrt{\pi}$, with $\nu=m_r Z\alpha/2$. The unperturbed atomic energy, $\epsilon_{\mu} = -m_r (Z \alpha)^2/8$, is degenerate in unperturbed $2S$ and $2P$ states, whose radial functions are respectively
\numparts
\begin{eqnarray}
\label{eq:R20}
R_{20}(r) &=& (1-\nu r) \exp\left(-\nu r\right) {,}
\\
\label{eq:R21}
R_{21}(r) &=& \nu r \exp\left(-\nu r\right) {.}
\end{eqnarray}
\endnumparts

$\Delta H$ in Eq.~\eqref{eq:H-muA} describes the correction to the muon-nucleus point Coulomb interaction from the charge distribution of the nucleus. In this section, the nucleus is approximated by a system consisting of point-like protons (charge $1$) and neutrons (charge $0$). Therefore, $\Delta H$ represents the sum of Coulomb interactions of the muon with each individual proton, located at a position $\bs R_a$ (or distance $R_a=|\bs{R_a}|$) from the nuclear center of mass, subtracted by the point Coulomb potential,
\begin{equation}
\label{eq:DVrR}
\eqalign{
\Delta H = \sum\limits_{a}^Z \Delta V(\bs{r},\bs{R}_a) {,}
\\ 
\Delta V(\bs{r},\bs{R}) = -\alpha\left(\frac{1}{|\bs{r}-\bs{R}|}-\frac{1}{r}\right) {.}
}
\end{equation}
The function $\Delta V$ is localized around the nucleus, and vanishes when the limit $r\gg R$ is approached. Eq.~\eqref{eq:DVrR} does not account for the internal nucleonic structure, whose correction to $\Delta H$ enters at higher orders. We will discuss the finite nucleon-size correction in Section~\ref{sec:NS-correct}.

Since $\Delta H$ scales with $Z\alpha$, which is small in light muonic atoms, we use perturbation theory to evaluate $\Delta H$'s correction to the muonic atom spectrum. The $Z\alpha$-dependence of $|\mu\ket$ and $\Delta H$ indicates that nuclear-structure corrections from the $k$th-order perturbation theory are sized with $(Z\alpha)^{k+3}$ for the $2S$ state and $(Z\alpha)^{k+5}$ for the $2P$ state.
The first-order perturbation theory leads to the expectation value $\bra N_0 \mu |\Delta H |N_0 \mu\ket$, and is represented by the muon-nucleus one-photon exchange process depicted in Figure~\ref{fig:ope}. As derived in~\cite{Friar:1978wv}, its contribution to the $2S$ state is approximately $m_r^3(Z\alpha)^4 r_{\rm nucl}^2/12$; while corrections to the $2P$-state are of order $(Z\alpha)^6$.

In this article, we limit the discussions to second-order perturbation theory. It is of order $(Z\alpha)^5$, and is characterized by the two-photon exchange process depicted in Figure~\ref{fig:tpe}.   
Depending on whether the nucleus remains in the ground state between the exchanged photons, corrections are further divided into elastic and inelastic parts.
The nuclear-elastic part is called {\it nuclear finite-size effect}, which was calculated by Friar~\cite{Friar:1978wv}~\footnote{Using perturbation theory up to third order, Friar calculated in \cite{Friar:1978wv} the nuclear finite-size effect through order $(Z\alpha)^6$.}. In the two-photon exchange process, the elastic part corresponds to the elastic nuclear Zemach contribution $\delta_{\rm Zem}^{A}$.
The nuclear-inelastic part is named {\it nuclear polarizability effect}, $\delta_{\rm pol}^{A}$, for which the nucleus is excited by absorbing a photon and is then de-excited by subsequently emitting another photon.

The nuclear polarizability $\delta_{\rm pol}^{A}$ is calculated in second-order perturbation theory by
\begin{equation}
\label{eq:dpol-00}
\delta_{\rm pol}^{A} = \bra N_0 {\mu} | \Delta H\, G\, \Delta H | N_0 {\mu}\ket {,}
\end{equation}
where $G$ is the Green's function represented in a complete basis except for the nuclear ground state, {i.e.}, $1 -|N_0\ket \bra N_0|$. Using closure, $G$ is given by 
\begin{equation}
\label{eq:G}
G = -\SumInt\limits_{N\neq N_0} \frac{|N\ket \bra N|}{ H_\mu+\omega_N-\epsilon_{\mu}} {,}
\end{equation}
where $\SumInt$ indicates the summation of both discrete and continuum nuclear states, and $\omega_N = E_N -E_{N_0}$.

As is proven in Section~\ref{sec:Coul-corr}, $\delta_{\rm pol}^{A}$ contributes equally to the two hyperfine states associated with the $2S$ state. Here we simply set the muonic-atom state $|N_0 \mu\ket$ with nuclear and muonic parts decoupled, {i.e.}, $|N_0 \mu\ket = |N_0\ket |\mu\ket$\footnote{In Section~\ref{sec:Coul-corr}, we use instead the nuclear-muonic coupled scheme to derive the Coulomb distortion corrections to $\delta_{\rm pol}^{A}$.}.
By substituting Eqs.~\eqref{eq:DVrR} and \eqref{eq:G} into Eq.~\eqref{eq:dpol-00}, and using closure in muon's coordinate-space, we have 
\begin{eqnarray}
\label{eq:dpol-eqn}
\delta_{\rm pol}^{A} &=& -\sum_{ab}^{Z} \SumInt_{N\neq N_0} \iint d^3r d^3 r' \bra N_0 | \Delta V(\bs{r},\bs{R}_a)|N\ket \bra {\mu}|\bs{r}\ket 
\nn
&&\times\bra \bs{r} |\frac{1}{H_\mu +\omega_N-\epsilon_{\mu}} | \bs{r}'\ket \bra \bs{r}' |{\mu}\ket \bra N| \Delta V(\bs{r}',\bs{R}_b) | N_0 \ket {.}
\end{eqnarray}

It is useful to define the point-proton transition density function
\begin{equation}
\label{eq:rhop_N}
\rho_N^{p}(\bs{R}) = \bra N | \frac{1}{Z}\sum_{a}^{A} \delta(\bs{R}-\bs{R}_a)\hat{e}_{p,a}|N_0\ket {,}
\end{equation}
with $\hat{e}_{p,a}$ defined in Eq.~\eqref{eq:e_pn}. $\rho_N^{p}$ satisfies the sum rules:
\numparts
\begin{eqnarray}
\label{eq:rhoN-sum-1}
\int d^3 R\, \rho_N^{p}(\bs{R}) \hat{O}(\bs{R}) = \frac{1}{Z}\sum_a^Z \bra N | \hat{O}(\bs{R}_a) | N_0\ket {,} 
\\
\label{eq:rhoN-sum-2}
\int d^3 R\, \rho_N^{p}(\bs{R}) = \frac{1}{Z}\sum_a^Z \bra N | N_0\ket  =\delta_{N N_0}{,}
\end{eqnarray}
\endnumparts
where $\hat{O}(\bs{R})$ is an arbitrary operator, and Eq.~\eqref{eq:rhoN-sum-2} indicates the orthonormal condition in nuclear states. One special case of Eq.~\eqref{eq:rhoN-sum-1} is the ground-state point-proton density $\rho_0^{p}(\bs R)$ defined in Eq.~\eqref{eq:rhoc_0},
which is normalized by $\int d^3 R \rho_0^{p}(\bs{R})=1$. Another case is that
\begin{equation}
\label{eq:DV-rhoN}
\sum\limits_a^Z \bra N | \Delta V(\bs{r},\bs{R}_a) |N_0\ket = Z \int d^3 R\, \rho^p_N(\bs{R})\, \Delta V(\bs{r},\bs{R} ){.}
\end{equation}

By substituting Eq.~\eqref{eq:DV-rhoN} into Eq.~\eqref{eq:dpol-eqn}, we have
\begin{equation}\label{eq:rho-W-rho}
  \delta_{\rm pol}^{A} = \SumInt_{N\neq N_0} \int d^3 R d^3 R'
     \rho_N^{p*}(\bs{R}){W}(\bs{R},\bs{R}',\omega_N) \rho_N^{p}(\bs{R}') {,}
\end{equation}
where ${W}$ is the muon matrix element
\begin{eqnarray}
\label{eq:W-1}
{W}(\bs{R},\bs{R}',\omega_N) 
&=& -Z^2 \int d^3r d^3 r' \Delta V(\bs{r},\bs{R}) \bra {\mu}|\bs{r}\ket 
\nn
&&\times
\bra \bs{r}|\frac{1}{H_\mu +\omega_N-\epsilon_{\mu}} | \bs{r}'\ket \bra \bs{r}'
  |{\mu}\ket \Delta V(\bs{r}',\bs{R}') {.} 
\end{eqnarray}

In the small $Z\alpha$ expansion,  $\Delta V \propto \alpha$,  $\bra \bs{r} |\mu_{20}\ket \sim (Z\alpha)^{3/2}$ and $\bra \bs{r} |\mu_{21}\ket \sim (Z\alpha)^{5/2}$. Therefore, ${W}$ in the atomic $2S$ state is of order $(Z\alpha)^5$; while that in the $2P$ state is of order $(Z\alpha)^7$. By considering the dominant polarizability contribution, which is of order $(Z\alpha)^5$, we neglect the atomic $2P$ state, and calculate $\delta_{\rm pol}^{A}$ only in the $2S$ state. In Eq.~\eqref{eq:W-1}, we also omit $Z\alpha$-dependent pieces in $(H_\mu+\omega_N-\epsilon_\mu)^{-1}$, whose contribution to $\delta_{\rm pol}^A$ is above $(Z\alpha)^5$. These approximations yield
\begin{eqnarray}
\label{eq:approx-Zalpha}
\eqalign{
\bra \bs{r} |\mu_{20}\ket \rightarrow \phi(0)~,\quad&
\bra \bs{r} |\mu_{21}\ket \rightarrow 0 {,}
\\
H_\mu \rightarrow \frac{q^2}{2m_r}~,\quad&
\epsilon_{\mu}\rightarrow 0 {.}
}
\end{eqnarray}
By inserting Eq.~\eqref{eq:approx-Zalpha} into Eq.~\eqref{eq:W-1}, and using closure in muon's momentum space, we have
\begin{eqnarray}
\label{eq:W-DV}
\fl
{W}(\bs{R},\bs{R}',\omega_N) &=&
-\phi^2(0) Z^2 \int \frac{d^3 q}{(2\pi)^3} d^3r d^3 r' \Delta V(\bs{r},\bs{R}) 
     e^{i\bs{q}\cdot \bs{r}} \frac{1}{\frac{q^2}{2m_r}+\omega_N}
     e^{-i\bs{q}\cdot \bs{r}'} \Delta V(\bs{r}',\bs{R}')
\nn
\fl
&=&-\phi^2(0) Z^2 \int \frac{d^3 q}{(2\pi)^3} \Delta \tilde{V}(\bs{q},\bs{R}) 
       \frac{1}{\frac{q^2}{2m_r}+\omega_N} \Delta\tilde{V}^*(\bs{q},\bs{R}') {,}
\end{eqnarray}
where $\Delta \tilde{V}(\bs{q},\bs{R})$ is the Fourier transform of $\Delta
V(\bs{r},\bs{R})$
\begin{equation}
\label{eq:DVqR}
\Delta \tilde{V}(\bs{q},\bs{R}) = \int d^3 r \Delta V(\bs{r},\bs{R})
    e^{i\bs{q}\cdot\bs{r}} = \frac{4\pi\alpha}{q^2}
    \left(1-e^{i\bs{q}\cdot\bs{R}}\right) {.}
\end{equation}
By inserting Eq.~\eqref{eq:DVqR} into Eq.~\eqref{eq:W-DV} and integrating over $d \hat q$, we have
\begin{equation}
\label{eq:W1}
{W}(\bs{R},\bs{R}',\omega_N) = 
-16 m_r (Z\alpha)^2 \phi^2(0) \int_0^\infty \frac{dq}{q^2} \frac{1}{q^2+2m_r\omega_N}
I(q,\bs{R},\bs{R}') {,}
\end{equation}
where the function $I(q,\bs{R},\bs{R}')$ is given by
\begin{equation}
\label{eq:I-qrr}
I(q,\bs{R},\bs{R}') = 1-\frac{\sin(qR)}{qR}-\frac{\sin(qR')}{qR'}
      +\frac{\sin(q|\bs{R}-\bs{R}'|)}{q|\bs{R}-\bs{R}'|} {.}
\end{equation}
The first three terms in Eq.~\eqref{eq:I-qrr}, which are independent on either $R$ or $R'$, do not contribute to $\delta_{\rm pol}^A$, since they lead to terms in Eq.~\eqref{eq:rho-W-rho} proportional to $\int d^3 R \,\rho_N^{p}(\bs{R})=0$ when $N\neq N_0$. Therefore, the irreducible part of ${W}$ yields
\begin{equation}
\label{eq:W2}
\fl
{W}(\bs{R},\bs{R}',\omega_N) = -16 m_r (Z\alpha)^2 \phi^2(0) \int_0^\infty \frac{d
  q}{q^2} \frac{1}{q^2+2m_r\omega_N}\left[ \frac{\sin(q|\bs{R}-\bs{R}'|)}{q|\bs{R}-\bs{R}'|}-1  \right] {,} 
\end{equation}
where the constant $-1$ is added to cancel the divergence of the integrand at $q=0$.

After integrating over $q$, ${W}$ becomes 
\begin{equation}
\label{eq:W-full}
{W}(\bs{R},\bs{R}',\omega_N)  = -  \frac{\pi}{m_r^2} (Z\alpha)^2 \phi^2(0) \left(\frac{2m_r}{\omega_N}\right)^{3/2}
\frac{1}{\eta} \left( e^{-\eta} -1 +\eta - \frac{1}{2}\eta^2\right) {,}
\end{equation}
where $\eta \equiv \sqrt{2m_r \omega_N} |\bs{R} -\bs{R}'|$ is a dimensionless operator.
The quantity $|\bs{R} -\bs{R}'|$ indicates
the ``virtual'' distance that a proton inside the nucleus travels in the two-photon exchange process. We argue qualitatively using uncertainty principle, that $|\bs{R} -\bs{R}'|$ scales inversely with the momentum boost $Q$ of the traveling proton. It is thus roughly related to the nuclear excitation energy by $\omega_N \sim Q^2/2m_p\sim (2m_p)^{-1}|\bs{R} -\bs{R}'|^{-2}$, with $m_p$ denoting the proton mass.
Therefore, the parameter $\eta$ in Eq.~\eqref{eq:W-full} is approximately of order $\sqrt{m_r/m_p}\approx \sqrt{m_\mu/m_p} \ll 1$, and becomes a small parameter. 
Now we expand Eq.~\eqref{eq:W-full} in powers of $\eta$ and obtain 
\begin{eqnarray}
\label{eq:W-exp}
\fl
{W}(\bs{R},\bs{R}',\omega_N) &=& \frac{2\pi}{3} (Z \alpha)^2 \phi^2(0) \sqrt{\frac{2m_r}{\omega_N}}
\nn
\fl
&\times&
 \left[ |\bs{R}-\bs{R}'|^2 -\frac{1}{4} \sqrt{2m_r
  \omega_N} |\bs{R}-\bs{R}'|^3 + \frac{1}{10}  m_r \omega_N
|\bs{R}-\bs{R}'|^4 +\cdots~\right] {,} 
\end{eqnarray}
where dots indicate higher-order terms omitted in the expansion. The three terms in the square brackets yield the leading $(0)$, sub-leading $(1)$, and sub-sub-leading $(2)$ non-relativistic contributions to $\delta_{\rm pol}^{A}$ in the $\eta$-expansion. Each term is further decomposed by
\begin{eqnarray}
\label{eq:dpol-NR-expand}
\eqalign{
\delta_{\rm pol}^{\rm NR} &= \delta_{\rm NR}^{(0)} + \delta_{\rm NR}^{(1)} + \delta_{\rm NR}^{(2)}
\\
&= [\delta^{(0)}_{D1}]+[\delta^{(1)}_{Z3}+\delta^{(1)}_{R3}]+[\delta^{(2)}_{R^2}+\delta^{(2)}_{Q}+\delta^{(2)}_{D1D3}]
{.}
}
\end{eqnarray}
In the remaining part of this section, we explain the derivation of each term and evaluate their contributions.

%=============================================================================
\subsubsection{Leading non-relativistic contributions}\label{sec:NR0}

In the $\eta$-expansion, the leading piece in Eq.~\eqref{eq:W-exp} is proportional to $|\bs{R}-\bs{R}'|^2$. It is expanded in spherical-harmonic basis by 
\begin{equation}
\label{eq:R2-expan}
|\bs{R}-\bs{R}'|^2 \rightarrow -\frac{8\pi}{3} R R'\, Y_1(\hat R) \cdot Y_1(\hat R') {,}
\end{equation}
where $Y_l$ denotes the rank-$l$ spherical harmonic tensor, and the scalar product is defined by $Y_l \cdot Y_l = \sum_{m=-l}^{l} Y_{lm}^* Y_{lm}$. $R^2$ and $R'^2$ terms are dropped since they do not contribute to $\delta_{\rm pol}^{A}$ due to the orthogonality condition in Eq.~\eqref{eq:rhoN-sum-2}.
We separate Eq.~\eqref{eq:R2-expan} from the remaining pieces of $W$ in Eq.~\eqref{eq:W-exp}, and insert it into Eq.~\eqref{eq:rho-W-rho}. By doing so, we obtain the leading non-relativistic polarizability contribution $\delta^{(0)}_{\rm NR}$, which equals to an electric dipole polarization, $\delta^{(0)}_{D1}$:
\begin{eqnarray}
\label{eq:D1-full}
\delta^{(0)}_{D1} 
&=& -\frac{16\pi^2}{9}\alpha^2 \phi^2(0) 
\sum_{ab}^Z \SumInt_{N\neq N_0,JM} \sqrt{\frac{2m_r}{\omega_N}} 
\nn
&&\times
\bra N_0 J_0 M_0 |R_a Y_1(\hat R_a) |N J M\ket \cdot \bra N J M |R_b Y_1(\hat R_b)|N_0 J_0 M_0\ket {.}
\end{eqnarray}
Where the full notation of a nuclear state $| NJM\ket$ specifies the total angular momentum $J$ and its $z$-component $M$. For a given multipolarity-$l$ operator, $Y_l$, the allowed transition is constrained by $|J_0-l|\leq J \leq J_0+l$.

By defining an electric-dipole operator $\hat D_1 = \frac{1}{Z} \sum_a^Z R_a Y_1(\hat R_a)$, we rewrite Eq.~\eqref{eq:D1-full} based on the Wigner-Eckart theorem in Eq.~\eqref{eq:App-I2} as
\begin{equation}
\label{eq:delta_D1}
 \delta^{(0)}_{D1} = -\frac{16\pi^2}{9}(Z\alpha)^2 \phi^2(0) \int^\infty_{0} d\omega\, \sqrt{\frac{2m_r}{\omega}} S_{D_1}(\omega) {,}
\end{equation}
which is proportional to an electric dipole sum rule with an energy weight $\omega^{-1/2}$.
$S_{D_1}$ is the electric dipole response function, defined in terms of reduced matrix elements
\begin{equation}
\label{eq:d1-response}
S_{D_1}(\omega) = \frac{1}{2J_0+1} \SumInt \limits_{N\neq N_0,J} |\bra N_0 J_0 ||\hat D_1 ||N J\ket|^2 \delta(\omega-\omega_N) {.}
\end{equation}

%=============================================================================
\subsubsection{Sub-leading non-relativistic contributions}\label{sec:NR1}

$|\bs{R}-\bs{R}'|^3$ in Eq.~\eqref{eq:W-exp} leads to the part independent of $\omega_N$. Inserting this piece of ${W}$ into Eq.~\eqref{eq:rho-W-rho} yields the sub-leading non-relativistic contribution $\delta_{\rm NR}^{(1)}$:
\begin{equation}
\label{eq:dpol-NR1}
\delta^{(1)}_{\rm NR} =  -\frac{\pi (Z\alpha)^2}{3}m_r\phi^2(0) 
 \int d^3 R \int d^3 R'\,|\bs{R}-\bs{R}'|^3 \, 
     \SumInt_{N\neq N_0} \rho_N^{p*}(\bs{R})\rho_N^{p}(\bs{R}') {.}
\end{equation}
Using closure $\SumInt_{N\neq N_0} |N\ket\bra N|=1-|N_0\ket\bra N_0|$, the product-sum of point-proton transition density functions are separated into two ground-state expectation functions:
\begin{equation}
  \SumInt_{N\neq N_0} \rho_N^{p*}(\bs{R})\rho_N^{p}(\bs{R}') 
 = \rho_0^{pp}(\bs{R},\bs{R}')-\rho_0^{p}(\bs{R})\rho_0^{p}(\bs{R}') {,}
\end{equation}
where $\rho_0^{pp}$ is the point proton-proton correlation function defined in Eq.~\eqref{eq:rhocc_0}.

Therefore, the sub-leading contribution is separated into two parts, $\delta^{(1)}_{\rm NR} =  \delta^{(1)}_{R3}+ \delta^{(1)}_{Z3}$, which are defined respectively as
\numparts
\begin{eqnarray}
\label{eq:d1-R3}
\delta^{(1)}_{R3} &=& -\frac{\pi}{3} m_r (Z\alpha)^2 \phi^2(0) 
       \iint d^3 R d^3 R' |\bs{R}-\bs{R}'|^3 \,
       \rho_0^{(pp)}(\bs{R},\bs{R})  {,}
\\
\label{eq:d1-Z3}
\delta^{(1)}_{Z3} &=& \frac{\pi}{3} m_r (Z\alpha)^2 \phi^2(0) 
       \iint d^3 R d^3 R' |\bs{R}-\bs{R}'|^3 \,
       \rho_0^{p}(\bs{R})\rho_0^{p}(\bs{R}') {.}
\end{eqnarray} 
\endnumparts
$\delta^{(1)}_{R3}$ is zero for hydrogen isotopes, but becomes finite for a nucleus with more than one proton.
In the point-nucleon limit, the full charge distribution $\rho_E(\bs R)$ is then replaced by the point-proton density $\rho_0^{p}(\bs R)$. Therefore, $-\delta^{(1)}_{Z3}$ becomes exactly the elastic Zemach term, which is the elastic two-photon exchange contribution defined in Eq.~\eqref{eq:zemach3}. This cancellation was shown in Ref.~\cite{Friar:1997ma} for TPE contributions in hydrogen-like atoms, and was later applied by Refs. \cite{Friar:2013rha, Pachucki:2011xr} to muonic atoms. However, when the internal nucleon charge density is considered, higher-order corrections to $\delta^{(1)}_{Z3}$ need to be evaluated. This is done in Section~\ref{sec:NS-correct}.

%================================================================
\subsubsection{Sub-sub-leading non-relativistic contributions}\label{sec:NR2}

The $|\bs{R}-\bs{R}'|^4$ term in Eq.~\eqref{eq:W-exp} yields the sub-sub-leading contribution in the $\eta$-expansion. It is represented in spherical-harmonic basis by
\begin{equation}
\label{eq:R4-expan}
\fl
|\bs{R}-\bs{R}'|^4  \rightarrow
\frac{10}{3}
R^2 R'^2 + \frac{32\pi}{15} R^2 R'^2 Y_2(\hat R)\cdot Y_2(\hat R')
-\frac{16\pi}{3}(R^2 +R'^2) R R' Y_1(\hat R)\cdot Y_1(\hat R') {,}
\nn
\end{equation}
where $R^4$ and $R'^4$ are dropped, since they do not contribute to $\delta_{\rm pol}^{A}$ due to the orthogonality condition in Eq.~\eqref{eq:rhoN-sum-2}. 
By inserting the corresponding components of ${W}$ into Eq.~\eqref{eq:rho-W-rho}, we have the sub-sub-leading non-relativistic contribution $\delta_{\rm NR}^{(2)}$
\begin{eqnarray}
\label{eq:d2-R4}
\fl
\delta^{(2)}_{\rm NR} &=& \frac{4\pi\alpha^2}{9}  m_r^2 \phi^2(0)  \sum_{ab}^Z \SumInt_{N\neq N_0,JM} \sqrt{\frac{\omega_N}{2m_r}} \Bigl[ \bra N_0 J_0 M_0 |R^2_a |N J M\ket \bra N J M |R^2_b |N_0 J_0 M_0 \ket
\nn
\fl
&&+ \frac{16\pi}{25} \bra N_0 J_0 M_0 |R^2_a Y_2(\hat R_a) |N J M\ket \cdot \bra N J M |R^2_b Y_2(\hat R_b)|N_0 J_0 M_0\ket
\nn 
\fl
&& 
-\frac{8\pi}{5} \left(\bra N_0 J_0 M_0 |R^3_a Y_1(\hat R_a) |N J M\ket \cdot \bra N J M |R_b Y_1(\hat R_b)|N_0 J_0 M_0\ket
+{\rm c.c.}\right)
\Bigr]  {.}
\end{eqnarray}
For simplicity, we define a monopole operator $\hat{R}^2\equiv \frac{1}{Z} \sum_a^Z R^2_a$, a quadrupole operator $\hat{Q}_2 =\frac{1}{Z} \sum_a^Z R^2_a  Y_2(\hat{R}_a)$, and a new rank-$1$ operator $\hat{D}_3 = \frac{1}{Z} \sum_a^Z R_a^3 Y_1(\hat{R}_a)$.
Based on the Wigner-Eckart theorem in Eqs.~(\ref{eq:App-I0}, \ref{eq:App-I2}), we rewrite $\delta^{(2)}_{\rm NR}$ as a combination of three electric multipole sum rules
\begin{eqnarray}
\label{eq:d2-R4-resp}
\fl
\delta^{(2)}_{\rm NR} &=& 
\delta^{(2)}_{R^2}+\delta^{(2)}_{Q}+\delta^{(2)}_{D1D3}
\nn
\fl
&=& \frac{4\pi}{9}  m_r^2 (Z\alpha)^2\phi^2(0)  \int^\infty_{0} d\omega\,\sqrt{\frac{\omega}{2m_r}} \left[ S_{R^2}(\omega)
+ \frac{16\pi}{25} S_{Q}(\omega)
- \frac{16\pi}{5}  S_{D_1 D_3}(\omega) \right] {,}
\end{eqnarray}
where $S_{R^2}$, $S_{Q}$, and $S_{D_1 D_3}$ are respectively the electric monopole, quadrupole, and $D_1 D_3$-interference response functions. They are defined by
\numparts
\begin{eqnarray}
\label{eq:S-R2}
\fl
S_{R^2}(\omega) &=& \frac{1}{2J_0+1} \SumInt_{N\neq N_0,J} |\bra N_0 J_0 ||\hat R^2 ||N J\ket|^2 \delta(\omega-\omega_N) {,}
\\
\label{eq:S-Q}
\fl
S_{Q}(\omega) &=& \frac{1}{2J_0+1} \SumInt_{N\neq N_0,J} |\bra N_0 J_0 ||\hat{Q}_2 ||N J\ket|^2 \delta(\omega-\omega_N) {,}
\\
\label{eq:S-D1D3}
\fl
S_{D_1 D_3}(\omega) &=& 
\frac{1}{2J_0+1} \operatorname{Re}\SumInt_{N\neq N_0,J}
\bra N_0 J_0||\hat D_1^\dagger ||N J\ket
\bra N J ||\hat D_3 ||N_0 J_0\ket \delta(\omega-\omega_N) {.}
\end{eqnarray}
\endnumparts

%================================================================
%================================================================

\subsection{Coulomb distortion corrections}
\label{sec:Coul-corr}

In Section~\ref{sec:polar-nonrel}, Eq.~\eqref{eq:approx-Zalpha}, the leading approximation made in the $Z\alpha$ expansion guarantees that the non-relativistic nuclear polarizability effect $\delta_{\rm pol}^{\rm NR}$ is of order $(Z\alpha)^5$. In general, corrections of order $(Z\alpha)^6$ and beyond emerge not only from high-order terms omitted in Eq.~\eqref{eq:approx-Zalpha}, but also from higher multi-photon exchanges (beyond two-photon). These higher-in-$Z\alpha$ contributions are expected to be small, and we do not evaluate $(Z\alpha)^6$ effects in this article.
However, as is shown in this section, the Coulomb distortion correction yields an additional contribution to $\delta_{\rm pol}^{A}$, which is of order $(Z\alpha)^6\ln(Z\alpha)$. This contribution is logarithmically enhanced compared to a generic $(Z\alpha)^6$ contribution, and thus needs to be included in our analysis.

The Coulomb distortion originates from the Coulomb attraction between the muon and nucleus in the intermediate stages of the two-photon exchange, during which the muon wave-function is distorted from the free-particle one. Instead of Eq.~\eqref{eq:approx-Zalpha}, we partially keep, to the end of this section, some higher-in-$Z\alpha$ terms, {i.e.}, $\bra \bs{r} |\mu_{20}\ket = \phi(0)(1-\nu r)\exp(-\nu r)$ and $H_\mu = p^2/2m_r-Z\alpha/r$. While we retain $\epsilon_{\mu}\rightarrow 0$ as in Eq.~\eqref{eq:approx-Zalpha}, because it is two orders higher in $Z\alpha$ and only enters beyond $(Z\alpha)^6\ln(Z\alpha)$. For the same reason, we also keep $\bra \bs{r} |\mu_{21}\ket \rightarrow 0$ and calculate Coulomb distortion corrections only to the $2S$ state.

Instead of performing the full $\eta$-expansion, as is done in Section~\ref{sec:polar-nonrel}, we focus on the Coulomb distortion correction to the leading dipole contribution, $\delta_{\rm NR}^{(0)} =\delta_{D1}^{(0)}$. Higher-order terms in $\delta_{\rm pol}^{\rm NR}$ are already small. Therefore, Coulomb distortion corrections to $\delta^{(1)}_{\rm NR}$ and $\delta^{(2)}_{\rm NR}$ are omitted in our analysis. We then approximate $\Delta V$ in Eq.~\eqref{eq:DVrR} by its dominant dipole part, 
\begin{equation}
\label{eq:DV-di}
\Delta V(\bs{r},\bs{R}) \approx -\frac{4\pi\alpha}{3} \frac{R}{r^2} Y_1 (\hat R) \cdot Y_1(\hat r) {,}
\end{equation}
where we assume the nuclear scale is much smaller than the atomic scale, {i.e.}, $R \ll r$.

The Green's function with Coulomb interaction satisfies that
\begin{equation}
\left(-\omega_N + \frac{1}{2m_r} \nabla^2_{\bs{r}} + \frac{Z\alpha}{r}\right) G_C(-\omega_N;\bs{r},\bs{r}') = \delta^{(3)}(\bs{r}-\bs{r}') {.}
\end{equation}
$G_C$ is expanded in the atomic hyperfine state basis as
\begin{equation}
\label{eq:Gc}
\fl
G_C = \SumInt_{N\neq N_0,J} \sum\limits_{\ell j F M_F} \left| NJ,\left(\ell \frac{1}{2}\right)j;F M_F \right\ket \frac{g_\ell(-\omega_N;r,r')}{rr'} \left\bra NJ,\left(\ell \frac{1}{2}\right)j;F M_F \right| {,}
\end{equation}
where $(\ell \frac{1}{2})j$ denotes the angular momentum, spin and total angular momentum of the intermediate-state muon. Considering also the muon-nucleus total-angular-momentum coupling, the atomic hyperfine states are labeled by $F M_F$. The radial Green's function $g_\ell$ satisfies
\begin{equation}
\left(-\omega_N + \frac{1}{2m_r} \frac{d^2}{dr^2} - \frac{\ell(\ell+1)}{2m_r r^2} + \frac{Z\alpha}{r}\right) g_\ell(-\omega_N;r,r') = \delta(r-r') {.}
\end{equation}

Inserting $\Delta V$ from Eq.~\eqref{eq:DV-di} and $G_C$ from Eq.~\eqref{eq:Gc} into Eq.~\eqref{eq:dpol-00}, we write the Coulomb-distorted dipole polarizability contribution to an unperturbed $2S$ (or $2P$) hyperfine state, $| N_0 J_0, (\ell_0 \frac{1}{2})j_0;F_0 M_{F_0} \ket$, as
\begin{eqnarray}
\label{eq:Ec-1}
\fl
\delta_C^{(0)} &=& \phi^2(0)\SumInt_{N\neq N_0,J} \sum_{\ell j} \left(\frac{4\pi\alpha}{3}\right)^2 \frac{4\pi}{2\ell_0+1} \int_0^\infty dr \int_0^\infty dr' R_{2\ell_0}(r) R_{2\ell_0}(r') \frac{g_\ell(-\omega_N;r,r')}{rr'}
\nn
\fl
&&\times \sum_{ab}^Z\; \left\bra N_0 J_0,\left(\ell_0 \frac{1}{2}\right)j_0;F_0 M_{F_0}\right| R_a  Y_1 (\hat R_a) \cdot  Y_1(\hat r) \left|NJ,\left(\ell \frac{1}{2}\right)j;F_0 M_{F_0} \right\ket
\nn
\fl
&&\times \left\bra NJ,\left(\ell \frac{1}{2}\right)j;F_0 M_{F_0}\right| R_b Y_1 (\hat R_b) \cdot  Y_1(\hat r') \left| N_0 J_0,\left(\ell_0 \frac{1}{2}\right) j_0;F_0 M_{F_0}\right\ket {.}
\end{eqnarray}
$F_0$ and $M_{F_0}$ are conserved in the two-photo exchange process, since the involved operator $Y_1(\hat R_a) \cdot Y_1(\hat r)$ is a scalar. 
Eq.~\eqref{eq:Ec-1} is simplified as
\begin{equation}
\label{eq:Ec-2-simp}
\fl
\delta_C^{(0)} 
= \frac{16\pi^2}{9} (Z\alpha)^2 \phi^2(0) \SumInt_{N\neq N_0,J} |\bra N_0 J_0|| \hat D_1 || N J\ket|^2
 \sum_{\ell} \mathcal{F}_{\ell_0 \ell}(\omega_N) \left(\sum_{j} \mathcal{K}_{\ell_0 j_0 \ell j}^{J_0  J F_0}\right) {.}
\end{equation}
The coefficient $\mathcal{K}_{\ell_0 j_0 \ell j}^{J_0  J F_0}$ is a result of Wigner-Eckart theorem in Eqs.~(\ref{eq:App-II}, \ref{eq:App-Tk}, \ref{eq:App-Tk2}): 
\begin{eqnarray}
\label{eq:Coloumb-spin}
\fl
\mathcal{K}_{\ell_0 j_0 \ell j}^{J_0  J F_0} &=& \frac{4\pi}{2\ell_0+1} \sixj{J_0}{1}{J}{j}{F_0}{j_0}^2  \left|\left\bra \left(\ell_0 \frac{1}{2}\right) j_0 \right\| Y_1 \left\| \left(\ell \frac{1}{2}\right) j \right\ket \right|^2
\nn
\fl
&=& \frac{3 (2j_0+1)(2j+1)}{2\ell_0+1} \sixj{J_0}{1}{J}{j}{F_0}{j_0}^2  \sixj{\ell_0}{j_0}{1/2}{j}{\ell}{1}^2
\nn
\fl
&&\times \left[(\ell_0+1)\delta_{\ell,\ell_0+1} + \ell_0 \delta_{\ell,\ell_0-1}\right] {,}
\end{eqnarray}
where $\{:::\}$ denotes the 6-$j$ symbol~\cite{Edmonds:1996}.
$\mathcal{F}_{\ell_0 \ell}$ represents the Coulomb integral:
\begin{equation}
\label{eq:coulomb-int}
\mathcal{F}_{\ell_0 \ell}(\omega_N) = \int dr\, dr' R_{2\ell_0}(r)\; R_{2\ell_0}(r')\; \frac{g_\ell(-\omega_N;r,r')}{rr'} {.}
\end{equation}

Since we consider polarizability contributions only to the $2S$ related hyperfine states, we take $\ell_0=0$, and $j_0=1/2$. Therefore, only terms with $\ell=1$ and $j=\frac{1}{2},\frac{3}{2}$ are non-zero in Eq.~\eqref{eq:Coloumb-spin}. The summation over $j$-dependent terms in Eq.~\eqref{eq:Ec-2-simp} yields $\sum_j \mathcal{K}_{0 \frac{1}{2} 1 j}^{J_0  J F_0} = 1/(2J_0+1)$, which is independent of $F_0$ and $J$. Therefore, the Coulomb distortion corrections contribute equally to the two hyperfine states associated with $2S$.

For the $2S$ state, only $\mathcal{F}_{01}$ is needed. As in Eq.~\eqref{eq:F201}, $\mathcal{F}_{01}$ is expanded in powers of $Z\alpha$ by
\begin{equation}
\label{eq:F2-01}
\mathcal{F}_{01}(\omega_N)
=-\sqrt{\frac{2m_r}{\omega_N}} -\frac{Z\alpha m_r}{\omega_N} \ln\frac{2(Z\alpha)^2 m_r}{\omega_N}
+\cdots {,}
\end{equation}
where dots indicate terms of higher orders in $Z\alpha$, which only contribute to $\delta_{\rm pol}^{A}$ at $(Z\alpha)^6$ and beyond\footnote{The higher-order terms in Eq.~\eqref{eq:F2-01}, which give small corrections to $\delta_C^{(0)}$, were included in Refs.~\cite{Ji13, Pachucki:2011xr}, but are omitted in this paper for a consistent evaluation of polarizability contributions at order $(Z\alpha)^5$.}.
The first term in Eq.~\eqref{eq:F2-01} reproduces the same energy weight as $\delta^{(0)}_{D1}$ in Eq.~\eqref{eq:delta_D1}, and is thus dropped to avoid double counting. The second term, which is logarithmically enhanced in the $Z\alpha$ expansion, makes a $(Z\alpha)^6\ln Z\alpha$ contribution to $\delta_{\rm pol}^{A}$.

By inserting $\mathcal{F}_{01}$'s logarithmic piece into Eq.~\eqref{eq:Ec-2-simp}, we have 
\begin{eqnarray}
\label{eq:dpol-C}
\fl
\delta_C^{(0)}
&=& \frac{16\pi^2}{9} (Z\alpha)^2\phi^2(0)\,  \frac{1}{2J_0+1} \SumInt_{N\neq N_0,J} |\bra N_0 J_0|| \hat D_1 || N J\ket|^2
  \left(-\frac{Z\alpha m_r}{\omega_N} \ln\frac{2(Z\alpha)^2 m_r}{\omega_N}\right)
\nn
\fl
&=& -\frac{16\pi^2}{9} (Z\alpha)^3\phi^2(0)  \int_{0}^{\infty} d\omega\, \frac{m_r}{\omega} \ln\frac{2(Z\alpha)^2 m_r}{\omega}\, S_{D_1}(\omega) {,}
\end{eqnarray}
which contains an electric-dipole sum rule with an unusual logarithmic energy weight.

%=============================================================================
%=============================================================================

\subsection{Relativistic corrections} \label{sec:relativstic}

\begin{figure}[htb]
\centerline{
\includegraphics[angle=0,width=0.8\linewidth]{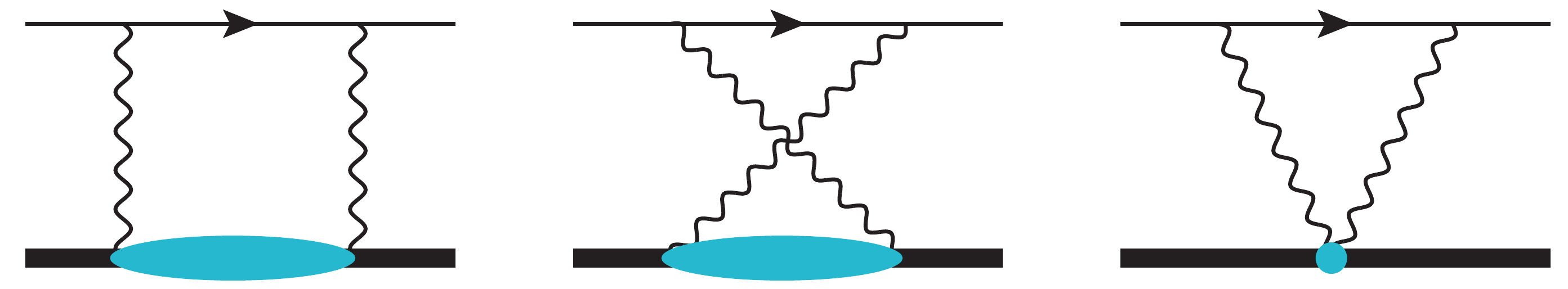}}
\caption{
Two-photon exchange direct, crossed, and seagull diagrams.
} 
\label{pic:compton} 
\end{figure}

The description of relativistic corrections to the nuclear polarizability is beyond the scope of Figure~\ref{fig:tpe}, where the muon in the two-photon loop is non-relativistic and does not obey time-reversal symmetry. This approximation is valid because the typical photon-energy scale, related to the first nuclear excitation $\omega_{\rm th}$, is much smaller than the muon mass. Relativistic corrections enter at higher orders in the $\omega_{\rm th}/m_\mu$ expansion. In this section, we work in the relativistic framework, and calculate $\delta_{\rm pol}^{A}$ using the two-photon exchange Feynman diagrams as depicted in Figure~\ref{pic:compton}. Besides the direct and crossed diagrams, an additional two-photon exchange counterterm (seagull diagram) is introduced to keep gauge invariance. As shown by Rosenfelder in \cite{Rosenfelder:1983aq}, the combination of these three forms a polarization potential $\Delta \mathcal{V}_{\rm pol}$, which is directly related to the two-photon loop amplitude. From $\Delta \mathcal{V}_{\rm pol}$, the nuclear polarizability corrections to the atomic spectrum are calculated in the relativistic limit as $\delta_{\rm pol}^{A} = \bra N_0 {\mu} | \Delta \mathcal{V}_{\rm pol} |N_0 {\mu}\ket$. 

We take into account only the relativistic corrections to
the electric dipole polarizability contributions, $\delta^{(0)}_{D1}$, which represents the leading contribution $\delta^{(0)}_{\rm NR}$ in the non-relativistic $\eta$-expansion. Since $\delta^{(1)}_{\rm NR}$ and $\delta^{(2)}_{\rm NR}$ are already small, relativistic corrections to these higher-order terms are thus neglected in our analysis.
Based on \cite{Rosenfelder:1983aq}, the evaluation of the polarizability contributions are given in the point-proton and the relativistic limits by
\begin{equation}
\label{eq:rel-LTS}
\delta_{\rm pol}^{\rm R} = -8\alpha^2 \phi^2(0) \int^{\infty}_{0} d q \left[\, \mathcal{R}_L(q) + \mathcal{R}_T(q) + \mathcal{R}_S(q) \,\right] {,}
\end{equation}
where $q$ denotes the photon-exchange transfer momentum, and $\mathcal{R}_L$ and $\mathcal{R}_T$ are related respectively to the longitudinal and transverse photon polarization. $\mathcal{R}_S$, the seagull term, is required by gauge invariance and cancels exactly the singularity at $q=0$ in $\mathcal{R}_T$. These kernel functions are given in \cite{Rosenfelder:1983aq} as
\numparts
\begin{eqnarray}
\label{eq:RL}
\mathcal{R}_L(q) &=& \int^{\infty}_{0} d \omega S_L(\omega,\bs{q}) g(\omega,q) {,}
\\
\label{eq:RT}
\mathcal{R}_T(q) &=& \int^{\infty}_{0} d \omega S_T(\omega,\bs{q}) \left[ -\frac{1}{4m_r q} \frac{\omega+2q}{(\omega+q)^2} + \frac{q^2}{4m_r^2} g(\omega,q) \right]  {,}
\\
\label{eq:RS}
\mathcal{R}_S(q) &=& \int^{\infty}_{0} d \omega S_T(\omega,0) \frac{1}{4m_r \omega} \left[\frac{1}{q} - \frac{1}{E_q}\right] \,,
\end{eqnarray}
\endnumparts
with $E_q = \sqrt{q^2+m_r^2}$ and
\begin{equation}
\label{eq:g-wq}
\fl
g(\omega,q) = \frac{1}{2E_q} \left[ \frac{1}{(E_q-m_r)(E_q-m_r+\omega)} - \frac{1}{(E_q+m_r)(E_q+m_r+\omega)} \right] {.}
\end{equation}

In the equations above, although we have taken the infinite-nuclear-mass approximation, the muon mass $m_\mu$ is replaced by the non-relativistic reduced mass $m_r$. By doing so, the leading term in the $\omega_N/m_r$-expansion of $\delta_{\rm pol}^{\rm R}$ in Eq.~\eqref{eq:rel-LTS} matches exactly to $\delta^{(0)}_{D1}$ in Eq.~\eqref{eq:delta_D1}, and is thus subtracted out to avoid double counting. The remaining contributions represent the relativistic corrections to $\delta^{(0)}_{\rm NR}$. This approximation naturally takes into account the dominant relativistic recoil effects; while higher-order recoil corrections only enter at higher orders in the $m_\mu/M_A$ expansion and are neglected in this paper\footnote{The calculation of higher-order relativistic recoil corrections to $\delta_{\rm pol}^{A}$ was performed by Pachucki in~\cite{Pachucki_2}. The effects turned out to be negligibly small.}.

$S_{L}$ and $S_{T}$ are respectively the longitudinal and transverse response functions, which are defined as~\cite{Rosenfelder:1983aq}\footnote{Here we follow the definition in Ref.~\cite{Rosenfelder:1983aq} and do not use the notation of reduced matrix elements. $S_{L}$ and $S_{T}$ are connected to dipole response functions in the following part of this section.}
\begin{equation}
\label{eq:SLT}
S_{L,T} (\omega,\bs{q}) = \SumInt\limits_{N\neq N_0,J} | \bra N J | \hat{O}_{L,T} (\bs{q}) | N_0 J_0\ket |^2 \delta(\omega-\omega_N) {,}
\end{equation}
where $\hat{O}_L(\bs{q}) = \hat{\mathcal{J}}_0(\bs{q})$ denotes the charge operator, and $\hat{O}_T(\bs{q}) =\hat{q} \times \hat{\bs{\mathcal{J}}} (\bs{q})$ indicates the transverse part of the current operator. By defining the longitudinal direction $\bs{e}_0$ along $\bs{q}$, and two circular transverse directions $\bs{e}_{\pm 1}$\footnote{The circular transverse vectors satisfy the relation $\bs{e}^{\dagger}_{\lambda} = (-1)^{\lambda} \bs{e}_{\lambda}$ and $\bs{e}^{\dagger}_{\lambda} \cdot \bs{e}_{\lambda'} = \delta_{\lambda \lambda'}$.}, we have $\hat{O}_T(\bs{q}) =\sum_{\lambda=\pm 1} \lambda (\bs{e}^{\dagger}_{\lambda}\cdot\hat{\bs{\mathcal{J}}}) \bs{e}_{-\lambda}$, which leads to
\begin{equation}
\label{eq:ST}
S_{T} (\omega,\bs{q}) =  \SumInt\limits_{N\neq N_0,J}  \sum_{\lambda=\pm 1} 
| \bra N,J | \bs{e}^{\dagger}_{\lambda}\cdot \hat{\bs{\mathcal{J}}}(\bs{q}) | N_0, J_0\ket |^2 \delta(\omega-\omega_N) {.}
\end{equation}

We then express $\hat{\mathcal{J}}_0(\bs{q})$ and $\bs{e}^{\dagger}_{\lambda}\cdot\hat{\bs{\mathcal{J}}}(\bs{q})$ in the plane-wave expansion as~\cite{Walecka:2004}
\numparts
\begin{equation}
\label{eq:rho-q-CJ}
\hat{\mathcal{J}}_0(\bs{q}) = \sum\limits_{l\geq 0} \sqrt{4\pi (2l+1)}\, i^l C_l(q) {,}
\end{equation}
\begin{equation}
\label{eq:J-q-T}
\bs{e}^{\dagger}_{\lambda}\cdot \hat{\bs{\mathcal{J}}}(\bs{q})
= - \sum_{l\geq 1} \sqrt{2\pi (2l+1)} i^l \left[ \hat{T}^{\rm el}_{l -\lambda} (q) + \lambda \hat{T}^{\rm mag}_{l -\lambda} (q) \right] {,}
\end{equation}
\endnumparts
where $C_l(q)$, $\hat{T}^{\rm el}_{l \lambda}$ and $\hat{T}^{\rm mag}_{l \lambda}$ denote respectively the $l$th moments of the electric-charge, electric-transverse-current, and magnetic-transverse-current operators. Since the nucleus is much heavier than the muon, these moments are approximated for small $q$ by
\numparts
\begin{eqnarray}
\fl
\label{eq:Cl-q0}
C_l(q) &= \frac{q^l}{(2l+1)!!} \int d^3 x\, \hat{\mathcal{J}}_0(\bs{x}) x^l Y_{l0} (\hat{x}) {,}
\\
\label{eq:T-el}
\fl
\hat{T}^{\rm el}_{l \lambda} (q) &= \frac{-i q^{l-1}}{(2l+1)!!}  \sqrt{\frac{l+1}{l}} 
\int d^3 x \left\lbrace \bs{\nabla} \cdot \hat{\bs{\mathcal{J}}_c} \, x^l Y_{l \lambda} 
+ \frac{q^2}{l+1} \hat{\bs{\mathcal{J}}_s} \cdot \left[\bs{x}\times \bs{\nabla} x^l  Y_{l \lambda}\right]
\right\rbrace {,}
\\
\fl
\label{eq:T-mag}
\hat{T}^{\rm mag}_{l \lambda} (q) &= \frac{i q^{l}}{(2l+1)!!} \sqrt{\frac{l+1}{l}} 
\int d^3 x \left[ \hat{\bs{\mathcal{J}}_s} + \frac{1}{l+1} \bs{x}\times\hat{\bs{\mathcal{J}}_c} \right] \cdot \bs{\nabla} x^l  Y_{l \lambda} {,}
\end{eqnarray}
\endnumparts
where $\hat{\bs{\mathcal{J}}}_c$ denotes the electric-convection current and $\hat{\bs{\mathcal{J}}_s}$ is the magnetic-spin current.

In the following, we separate the response functions into electric longitudinal, electric transverse and magnetic transverse parts, {i.e.}, $\delta_{\rm pol}^{\rm R} = \delta_{L}^{(0)}+\delta_{T}^{(0)} +\delta_{M}^{(0)} $, and study their contributions respectively.

%====================================================================
\subsubsection{Electric longitudinal polarizability corrections} \label{sec:dpol-L}

In the point-nucleon approximation, $\hat{\mathcal{J}}_0(\bs{x}) = Z\hat{\rho}^p(\bs{x})=\sum_{a}^{Z} \delta^{(3)}(\bs{x}-\bs{R}_a)$, so we have
\begin{equation}
C_l(q) = \frac{q^l}{(2l+1)!!} \sum_{a}^{Z} \bs{R}_a^l Y_{l0} (\hat{\bs{R}}_a) {.}
\end{equation}
Since $C_0(q) = Z/\sqrt{4\pi}$ is constant, it does not contribute to the transition matrix element in Eq.~\eqref{eq:SLT}. Therefore, the leading term contributing to the longitudinal polarizability effect is from $C_1(q)$. It is related to the $z$-component of the electric-dipole operator by $C_1(q) =  (Zq/3) \hat{D}_{1z}$. 
In this approximation, $\hat{\mathcal{J}}_0(\bs{q}) \approx i  Z q \sqrt{4\pi/3}\hat{D}_{1z}$.
By substituting $\hat{\mathcal{J}}_0(\bs{q})$'s low-momentum expression into Eq.~\eqref{eq:SLT}, we obtain the approximated electric longitudinal response function, which is related to the electric dipole response function by:
\begin{eqnarray}
\label{eq:SLq00}
S_L^{(0)}(\omega,\bs{q}) &=& \frac{4\pi}{3} Z^2 q^2 \SumInt\limits_{N\neq N_0, J} | \bra NJ| \hat{D}_{1z} | N_0 J_0 \ket|^2 \delta(\omega-\omega_N)
\nn
&=& \frac{4\pi}{9} Z^2 q^2 S_{D_1} (\omega) {,}
\end{eqnarray}
where $ |\bra NJ| \hat{D}_{1z} | N_0 J_0 \ket|^2 = \frac{1}{3(2J_0+1)} |\bra NJ|| \hat{D}_1 || N_0 J_0 \ket|^2$ is used by averaging the full photon angle, which is arbitrary to the direction of the nuclear quantization~\cite{Walecka:2004}.

We substitute Eq.~\eqref{eq:SLq00} into Eqs.~\eqref{eq:rel-LTS} and \eqref{eq:RL}, and obtain the electric longitudinal polarizability contribution as
\begin{equation}
\label{eq:delta-Lq00}
\delta_{L}^{(0)} 
= \frac{32\pi}{9} (Z\alpha)^2 \phi^2(0) \int^{\infty}_{0} d\omega\,  S_{D_1} (\omega)\, \mathcal{F}_L(\omega/m_r) {,}
\end{equation}
where $\mathcal{F}_L$ represents a $q$-integration, {i.e.}, $\mathcal{F}_L(\omega/m_r) \equiv -\int^\infty_0 dq\, q^2 g(\omega, q)$, whose evaluation yields
\begin{equation}
\label{eq:mathFL}
\mathcal{F}_L(\lambda) = \sqrt{\frac{\lambda-2}{\lambda}} \arctanh\sqrt{\frac{\lambda-2}{\lambda}}
-\sqrt{\frac{\lambda+2}{\lambda}} \arctanh\sqrt{\frac{\lambda}{\lambda+2}} {,}
\end{equation}
where $\lambda = \omega/m_r$. We note that $\mathcal{F}_L$ is real for $\lambda>0$ when analytic continuation at $\lambda=2$ is applied.

Eq.~\eqref{eq:delta-Lq00} contains relativistic corrections to only the electric dipole polarizability contribution; while relativistic corrections to higher-multipole contributions are neglected in the low-$q$ approximation made in Eq.~\eqref{eq:SLq00}.
If we expand $\mathcal{F}_L$ for small $\lambda=\omega/m_r$, the leading and sub-leading terms are
\begin{equation}
\mathcal{F}_L(\omega/m_r\rightarrow 0) \approx 
-\frac{\pi}{2}\sqrt{\frac{2m_r}{\omega}} \left(1-\frac{\omega}{4 m_r}\right)+\cdots
\end{equation}
The first term in $\mathcal{F}_L$ matches exactly to $\delta^{(0)}_{D1}$ in Eq.~\eqref{eq:delta_D1}. Therefore, we subtract the leading term to avoid double counting. We then obtain the relativistic longitudinal polarizability correction as
\begin{equation}
\label{eq:delta-eL}
\delta_{L}^{(0)} = \frac{32\pi}{9} (Z\alpha)^2 \phi^2(0) \int^{\infty}_{0} d\omega\,  S_{D_1} (\omega) \left[\mathcal{F}_L\left(\frac{\omega}{m_r}\right)+\frac{\pi}{2}\sqrt{\frac{2m_r}{\omega}}\right] {.}
\end{equation}
One can use dimensional analysis on the sub-leading term in $\mathcal{F}_L$ to roughly estimate the size of $\delta_{L}^{(0)}$. The typical scale of $\omega$ is represented by the first nuclear excitation energy $\omega_{\rm th}$. Therefore, $\delta_{L}^{(0)}$ is approximately smaller than $\delta_{D1}^{(0)}$ by order $\omega_{\rm th}/m_\mu$.

%===================================================================
\subsubsection{Electric transverse polarizability corrections}\label{sec:dpol-T}

In the $q\rightarrow 0$ limit, the electric transverse current is dominated by its first moment, which is related to the electric-dipole operator by 
\begin{equation}
\hat{T}^{\rm el}_{1 \lambda} (q\rightarrow 0)= -i \frac{\sqrt{2}}{3} \int d^3 x \bs{\nabla}\cdot\hat{\bs{\mathcal{J}}}_c\, x Y_{1 \lambda}(\hat{x})
=\frac{\sqrt{2}}{3} Z \omega \hat{D}_{1\lambda} {,}
\end{equation}
where the Siegert's theorem $\bs{\nabla}\cdot\hat{\bs{\mathcal{J}}}_c = i\omega \mathcal{J}_0$ is used.
This leads to the dominant component of the electric transverse current in low-$q$ expansion as $\bs{e}^{\dagger}_{\lambda}\cdot \hat{\bs{\mathcal{J}}}^{\rm el} \approx -i\sqrt{4\pi/3}\, Z\omega\, \hat{D}_{1 -\lambda}$.
By substituting the low-$q$ approximation of $\bs{e}^{\dagger}_{\lambda}\cdot \hat{\bs{\mathcal{J}}}^{\rm el}$ into Eq.~\eqref{eq:ST}, we obtain the dominant component of the electric transverse response function as
\begin{eqnarray}
\label{eq:ST-q00}
S_{T}^{{\rm el}\,(0)} (\omega, \bs{q}) 
&=& \frac{4\pi}{3} Z^2 \omega^2 \SumInt\limits_{N\neq N_0,J} \sum_{\lambda=\pm 1}
|\bra NJ| \hat{D}_{1\lambda} | N_0 J_0 \ket|^2 \delta(\omega-\omega_N)
\nn
&=& \frac{8\pi}{9} Z^2 \omega^2 S_{D_1} (\omega) {,}
\end{eqnarray}
where $\sum_{\lambda=\pm 1}
|\bra NJ| \hat{D}_{1\lambda} | N_0 J_0 \ket|^2 =  \frac{2}{3(2J_0+1)} |\bra NJ|| \hat{D}_1 || N_0 J_0 \ket|^2$ after averaging the photon angle.
By substituting Eq.~\eqref{eq:ST-q00} into Eqs.~\eqref{eq:rel-LTS}, \eqref{eq:RT} and \eqref{eq:RS}, we obtain the relativistic electric transverse polarizability correction as
\begin{eqnarray}
\label{eq:delta-eT}
\delta_{T}^{(0)} &=& -8\alpha^2 \phi^2(0) \int^\infty_0 dq \left[\mathcal{R}_T+\mathcal{R}_S\right]^{\rm el}
\nn
&=& \frac{16\pi}{9} (Z\alpha)^2 \phi^2(0) \int^{\infty}_{0} d\omega\, S_{D_1} (\omega) 
\mathcal{F}_T (\omega/m_r) {,}
\end{eqnarray}
where $\mathcal{R}_S^{el}$ cancels the infrared divergence in $\mathcal{R}_T^{el}$ at $q=0$. $\mathcal{F}_T$ represents a $q$-integration: 
\begin{equation}
\fl
\mathcal{F}_T \left(\frac{\omega}{m_r}\right) 
= -4\omega^2 \int^\infty_0 dq \left\lbrace 
-\frac{1}{4m_r q} \frac{\omega+2q}{(\omega+q)^2} +\frac{q^2}{4m_r^2} g(\omega,q)
+\frac{1}{4m_r\omega} \left[\frac{1}{q}-\frac{1}{E_q}\right]
\right\rbrace {.}
\end{equation}
The evaluation of the integration above yields
\begin{equation}
\label{eq:mathFT}
\mathcal{F}_T(\lambda) = \lambda +\lambda\ln (2\lambda)  + \lambda^2\mathcal{F}_L(\lambda) {.}
\end{equation}

We roughly estimate the size of $\delta_{T}^{(0)}$ by taking the dominant term of $\mathcal{F}_T$ in the small-$\omega/m_r$ expansion, {i.e.}, $\mathcal{F}_T \approx (\omega/m_r) \ln (2\omega/m_r)$.
Comparing the energy weights in Eq.~\eqref{eq:delta_D1} and \eqref{eq:delta-eT}, the contribution of $\delta_{T}^{(0)}$ is approximately smaller than $\delta_{L}^{(0)}$ by order $\sqrt{\omega_{\rm th}/m_\mu } \ln(\omega_{\rm th}/m_\mu)$.

%======================================================================
\subsubsection{Magnetic transverse polarizability corrections} \label{sec:dpol-M}

In the limit $q\rightarrow 0$, $\hat{T}^{\rm mag}_{1 \lambda}$ plays the dominant role in the magnetic transverse current. In the point-nucleon limit, we express $\hat{\bs{\mathcal{J}}_s}$ and $\bs{x}\times\hat{\bs{\mathcal{J}}_c}$ respectively as
\numparts
\begin{eqnarray}
\hat{\bs{\mathcal{J}}_s}(\bs{x}) &=& \frac{1}{2m_p} \sum_{i}^{A} (g_p\hat{e}_{p,i}+g_n\hat{e}_{n,i}) \bs{s}_i\, \delta^{(3)} \left(\bs{x} - \bs{R}_i \right) {,}
\\
\bs{x}\times\hat{\bs{\mathcal{J}}_c}(\bs{x}) &=& \frac{1}{m_p} \sum_i^{Z} \bs{l}_i \hat{e}_{p,i}\, \delta^{(3)}(\bs{x}-\bs{R}_i) {,}
\end{eqnarray}
\endnumparts
where $g_p=5.586$ and $g_n=-3.826$. $\bs{s}_i$ and $\bs{l}_i$ indicate the spin and angular momentum of the $i$th nucleon. Using these equations above, we obtain that $\hat{T}^{\rm mag}_{1 \lambda}= i Zq \hat{M}_{1\lambda} /(2m_p\sqrt{6\pi})$~\cite{Walecka:2004}, with $\hat{M}_1$ denoting a magnetic-dipole operator:
\begin{equation}
\hat{M}_1 \equiv \frac{1}{Z}\sum_{i}^{A}  \left[ (g_p\hat{e}_{p,i}+g_n\hat{e}_{n,i}) \bs{s}_i + \hat{e}_{p,i} \bs{l}_i \right] {.}
\end{equation}
In the low-$q$ limit, the magnetic current operator is dominated by the magnetic-dipole part, {i.e.}, $[\bs{e}^{\dagger}_{\lambda}\cdot \hat{\bs{\mathcal{J}}}]_{\rm mag} \approx \lambda Z q \hat{M}_{1 -\lambda}/(2m_p)$, and the magnetic transverse response function is approximated by
\begin{equation}
\label{eq:SM-q00}
S_T^{{\rm mag}\,(0)} (\omega,\bs{q}) 
=\frac{Z^2 q^2}{6 m_p^2} S_{M_1} (\omega) {,}
\end{equation}
where $S_{M_1}$ is the magnetic-dipole structure function
\begin{equation}
\label{eq:S-MSL}
S_{M_1} (\omega) = \frac{1}{2J_0+1} \SumInt\limits_{N\neq N_0, J} |\bra NJ|| \hat{M}_1 || N_0 J_0 \ket|^2 \delta(\omega-\omega_N) {.}
\end{equation}

Combining Eqs.~\eqref{eq:SM-q00}, \eqref{eq:rel-LTS} and \eqref{eq:RT}, we obtain the magnetic transverse polarizability contribution:
\begin{eqnarray}
\label{eq:delta-M-0}
\delta_{M}^{(0)} &=& -8\alpha^2 \phi^2(0) \int^\infty_0 dq\, \mathcal{R}_T^{\rm mag}
\nn
&=& \frac{(Z\alpha)^2 }{3 m_p^2}  \phi^2(0) \int^{\infty}_{0} d\omega\, S_{M_1} (\omega)
\mathcal{F}_M (\omega/m_r ) {,}
\end{eqnarray}
A seagull term $\mathcal{R}_S^{\rm mag}$ is not needed in Eq.~\eqref{eq:delta-M-0}, since $R_T^{\rm mag}$ is finite at $q=0$. $\mathcal{F}_M$ defines a $q$-integration:
\begin{equation}
\label{eq:FM}
\mathcal{F}_M \left(\frac{\omega}{m_r}\right) = -4\int^\infty_0 dq\, q^2 \left[
-\frac{1}{4m_rq} \frac{\omega+2q}{(\omega+q)^2} +\frac{q^2}{4m_r^2}g(\omega,q)
\right] {,}
\end{equation}
whose evaluation yields
\begin{equation}
\label{eq:mathFM}
\fl
\mathcal{F}_M(\lambda) 
= \sqrt{\lambda} \left[
(\lambda-2)^{\frac{3}{2}} \arctanh\sqrt{\frac{\lambda-2}{\lambda}} -
(\lambda+2)^{\frac{3}{2}} \arctanh\sqrt{\frac{\lambda}{\lambda+2}}
\right]
+\lambda  +3\lambda \ln (2\lambda) {.}
\end{equation}
Similarly, $\mathcal{F}_M$ is real for $\lambda>0$ when analytic continuation is applied at $\lambda=2$. When $\lambda=\omega/m_r\ll 1$, $\mathcal{F}_M\approx\pi\sqrt{2\omega/m_r}$, which is an approximated energy-weight used in our previous estimates of the magnetic-dipole polarizability contribution~\cite{Nevo_Dinur_2016,Hernandez_2014}. In this paper, we use the complete expression~\eqref{eq:mathFM} as a more accurate energy-weight for the magnetic-dipole sum rule.
Different from $\delta_{L,T}^{(0)}$, the magnetic-dipole sum rule is characterized by a different threshold energy, because the magnetic excitation in light nuclei normally involves higher lying states than does the electric-dipole excitation. Suppressed by the $1/m_p^{2}$ factor, we expect $\delta_{M}^{(0)}$ to be much smaller than $\delta^{(0)}_{D1}$.

%============================================================================
%=============================================================================
\subsection{Nucleon-size corrections} \label{sec:NS-correct}

When considering the intrinsic charge distribution of nucleons, the position of proton in Eq.~\eqref{eq:DVrR} needs to be replaced by a convolution over the proton and neutron charge densities. Therefore, Eq.~\eqref{eq:DVrR} is modified by
\begin{equation}
\label{eq:DVrRm}
\eqalign{
  \Delta H = \sum_{a}^A \Delta V_a(\bs{r},\bs{R}_a) {,}
\\
\Delta V_a(\bs{r},\bs{R}_a) = \Delta V_{p}(\bs{r},\bs{R}_a) \hat{e}_{p,a}
                            + \Delta V_{n}(\bs{r},\bs{R}_a) \hat{e}_{n,a} {,}
}
\end{equation}
where $\Delta V_{p}$ and $\Delta V_{n}$ are defined respectively as
\numparts
\begin{eqnarray}
\Delta V_{p}(\bs{r},\bs{R}_a) &\equiv & 
       \alpha\left(\frac{1}{r}-\int d^3 R'\,
       \frac{n_p(\bs{R}'-\bs{R}_a)}{|\bs{r}-\bs{R}'|}\right) {,}
\\
\Delta V_{n}(\bs{r},\bs{R}_a) &\equiv & 
       - \alpha \int d^3 R'\,
       \frac{n_n(\bs{R}'-\bs{R}_a)}{|\bs{r}-\bs{R}'|} {,}
\end{eqnarray}
\endnumparts
with $n_p$ and $n_n$ indicating the intrinsic proton and neutron charge densities.

Besides $\rho^p_N$ in Eq.~\eqref{eq:rhop_N}, we also define at this point the point-neutron transition density function
\begin{eqnarray}\label{eq:rhon_N}
 \rho^n_N(\bs{R}) &=& \bra N | \frac{1}{Z}\sum_{a}^A 
                           \delta(\bs{R}-\bs{R}_a)\hat{e}_{n,a}|N_0\ket {,}
\end{eqnarray}
with $\rho^n_0(\bs{R})$ denoting the point-neutron densities defined in Eq.~\eqref{eq:rhoc_0}.
Using this function we write
\begin{eqnarray}
\fl
   \sum_a^A \bra N | \Delta V_a(\bs{r},\bs{R}_a) | N_0\ket &=&
   Z \int d^3 R \left[\rho^p_N(\bs{R}) \Delta V_p(\bs{r},\bs{R}) 
                       +\rho^n_N(\bs{R}) \Delta V_n(\bs{r},\bs{R})\right] {.}
\end{eqnarray}
$\delta_{\rm pol}^{A}$ is then expressed as
\begin{equation}\label{eq:rhoPm}
  \delta_{\rm pol}^{A} =  \sum_{c,c'=n,p}\SumInt_{N\neq N_0} \int d^3 R d^3 R'
     \rho_N^{c*}(\bs{R}){W}^{c c'}(\bs{R},\bs{R}',\omega_N) \rho^{c'}_N(\bs{R}') {,}
\end{equation}
where the muon matrix elements ${W}^{c c'}$ (with $c,c'=n,p$) are defined as
\begin{eqnarray}\label{eq:P-NR-3}
\fl
  {W}^{cc'}(\bs{R},\bs{R}',\omega_N) &=& -Z^2 \phi^2(0)\int d^3r d^3 r'
     \Delta V_{c}(\bs{r},\bs{R})  \bra \bs{r}
       |\frac{1}{\frac{q^2}{2 m_r} +\omega_N} | \bs{r}'\ket \Delta V_{c'}(\bs{r}',\bs{R}') {.}
\end{eqnarray}

Now we use the Fourier transform of $\Delta V_p(\bs{r},\bs{R})$ and $\Delta V_n(\bs{r},\bs{R})$ with respect to the muon-coordinates and have
\begin{eqnarray}\label{eq:DVqRm}
\Delta \tilde{V}_p(\bs{q},\bs{R})&=& \int d^3 r V_p(\bs{r},\bs{R})
    e^{i\bs{q}\cdot\bs{r}} = \frac{4\pi\alpha}{q^2}
    \left(1-\tilde{n}_p(q) e^{i\bs{q}\cdot\bs{R}}\right)  {,}
\cr
\Delta \tilde{V}_n(\bs{q},\bs{R})&=& \int d^3 r V_n(\bs{r},\bs{R})
    e^{i\bs{q}\cdot\bs{r}} =-\frac{4\pi\alpha}{q^2}
                    \tilde{n}_n(q) e^{i\bs{q}\cdot\bs{R}}  {,}
\end{eqnarray}
where $\tilde{n}_{p/n}(q) = \int d^3 R\, n_{p/n}(\bs{R}) e^{i\bs{q}\cdot\bs{R}}$ is the Fourier transform of the nucleon charge density, which depends only on $q=|\bs{q}|$.  In the non-relativistic limit, $\tilde{n}_{p}(q)$ and $\tilde{n}_{n}(q)$ represent the nucleon electric form factors, with $\tilde{n}_p(0)=1$ and $\tilde{n}_n(0)=0$ at $q=0$. Inserting the above expressions into Eq.~\eqref{eq:P-NR-3}, we have
\numparts
\begin{eqnarray}
\label{eq:PNRpp}
\fl
{W}^{pp} &=& -Z^2 \phi^2(0)  \int \frac{d^3 q}{(2\pi)^3}
        \left(\frac{4\pi\alpha}{q^2}\right)^2 \frac{1}{\frac{q^2}{2m_r}+\omega_N}
        \left[\tilde{n}_p^2(q) e^{i\bs{q}\cdot(\bs{R}-\bs{R}')}-1\right] {,} 
\\
\label{eq:PNRpn}
\fl
{W}^{np/pn} &=& -Z^2 \phi^2(0) \int \frac{d^3 q}{(2\pi)^3}
        \left(\frac{4\pi\alpha}{q^2}\right)^2 \frac{1}{\frac{q^2}{2m_r}+\omega_N}
        \tilde{n}_p(q) \tilde{n}_n(q)  e^{i\bs{q}\cdot(\bs{R}-\bs{R}')} {,}
\\
\label{eq:PNRnn}
\fl
{W}^{nn} &=& -Z^2 \phi^2(0) \int \frac{d^3 q}{(2\pi)^3}
        \left(\frac{4\pi\alpha}{q^2}\right)^2 \frac{1}{\frac{q^2}{2m_r}+\omega_N}
        \tilde{n}_n^2(q) e^{i\bs{q}\cdot(\bs{R}-\bs{R}')} {.} 
\end{eqnarray}
\endnumparts
Similarly to Eq.~\eqref{eq:W2}, here we have omitted terms that depend on $R$ or $R'$ alone. Such terms yield matrix elements proportional to
$\bra N_0 |R^n |N\ket \bra N|N_0\ket$, which are zero due to the orthogonality between nuclear states $|N_0\ket$ and $|N\ket$.

For convenience of calculations, we take low-$q$ approximation of $\tilde{n}_p(q)$ and $\tilde{n}_n(q)$, and expand them up to $q^2$:
\numparts
\begin{eqnarray}
\label{eq:np-q}
  \tilde{n}_p(q) &=& \frac{1}{(1+q^2/\beta^2)^2} \approx 1-2q^2/\beta^2+\cdots {,}
\\
\label{eq:nn-q}
  \tilde{n}_n(q) &=& \frac{\lambda q^2}{(1+q^2/\beta^2)^3} \approx \lambda q^2+\cdots {,}
\end{eqnarray}
\endnumparts
where $\beta=\sqrt{12/r_p^2}$ and $\lambda = -r_n^2/6$ are parameterized by the proton and neutron charge radius squared. We take $r_p=0.8409\;{\rm fm}$ from the $\mu$H measurement, and $r_n^2=-0.1161\;{\rm fm}^2$ from Particle Data Group~\cite{PDG2016}, and obtain $\beta=4.120\;{\rm fm}^{-1}$ and $\lambda = 0.01935\;{\rm fm}^{2}$. Therefore, $\tilde{n}_c(q)\tilde{n}_{c'}(q)$ is expanded up to $q^2$ as
\numparts
\begin{eqnarray}
\label{eq:np-np}
\tilde{n}_p^2(q) \approx 1-4q^2/\beta^2 {,}
\\
\label{eq:nn-np}
\tilde{n}_n(q)\tilde{n}_p(q) \approx \lambda q^2 {,}
\\
\label{eq:nn-nn}
\tilde{n}_n^2(q) \approx 0 {.}
\end{eqnarray}
\endnumparts
The low-$q$ truncation leads to an approximated treatment of proton-proton correction and proton-neutron overlap contribution to $\delta_{\rm pol}^{A}$. As indicated by Eq.~\eqref{eq:nn-nn}, the neutron-neutron contribution enters at one order higher and is thus omitted.

By inserting Eq.~\eqref{eq:np-np} into Eq.~\eqref{eq:PNRpp}, we rewrite ${W}^{pp}$ as
\begin{eqnarray}
\label{eq:PNRppfsa}
\fl
{W}^{pp} &\approx& 
         -Z^2 \phi^2(0) \left[ 1 + \frac{2}{\beta^2}\nabla_R^2 
                                 + \frac{2}{\beta^2}\nabla_{R'}^2 \right]
        \int \frac{d^3 q}{(2\pi)^3}
        \left(\frac{4\pi\alpha}{q^2}\right)^2 \frac{1}{\frac{q^2}{2m_r}+\omega_N}
        \left(e^{i\bs{q}\cdot(\bs{R}-\bs{R}')}-1 \right) {.} 
\cr
\fl &&
\end{eqnarray}
After integrating over the $q$ dependence in Eq.~\eqref{eq:PNRppfsa}, the term $1$ in the leftmost square bracket of Eq.~\eqref{eq:PNRppfsa} yields exactly the point-nucleon result $\delta_{\rm pol}^{\rm NR}$ as indicated by Eq.~\eqref{eq:W2}. By dropping the ``1'' term, the remaining expression in Eq.~\eqref{eq:PNRppfsa} results in the proton-proton correction $\Delta {W}^{pp}$. Utilizing the $\eta =\sqrt{2m_r\omega_N} |\bs{R}-\bs{R}'|$ expansion up to the fourth order, we have 
\begin{eqnarray}
\label{eq:P-NR-pp-series}
\fl
\Delta {W}^{pp} &=& \frac{4\pi}{3 \beta^2} (Z\alpha)^2 \phi^2(0) \sqrt{\frac{2m_r}{\omega_N}}
       (\nabla_R^2 + \nabla_{R'}^2 )
       \cr 
\fl 
&&\times
       \left\lbrace |\bs{R}-\bs{R}'|^2 -\frac{1}{4} \sqrt{2m_r
             \omega_N} |\bs{R}-\bs{R}'|^3 + \frac{1}{10}  m_r \omega_N
       |\bs{R}-\bs{R}'|^4\right] 
\nn
\fl
&=& 
       \frac{8\pi}{3\beta^2} (Z\alpha)^2 \phi^2(0) \sqrt{\frac{2m_r}{\omega_N}}
       \left[ 6 - 3 \sqrt{2m_r
             \omega_N} |\bs{R}-\bs{R}'| + 2 m_r \omega_N
       |\bs{R}-\bs{R}'|^2 \right] {.} 
\end{eqnarray}
The first term in the bracket does not contribute to the nuclear polarizability, since it does not depend on $\bs{R}$ and $\bs{R}'$, Consequently the leading proton-size correction to $\delta_{\rm pol}^{A}$ is
\begin{equation}
\label{eq:d1-pp}
\fl
\delta^{(1)}_{pp} =\ -\frac{16\pi m_r}{\beta^2}  (Z\alpha)^2 \phi^2(0)        
       \iint d^3 R d^3 R' |\bs{R}-\bs{R}'| \,
       \left(\rho_0^{pp}(\bs{R},\bs{R}')-\rho_0^p(\bs{R})\rho_0^p(\bs{R}')\right) {.}
\end{equation}

Similarly, the sub-leading proton-proton correction is analyzed from the last term in Eq.~\eqref{eq:P-NR-pp-series}, which yields
\begin{equation}
\label{eq:d2-pp}
\delta^{(2)}_{pp} = -\frac{256\pi^2 m_r^2}{9\beta^2} (Z\alpha)^2 \phi^2(0) \int^\infty_{0} d\omega\, \sqrt{\frac{\omega}{2m_r}} S_{D_1}(\omega) {.}
\end{equation}

The neutron-proton overlap muonic matrix element $\Delta {W}^{np}$ is calculated in the $\eta$-expansion up to the fourth order:
\begin{eqnarray}
\fl
\Delta {W}^{np} &=& {W}^{np}+{W}^{pn}
\nn
\fl
&=& -\frac{2\lambda \pi}{3} (Z\alpha)^2 \phi^2(0) \sqrt{\frac{2m_r}{\omega_N}}
       (\nabla_{R}^2+\nabla_{R'}^2)
\nn
\fl
&&\times
       \left[ |\bs{R}-\bs{R}'|^2 -\frac{1}{4} \sqrt{2m_r
             \omega_N} |\bs{R}-\bs{R}'|^3 + \frac{1}{10}  m_r \omega_N
       |\bs{R}-\bs{R}'|^4 \right]
\nn
\fl
&=& -\frac{4\lambda\pi}{3} (Z\alpha)^2 \phi^2(0) \sqrt{\frac{2m_r}{\omega_N}}
       \left[ 6 - 3 \sqrt{2m_r
             \omega_N} |\bs{R}-\bs{R}'| + 2 m_r \omega_N
       |\bs{R}-\bs{R}'|^2\right] {.} 
\end{eqnarray}

Similarly, we drop the constant term in the last bracket using orthogonality condition. Therefore, the leading neutron-proton overlap correction to $\delta_{\rm pol}^{A}$ is
\begin{equation}
\label{eq:d1-np}
\fl
\delta^{(1)}_{np} = 8\lambda \pi m_r (Z\alpha)^2 \phi^2(0) 
       \iint d^3 R d^3 R' |\bs{R}-\bs{R}'| \,
       \left[\rho_0^{np}(\bs{R},\bs{R}')-\rho_0^n(\bs{R})\rho_0^p(\bs{R}')\right] {,}
\end{equation}
where the neutron-proton two-body density $\rho_0^{np}$ is given in Eq.~\eqref{eq:rhocc_0}.

The sub-leading n-p overlap contribution is written as
\begin{equation}
\delta^{(2)}_{np} = \lambda  \frac{128\pi^2 m_r^2 }{9}(Z\alpha)^2 \phi^2(0) \int^\infty_{0} d\omega\, \sqrt{\frac{\omega}{2m_r}} S_{D_{1}}^{(np)}(\omega) {,}
\end{equation}
where $S_{D_{1}}^{(np)}$ denotes a neutron-proton overlapping dipole response function:
\begin{equation}
\label{eq:Sd1-np}
\fl
S_{D_{1}}^{(np)}(\omega) = \frac{1}{2J_0+1} {\rm Re} \SumInt \limits_{N\neq N_0,J} \bra N_0 J_0 ||\hat D_1^{n\dagger} ||N J\ket  \bra N J ||\hat D_1^p ||N_0 J_0 \ket \delta(\omega-\omega_N) {,}
\end{equation}
with $D_1^{n/p}\equiv\frac{1}{Z} \sum_a^A R_a Y_1(\hat R_a)\hat{e}_{n/p,a}$ defining the neutron/proton dipole operator. 

For both $D_1^{n}$ and $D_1^{p}$, the iso-scalar part of both operators induces a nuclear center-of-mass motion and vanishes between the nuclear ground and excited states. The remaining iso-vector parts of both operators have an opposite sign, this yields the following condition
\begin{equation}
\bra N_0 J_0 ||\hat D_1^{n} ||N J\ket = - \bra N_0 J_0 ||\hat D_1^p ||N J \ket {.}
\end{equation}
Therefore, we have $S_{D_{1}}^{(np)}=-S_{D_1}$, which holds for all nuclei. We then rewrite the sub-leading neutron-proton overlap contribution as
\begin{equation}
\label{eq:d2-np}
\delta^{(2)}_{np} = -\lambda  \frac{128\pi^2 m_r^2 }{9}(Z\alpha)^2 \phi^2(0) \int^\infty_{0} d\omega\, \sqrt{\frac{\omega}{2m_r}} S_{D_1}(\omega) {.}
\end{equation}

Combining Eqs.~\eqref{eq:d1-pp} and \eqref{eq:d1-np}, we have the dominant nucleon-size correction, $\delta^{(1)}_{R1} +\delta^{(1)}_{Z1}$,
\numparts
\begin{eqnarray}
\label{eq:dns-R1}\fl
\delta^{(1)}_{R1} 
&=& -8\pi m_r (Z\alpha)^2 \phi^2(0)        
    \int\int d^3 R d^3 R' |\bs{R}-\bs{R}'| 
    \left[\frac{2}{\beta^2}\rho_0^{pp}(\bs{R},\bs{R}')-\lambda\rho_0^{np}(\bs{R},\bs{R}')\right] {,}
\\
\label{eq:dns-Z1}\fl
\delta^{(1)}_{Z1} 
&=& 8\pi m_r (Z\alpha)^2 \phi^2(0)        
    \iint d^3 R d^3 R' |\bs{R}-\bs{R}'| \rho_0^p(\bs{R})
    \left[\frac{2}{\beta^2} \rho_0^p(\bs{R}')-\lambda\rho_0^n(\bs{R}')\right] {.}
\end{eqnarray}
\endnumparts
Adding $\delta^{(1)}_{Z1}$ in Eq.~\eqref{eq:dns-Z1} to $\delta^{(1)}_{Z3}$ in Eq.~\eqref{eq:d1-Z3}, we obtain the nuclear part of the elastic Zemach contribution as
\begin{equation}
\label{eq:Zem_A}
\delta_{\rm Zem}^{A} = - \left[ \delta^{(1)}_{Z3} +\delta^{(1)}_{Z1}\right] {.}  
\end{equation}

The combination of Eqs.~\eqref{eq:d2-pp} and \eqref{eq:d2-np} gives the sub-dominant nucleon-size correction
\begin{equation}
\label{eq:dns-d1}
\delta^{(2)}_{NS} 
= -\frac{128}{9}\pi^2 m_r^2 (Z\alpha)^2 \phi^2(0) \left[\frac{2}{\beta^2}+\lambda\right]
    \int^\infty_{0} d\omega \sqrt{\frac{\omega}{2m_r}} S_{D_1}(\omega) {.}
\end{equation}

\subsection{Intrinsic nucleon two-photon exchange}
Besides the two-photon exchange contribution which probes the nuclear structures, the intrinsic nucleon TPE effects also make corrections to the muonic atom spectrum. When the muon exchanges two photons with a single nucleon at a short-time scale, it probes only the internal structure of a single proton (or neutron), which is independent of the nuclear wave function.

\subsubsection{Nucleon elastic Zemach contribution}
The inclusion of nucleon-size correction in Section~\ref{sec:NS-correct} is based on a low-$q$ expansion of the proton (or neutron) electric form factors. In other words, the expansion is done around the point-nucleon limit. Therefore, $\bs{R}$ and $\bs{R}'$ still represent the positions of point-like nucleons. When $\bs{R}=\bs{R}'$, the muon exchanges two photons with a single proton (or neutron) due to its intrinsic nucleon charge distribution. However, such contributions are not included in Section~\ref{sec:NS-correct}.

In order to consider the missing intrinsic nucleon contribution, we rewrite the muon matrix elements by introducing the convolution of nucleon charge density
\begin{equation}
\label{eq:muon_matrix}
\tilde{W}^{cc'}(\bs {\Delta R}, \omega_N)
\propto
\int d^3 x d^3 x' n_c(x) n_{c'}(x') W(\bs {\Delta R} + \bs {\Delta x}, \omega_N) {,}
\end{equation} 
where $W$ on the right hand side has the same functional form as in Eq.~\eqref{eq:W2}, but the argument $\bs {\Delta R}\equiv \bs {R}-\bs {R}'$ of the point-proton case is shifted by $\bs {\Delta x}\equiv \bs {x} - \bs {x}'$. The expansion of $\Delta x/\Delta R$ reproduces all the non-relativistic contributions in the point-nucleon limit (Section~\ref{sec:polar-nonrel}), plus the finite-nucleon-size corrections (Section~\ref{sec:NS-correct}). However, by taking the limit $\bs {R}=\bs {R}'$, we obtain additional non-vanishing parts $\tilde{W}^{cc}(0, \omega_N)$ with $c=p,n$. Since $\tilde{W}^{cc}(0, \omega_N)$ is independent of the nuclear coordinates, the corresponding nuclear transition amplitude $\bra N_0 | \tilde{W}^{cc}(0, \omega_N) |N\ket=0$ vanishes due to orthogonality condition. Therefore, it does not yield corrections to $\delta_{\rm pol}^{A}$.

As shown in Eq.~\eqref{eq:Zem_A}, the nuclear elastic Zemach moment $\delta_{\rm Zem}^{A}$ cancels exactly an inelastic term of $\delta_{\rm pol}^{A}$, i.e., $\delta^{(1)}_{Z3} +\delta^{(1)}_{Z1}$. However, this cancellation is derived in the $\Delta x/\Delta R$ expansion. When using closure, $\delta^{(1)}_{Z3}$ and $\delta^{(1)}_{R3}$ are both corrected by an additional muon matrix element at $\Delta R=0$:
\begin{equation}
\label{eq:Wcc-np}\fl
\Delta \tilde{W}^{cc(1)} = -\frac{\pi m_r}{3} \alpha^2 \phi^2(0) \sum_a^A \iint d^3 x d^3 x' 
       |\bs{x}-\bs{x}'|^3 n_c(\bs{x}) n_c(\bs{x}') \hat e_{c,a} {.}
\end{equation}
The double-integrals lead to the intrinsic third Zemach moments of the proton ($c=p$), $\bra r^3_p\ket_{(2)}$, and of the neutron ($c=n$), $\bra r^3_n\ket_{(2)}$.
As shown by Friar~\cite{Friar:2013rha}, their combination gives an additional correction to $\delta_{R3}^{(1)}$:
\begin{eqnarray}
\label{eq:ZemN}
\delta_{\rm Zem}^{N}  &=& -\frac{\pi}{3}m_r\alpha^2 \phi^2(0) \left[ Z \bra r^3_p\ket_{(2)} + (A-Z) \bra r^3_n\ket_{(2)} \right]~.
\end{eqnarray}
where the neutron third Zemach moment is much smaller than the proton one.

Similarly, an opposite contribution, {i.e.}, $-\delta_{\rm Zem}^{N}$ enters as an additional nucleon-size correction to $\delta_{Z3}^{(1)}$, which cancels exactly the part in $\delta_{R3}^{(1)}$. Therefore, the overall effects of the nucleon elastic Zemach contribution $\delta_{\rm Zem}^{N}$ does not make corrections to $\delta_{\rm pol}^{A}$, which is consistent with our statement above based on nuclear orthogonality. Note that $\pm \delta_{\rm Zem}^{N}$ is not included by either $\delta_{R1}^{(1)}$ or $\delta_{Z1}^{(1)}$, which is derived as a subleading term in the $\Delta x/\Delta R$ expansion.

Now we turn to the elastic two-photon exchange contribution. The elastic Zemach contribution $\delta_{\rm Zem}$ defined in Eq.~\eqref{eq:zemach3} involves the full normalized charge distribution of a nucleus. Here we calculate $\delta_{\rm Zem}$ using the expansion around the point-nucleon limit, and also include the intrinsic nucleon contribution. It is then straightforward to show that, $\delta_{\rm Zem}$ is calculated as
\begin{eqnarray}
\delta_{\rm Zem} &= \delta_{\rm Zem}^A + \delta_{\rm Zem}^N
= - \left[\delta^{(1)}_{Z3} + \delta^{(1)}_{Z1} \right] + \delta_{\rm Zem}^N {.}
\end{eqnarray}
Combining the elastic and inelastic pieces, $\delta_{\rm Zem}^N$ enters as a non-vanishing correction to the two-photon exchange contribution.

By omitting the tiny contribution of $\bra r^3_n\ket_{(2)}$ in Eq.~\eqref{eq:ZemN}, $\delta_{\rm Zem}^N$ is proportional to $(Z m_r)^4\alpha^5$. Therefore, $\delta_{\rm Zem}^N$ in a muonic atom $\mu{}{\rm X}$ is scaled to the elastic two-photon exchange contribution in $\mu{\rm H}$ by
\begin{equation}
\label{eq:Zem_N_scal}
\delta_{\rm Zem}^N (\mu {\rm X}) = \left[\frac{Z m_r(\mu{}{\rm X})}{m_r(\mu {\rm H})}\right]^4 \delta_{\rm Zem}^{N} (\mu {\rm H}) {.}
\end{equation}

\subsubsection{Nucleon polarizability} \label{sec:polN}
When the muon exchanges two photons with a single nucleon, the nucleon itself is virtually excited in this process. This yields the intrinsic nucleon polarizability contribution $\delta_{\rm pol}^{N}$.
Each nucleon's $\delta_{\rm pol}^{N}$ is scaled with the muonic-atom wave function squared $\phi^2(0)$. By assuming the neutron polarizability is approximately of the same size as the proton polarizability, we relate the intrinsic nucleon polarizability effects in a muonic atom $\mu{}{\rm X}$, to that in $\mu{\rm H}$ by
\begin{equation}
\label{eq:pol_N_scal}
\delta_{\rm pol}^N (\mu{}{\rm X})
= A \left[\frac{\phi_{\mu{}{\rm X}}(0)}{\phi_{\mu {\rm H}}(0)}\right]^2 \delta_{\rm pol}^N (\mu {\rm H})
= A \left[\frac{Z m_r(\mu{}{\rm X})}{m_r(\mu {\rm H})}\right]^3 \delta_{\rm pol}^N (\mu {\rm H}) {.}
\end{equation}

\section{Numerical Methods}
\label{sec:numerics}
In order to evaluate the two-photon exchange nuclear polarizability effects 
on the spectrum of light muonic atoms one needs to calculate various
moments of the nuclear densities and weighted integrals over different
response functions. In this section we present the numerical methods we have
used to calculate these quantities. 

Nuclear densities, such as the charge density in Eq.~\eqref{eq:rhoc_0},
are ground state expectation values. For their evaluations 
an accurate solution of the nuclear ground state wave function is needed. 
Nowadays, mainly due to the increase in available computing power, solving the nuclear 
Hamiltonian for the  
ground state of light nuclei $A\leq 4$ is not that demanding,
and an array of available techniques are up to the task, see, e.g., Refs.~\cite{4HeBenchmarkAV8,Leidemann:2012hr}.

In contrast, calculating the response functions is a completely different matter. Considering for 
example the dipole response function in Eq.~\eqref{eq:d1-response}, we see that to evaluate this expression one must
sum over the full nuclear excitation spectrum, which for light 
nuclei consists of continuum states. Consequently, variational techniques which are very efficient at calculating the ground state may not suffice, and an expansion over
local, square-integrable, basis functions is not even formally correct as continuum states are  
non square-integrable. 
Obtaining an  ab initio solution for all the continuum spectrum is a challenging task, often out of reach. 
Ergo, indirect methods, such as the Lorentz integral transform method (LIT)~\cite{EFROS94,REPORT07}, 
are presently among the few viable ways to calculate response functions.
Even so, obtaining accurate results from 
an explicit integration of the response function which we need for evaluating the
two-photon exchange effects, see, e.g., Eq. \eqref{eq:delta_D1},
may be a rather demanding task.

In our study of the two-photon exchange contributions to the muonic atom
spectrum we have used two methods to calculate the generalized sum-rules (GSR) $I$
of a response function $S_{O}(\omega)$ 
\be\label{eq:GSR}
   I =\int_{\rm Threshold}^\infty d\omega g(\omega)S_{O}(\omega),
\ee
with an arbitrary weight function $g(\omega)$.
At first, we have used the LIT method to calculate the response functions $S_{O}(\omega)$ 
and then used numerical integration over $\omega$ to evaluate the GSRs.
Later on we have realized that for smooth weight functions $g(\omega)$, the 
GSRs can be evaluated directly and more efficiently without explicit calculation of the response functions.
We have dubbed this technique for evaluating GSRs, 
 `the Laczos sum rule (LSR) method' \cite{Nevo14}. It can be used with any
diagonalization method and is very similar to the moments method often used 
in the frame work of shell model calculations (see, e.g., Ref.~\cite{WHITEHEAD1980313}).

The main advantage of both the LSR and the LIT methods stems from the fact that
the GSRs can be calculated numerically using a  set of localized square-integrable
basis functions ~\cite{EFROS94,REPORT07}. 
Taking advantage of this fact, we have used the harmonic oscillator (HO) basis functions
to solve the two-body problem and the hyperspherical harmonics (HH) expansion to solve
the three- and four-body problems. In the latter case, we have used the effective interaction
hyperspherical harmonics (EIHH) \cite{EIHH1,EIHH2} to accelerate the convergence.
 
In the following sections we will first briefly  present 
the LSR technique and then the HO and the EIHH methods.

%=============================================================================
\subsection{The Laczos sum-rule technique}
\label{LSR-Derivation}

The derivation of the LSR method and the full discussion of its merits and subtleties
is given in Ref. \cite{Nevo14}. For completeness, we repeat here the 
principal derivation of the method.

The starting point of our discussion is a generic response function given by
\begin{equation}\label{eq:S_omg}   
   S_{O}(\omega)=\SumInt_{N} |\bra N_0 |\hat{O}| N \ket|^2
    \delta\left(E_{N}-E_{N_0}-\omega\right)\;,
\end{equation}
and the Lorentz integral transform (LIT) function ~\cite{EFROS94}
\begin{equation} \label{eq:l_sgm}
   {\cal L}(\sigma,\Gamma)=\frac{\Gamma}{\pi}\int d\omega
    \frac{S_{O}(\omega)}{(\omega-\sigma)^2+\Gamma^2}\;,
\end{equation}
which is the integral transform of the response function with a Lorentzian kernel.
If $\hat{O}$ is a spherical tensor, and if we sum over all its projections
then $S_{O}(\omega)$ in Eq.~\eqref{eq:S_omg} corresponds to the response function 
$S_{O}(\omega)$ defined in Eq.~\eqref{eq:response-O}. Here, to simplify the notation we just omit the
angular momentum  from the bra and the ket and we work with matrix elements, as opposed to reduced matrix elements,
the difference being a trivial factor.
As detailed in Refs.~\cite{EFROS94,REPORT07}, the LIT function is the norm
\be
   {\cal L}=\bra \tilde\Psi|\tilde\Psi\ket
\ee
of the square-integrable solution $|\tilde \Psi\ket$ of the Schr\"odinger-like equation,
\be
\label{liteq}
   \left( H_{\rm nucl}-E_{N_0}-\sigma+i\Gamma \right)|\tilde\Psi\ket = \hat{O}|N_0\ket \;.
\ee 
Because of the spatial fall-off  of the ground state at large distances $ | N_0\ket \longrightarrow 0$, the r.h.s. of Eq.~(\ref{liteq}) vanishes.
Thus, the solution $|\tilde \Psi\ket$ must follow the same behavior and,
 for the operators of concern here, $|\tilde \Psi\ket$ is indeed a square-integrable function.

In order to derive the LSR formula, let us assume that there exists a function $h(\sigma,\Gamma)$
such that the weight function $g(\omega)$ in Eq.~\eqref{eq:GSR} can be 
written as
\begin{equation}\label{eq:h_sgm}
    g(\omega)=\frac{\Gamma}{\pi}\int d\sigma
              \frac{h(\sigma,\Gamma)}{(\omega-\sigma)^2+\Gamma^2}\;.  
\end{equation}
Comparing this ansatz \eqref{eq:h_sgm} with Eq.~\eqref{eq:l_sgm} 
it is evident that the relation
between $g(\omega)$ and $h(\sigma,\Gamma)$ is similar to the relation between
${\mathcal L(\sigma, \Gamma)}$ and $S_{O}(\omega)$. 
There is, however, one important difference: 
for any physical response function, the LIT integral ${\cal L}(\sigma,\Gamma)$ is well defined. 
 In contrast, the existence of $h(\sigma,\Gamma)$ is not self evident, but for a smooth 
 enough $g(\omega)$ or small enough $\Gamma$, Eq.~\eqref{eq:h_sgm} holds true,
 see Ref.~\cite{Nevo14}.

Inserting the weight function (\ref{eq:h_sgm}) into
the GSR of  Eq.~(\ref{eq:GSR}) and changing the order of  
integration, we can rewrite the GSR in terms of 
${\cal L}(\sigma,\Gamma)$ and $h(\sigma,\Gamma)$ instead of $S_{O}(\omega)$ and $g(\omega)$ as
\begin{eqnarray}\label{eq:I_Lh}
  I &=& \int d\omega\, \int d\sigma \,S_{O}\left(\omega\right)
        \frac{\Gamma}{\pi}
        \frac{h(\sigma,\Gamma)}{(\omega-\sigma)^2+\Gamma^2}     
\cr &=& \int d\sigma \, {\cal L}(\sigma,\Gamma) h(\sigma,\Gamma)\;.
\end{eqnarray}

The advantage of introducing the LIT function ${\cal L}(\sigma,\Gamma)$ stems 
from the fact that, as we have seen, it can be
calculated using square-integrable basis functions. 
Utilizing this property 
we expand ${\cal L}(\sigma,\Gamma)$ over a set of localized basis functions. 
Using $M$ such basis states and diagonalizing the Hamiltonian matrix, the 
resulting eigenvalues and eigenvectors $\{E_{N_m}, |N_m\ket\}$ can be used to 
evaluate the LIT function
\begin{eqnarray}\label{L_epsmu_diag}
{\cal L}_M(\sigma,\Gamma)
                      & =&\frac{\Gamma}{\pi}\sum_{m \ne 0}^M 
                     \frac{ |\bra N_m |  \hat{O} | N_0\ket|^2}
                     {(\omega_{m}-\sigma)^2+\Gamma^2}\;.
\end{eqnarray}
where $\omega_m = E_{N_m}-E_{N_0}$.
Substituting the calculated ${\cal L}_M(\sigma,\Gamma)$ 
into \eqref{eq:I_Lh} we finally get, 
\begin{equation}\label{eq:LSR_diag}
  I_M  = \sum\limits_{m\neq 0}^{M} |\bra N_m |  \hat{O} | N_0 \ket|^2 g(\omega_{m})\;,
\end{equation}
which is the LSR formula with full diagonalization.
To some extent this result is an intuitive discrete representation of the GSR. 
Nevertheless, the above derivation justifies the use of 
a localized basis.

Due to large-model-space, in many calculations 
a complete diagonalization of the Hamiltonian is computationally impractical.
To handle this problem, one often uses the Lanczos algorithm~\cite{Lanczos}  
that maps the full $M\times M$ Hamiltonian matrix into a 
tridiagonal matrix $T_{M}$
using the recursive Krylov subspace 
$\{\hat O | N_0\ket,H \hat O | N_0\ket,H^2\hat O | N_0\ket,\ldots, H^M\hat O | N_0\ket \}$.
The power of the Lanczos algorithm lays with its convergence properties.
The low-lying eigenstates and spectral moments converge after a
relatively small number of recursion steps $M'$,
where $M'$ is often much smaller than $M$ (see, e.g., Refs.~\cite{Haxton05,Caurier95}).

Using the Lanczos algorithm, the GSR in Eq.~\eqref{eq:GSR} becomes
\begin{equation}\label{eq:LSR_lanc}
  I_{M'} = \bra N_0 | \hat{O}^\dagger  \hat{O} | N_0 \ket 
            \sum_{m\neq 0}^{M'}| Q_{m 0}|^2 g(\omega_m) \;.
\end{equation}
Here the index $M'$ denotes the number of Lanczos iterations, 
$Q$ is the unitary transformation matrix that diagonalizes ${T}_{M'}$,
and \mbox{$\omega_m \equiv E^{(M')}_{N_m} - E_{N_0}$}, where in this case
 $E^{(M')}_{N_m}$ is the $m$-th eigenvalue of ${T}_{M'}$.

If we consider an expansion on a basis of size $M$, such that the accuracy of the
calculated function ${\cal L}_M(\sigma,\Gamma)$
is within $\varepsilon_M$,
\begin{equation} 
    |{\cal L}(\sigma,\Gamma)-{\cal L}_M(\sigma,\Gamma)| \leq 
    \varepsilon_M\;,
\end{equation} 
then the accuracy of  $I_M$ calculated using the same basis is bounded by
\begin{eqnarray}\label{convcond}
  |I-I_M| 
          &\leq& \int d\sigma \, 
                 \left|{\cal L}(\sigma,\Gamma) - {\cal L}_M(\sigma,\Gamma)\right| 
                 |h(\sigma,\Gamma)|    
\cr       &\leq& 
                 \varepsilon_M 
                 \int d\sigma \,
                 |h(\sigma,\Gamma)|   \;.
\end{eqnarray} 
Therefore, if the function $h(\sigma,\Gamma)$ exists and 
the integral $\int \! d\sigma|h(\sigma,\Gamma)|$ on
the right-hand-side of Eq.~\eqref{convcond} is finite, then 
the discretized GSR in Eq.~\eqref{eq:LSR_lanc}
converges to the exact sum rule $I$ at the same rate as ${\cal L}_M(\sigma,\Gamma)$ converges to ${\cal L}(\sigma,\Gamma)$.
In other words, the discrete representation becomes exact when the LIT function converges to its exact value without any need to recover the continuum limit. 

Eqs. (\ref{eq:LSR_diag}, \ref{eq:LSR_lanc}) summarize the LSR technique which we
have used to calculate the contribution of two-photon exchange to the spectrum of
$\mu^2{\rm H}$, $\mu^3{\rm H}$, $\mu^3{\rm He}^+$, and $\mu^4{\rm He}^+$. 
For the deuteron $A=2$ case, dealing with small model spaces, we have used the full diagonalization variant. For the larger $A=3,4$ nuclei, where we have encountered larger
model spaces, the Lanczos variant 
Eq. \eqref{eq:LSR_lanc} was used.
For $^4{\rm He}$ we have made a detailed comparison between the LSR and the LIT method
and we found a very good agreement between the two approaches~\cite{Ji13}.

%=============================================================================
\subsection{The harmonic oscillator basis ($A=2$)}

To calculate the deuteron ground state wave-function and excitation spectrum
we have used the HO basis ~\cite{messiah}.
After removing the center of mass coordinate, the basis states for the
relative part of the wave-function coupled with the spin-isospin degrees of freedom
is labeled by the following set of quantum numbers
\be \label{eq:HO_basis}
  | N \ket = | {\mathfrak n} ({\mathfrak l},S)J M,\; T T_z\ket \;,
\ee
where ${\mathfrak n}$ is the principle HO quantum number, ${\mathfrak l}$ is the relative orbital
angular momentum with $z$-projection $\mathfrak m$, $S$ the spin, $J$ and  $M$ the total angular momentum and its $z$-projection,
$T$ the isospin and $T_z$ its $z$-component.
In the coordinate representation the HO basis functions are given by
\be
\label{spacial}
    \bra \bs{r} | {\mathfrak n}{\mathfrak l} {\mathfrak m} \ket = \frac{1}{\sqrt{b^{3}}}{\cal N}_{{\mathfrak n}{\mathfrak l}}\, 
            L_{\mathfrak n}^{{\mathfrak l}+\frac{1}{2}}\left(\frac{r^2}{b^2}\right)
            e^{-\frac{r^2}{2 b^2}}
    \left(\frac{r}{b}\right)^{\mathfrak l} Y_{{\mathfrak l \mathfrak m}}(\hat{\bs{r}}),
\ee
where
\be
   {\cal N}_{\mathfrak n \mathfrak l} = \sqrt{\frac{2{\mathfrak n}!}{\Gamma({\mathfrak n}+\mathfrak l +\frac{3}{2})}}
\ee
is the normalization constant and  $b=\sqrt{\hbar/M_r\Omega_{\rm{HO}}}$
the characteristic length, defined by the reduced mass $M_r$ of the proton-neutron system and the oscillator
frequency $\Omega_{\rm{HO}}$. The spatial component of the wave function in Eq.~(\ref{spacial}) will then be coupled to the spin-wave function
and multiplied by the isospin component.

\begin{figure}[h]
  \centering
    \includegraphics[width=0.75\textwidth]{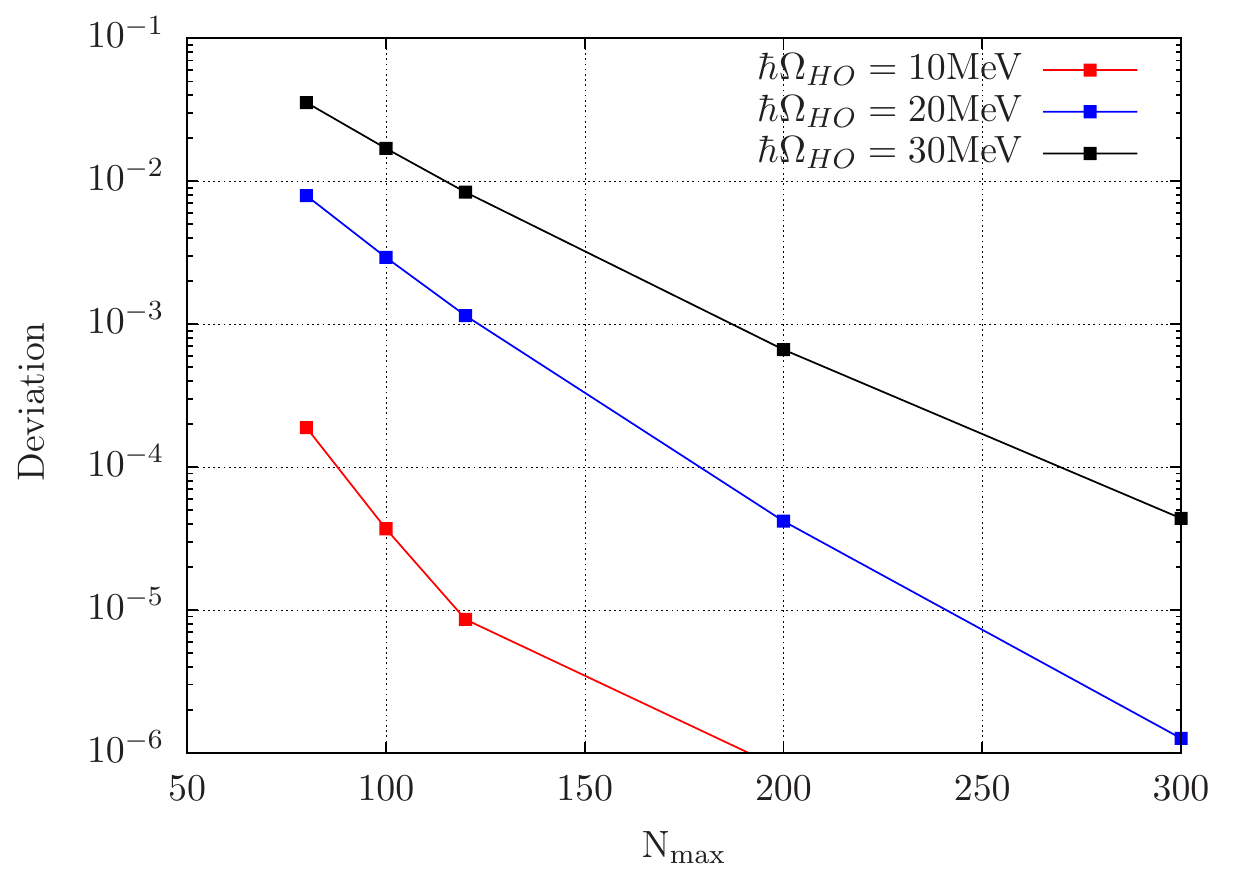}
      \caption{ The convergence of  $\delta_{D1}^{(0)}$ for $\mu ^2$H 
        as a function of $N_{\max}$ at different oscillator frequencies.
      The deviation is the fractional difference from the best estimate as in Eq. \eqref{eq:deviation}. Calculations are performed with the nuclear force from~\cite{AV18}.
      }
\label{fig:2H_D1_error}
\end{figure}

The size of the model-space is set by the harmonic oscillator levels with quantum numbers $\mathfrak n$ and $\mathfrak l$, so that 
$2\mathfrak n +\mathfrak l \le N_{\rm max}$.
When calculating the Lamb shift we have increased the value of $N_{\max}$ until satisfactory
convergence was achieved. In practice, few hundreds of basis states are sufficient.
To demonstrate this point, the convergence of $\delta_{D1}^{(0)}$  in $\mu^2{\rm H}$ is presented 
in Fig. \ref{fig:2H_D1_error}. In the figure we plot 
the deviation 
\be\label{eq:deviation}
   \rm{Deviation} = \left| 
              \frac{\delta_{D1}^{(0)}(N_{\max})}{\delta_{D1}^{(0)}(\infty)}-1 \right|
\ee
in logarithmic scale as a function of $N_{\max}$ for different oscillator frequencies,
where $\delta_{D1}^{(0)}(\infty)$ is our best estimate for
$\delta_{D1}^{(0)}$. 
Inspecting the figure one can observe an exponential convergence of $\delta_{D1}^{(0)}$ with 
the principal harmonic oscillator quantum number $N_{\max}$, and that the convergence is faster for   
$\hbar\Omega_{\rm{HO}}=10$ MeV. This fast convergence make the numerics
a negligible source of error in this case.

%=============================================================================
\subsection{The effective interaction hyperspherical harmonic method ($A=3,4$)}
%===========================================================================
To calculate the nuclear polarizability contribution to the spectrum of
muonic atoms for nuclei with mass number $A=3,4$ we have used the EIHH method.
The latter is a solver of the Schr\"odinger equation that expands the nuclear wave function on HH basis functions and
utilizes an ``effective interaction'' to accelerate convergence.
In the following subsections we first present the hyperspherical coordinates and
hyperspherical harmonics, then outline the method of effective interaction.
Full details of the method can be found in Refs.~\cite{EIHH1,EIHH2}.

%-----------------------------------------------------------------------------
\subsubsection{Hyperspherical coordinates and hyperspherical harmonics}
To separate the internal motion from the center of mass motion, 
the $A$-particle hyperspherical coordinates are defined by transformation of
the relative Jacobi coordinates $\bs{\eta}_1, \bs{\eta}_2,\ldots \bs{\eta}_{A-1}$.
In analogy with the spherical coordinates 
$(r,\hat\Omega_2=\hat{\bs{r}})$ in the two-body case, the $(3A-3)$ hyperspherical
coordinates for  $A$-particles
 are composed of one hyperradius,
\be
   \rho = \sqrt{\bs{\eta}_1^2+\bs{\eta}_2^2+\ldots +\bs{\eta}_{A-1}^2}\;, 
\ee 
and $(3A-4)$ hyperangular coordinates. Of the latter, $(2A-2)$ angular coordinates
 can be chosen to retain the 
Jacobi vector angles $\hat{\bs{\eta}}_1, \hat{\bs{\eta}}_2,\ldots \hat{\bs{\eta}}_{A-1}$. 
The remaining $(A-2)$ hyperangles are obtained by relating the norms of the Jacobi vector 
to $\rho$. For example, in the four-particle system we have three Jacobi vectors
and two such hyperangles 
$\alpha_1,\alpha_2$, defined through the relations
\begin{align}
    \eta_1 &= \rho \sin\alpha_1\,, \cr
    \eta_2 &= \rho \cos\alpha_1\sin\alpha_2\,, \cr
    \eta_3 &= \rho \cos\alpha_1\cos\alpha_2 \,.
\end{align}
In short, the hyperspherical coordinates 
include one hyperradius $\rho$ and $(3A-4)$ hyperangles, which we collectively denote 
by $\hat\Omega_A$. Written in hyperspherical coordinates, any function of the Jacobi coordinates $f(\bs\eta_1,\bs\eta_2,...\bs\eta_{A-1})$ becomes 
$f(\rho,\hat\Omega_{A})$.

In analogy to the 3-dimensional case, 
the kinetic energy operator written in these coordinates
 is separated into a hyperradial part $ \Delta_\rho $
and a hyper-centrifugal barrier $\hat{\bf K}_A^2/\rho^2$.
Here, $\hat{\bf K}_A^2$ is the  hyperangular 
momentum operator and depends on all the hyperangles. 
Accordingly, the internal
Hamiltonian for an $A$-particle system reads~\footnote{With respect to Eq.~(\ref{eq:H-nucl}) here we add the particle number $A$ in the notation.},
\begin{equation}\label{eihh_h_a}
  H_{\rm nucl}^{[A]} = - \frac{1}{2m}\Delta_{\rho}+ \frac{1}{2 m} \frac{\hat{\bf K}_A^2}{\rho^2}
           + V^{[A]}(\rho,\hat\Omega_A)\,,
\end{equation}
where $m$ is the mass of a single nucleon.

The hyperspherical harmonics $\cal Y_{[\it K]}$ are eigenfunctions of 
$\hat{\bf K}^2_A$ with eigenvalues
$K(K + 3A-5)$. They constitute a complete basis where one can expand the $A$-particle  wave function. 
For the hyperradial part we use an expansion into Laguerre polynomials $ L^\alpha_n(\rho)$
so that one has 
\begin{equation}
   \Psi(\rho,\hat\Omega_A) = \sum_{n[K]} C_{n [K]} L^\alpha_n(\rho) 
       {\cal Y}_{[K]}(\hat\Omega_A) \,.
\end{equation}
Of course the nuclear wave function must be complemented by the spin-isospin parts. 
The whole function must be antisymmetric.
This is a non-trivial task, that, however, has been solved in Refs. \cite{Nir97,Nir98}.

%-----------------------------------------------------------------------------
\subsubsection{The HH effective interaction}
\label{EIHHsubsection}
To accelerate the convergence of the HH expansion we substitute the bare
nucleon-nucleon interaction with an effective interaction \cite{NCSM,EIHH1,EIHH2,Morten94}. 
To derive the effective interaction, the Hilbert space of 
the $A$-body Hamiltonian $H^{[A]}_{\rm nucl}=H^{[A]}_0+V^{[A]}$
is divided into a model space and a residual space,
defined by the  
eigenprojectors $P$ and $Q$ of $H_0^{[A]}$,
\begin{equation}\label{eihh_pq}
   [H_0^{[A]},P]=[H_0^{[A]},Q]=0\,;\,\,\,\,QH^{[A]}_0P=PH^{[A]}_0Q=0\,;\,\,\,\,P+Q=1\,.
\end{equation}
The Hamiltonian $H^{[A]}_{\rm nucl}$ is then replaced by the 
effective model space Hamiltonian
\begin{equation}\label{eihh_heff}
   H^{[A] {\rm eff}} = PH_0^{[A]} P + P V^{[A] {\rm eff}} P \;
\end{equation}
that by construction has the same energy levels as the low-lying 
spectrum of $H^{[A]}_{\rm nucl}$. 
In general, the effective interaction defined this way
is an $A$-body interaction. Its 
construction is as difficult as finding the full-space 
solutions.
Therefore, one has to approximate $V^{[A] {\rm eff}}$. However, one must build 
the approximate effective potential in such a way that it coincides with the 
bare one for $P\longrightarrow 1$, so that 
 increasing the model space 
leads to a convergence of the eigenenergies and other observables
to the {\it true} values.
The EIHH method was developed along these lines.

In the EIHH approach we treat $\rho$ as parameter, and identify
$H_0^{[A]}$ with the hyperspherical kinetic energy  operator $\hat{\bf K}_A^2/\rho^2$.
Therefore, the model space $P$ is spanned by all the $A$-body HH with $K\leq K_{\rm max}$. 
In order to construct the effective interaction we truncated it at the
two body level, which we can easily solve, and calculate 
the two-body effective interaction $V^{[2]{\rm eff}}$ via
the Lee-Suzuki similarity transformation method \cite{LeeSuzuki80,SuzukiOkamoto83}. 
The total effective interaction is then approximated as
$V^{[A]{\rm eff}}\approx \sum_{i<j}^A V^{[2]{\rm eff}}_{ij}$.
It should be noted that $V^{[A]{\rm eff}}$ is tailored for the HH model space and 
is constrained to coincide with the bare interaction in the limit
$P\longrightarrow 1$.

\begin{figure}[h]
  \centering
    \includegraphics[width=0.75\textwidth]{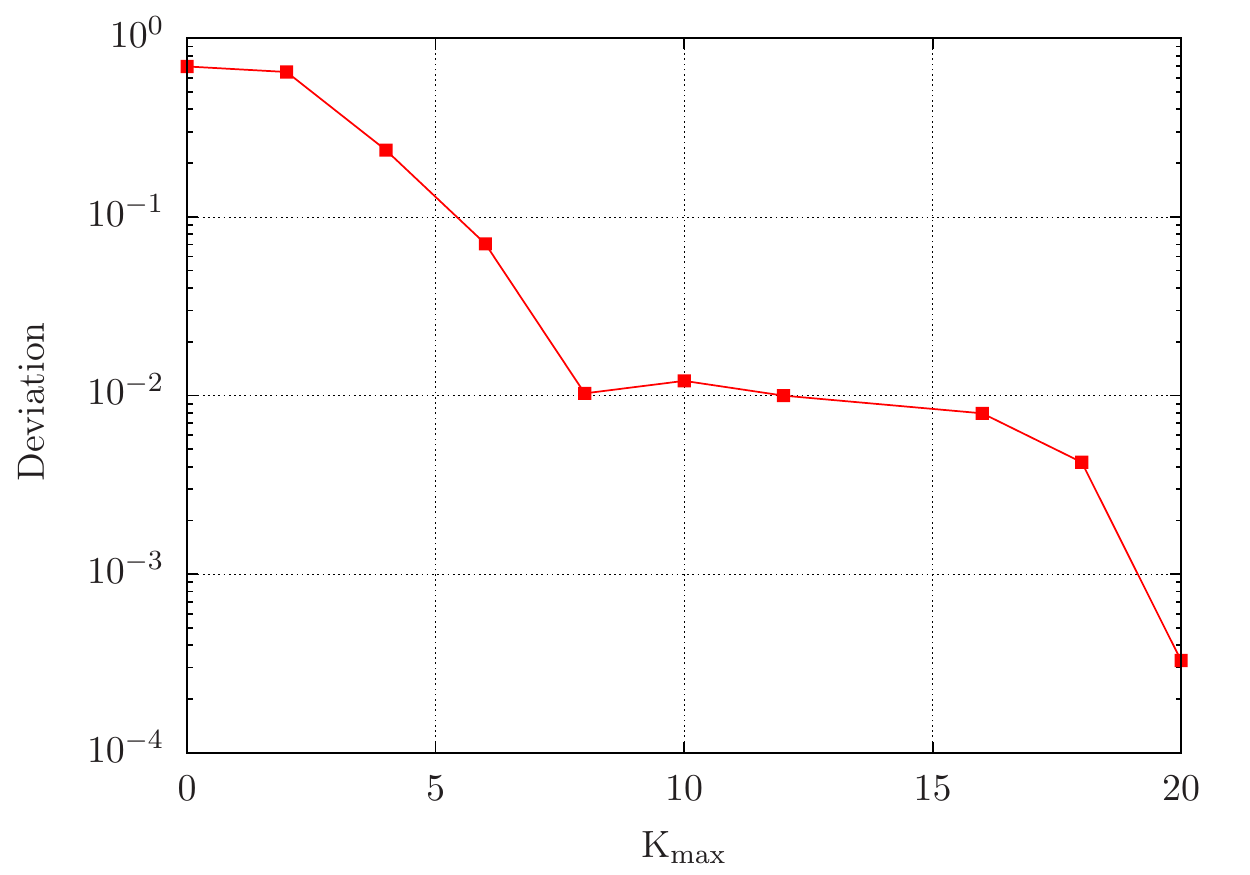}
      \caption{ The convergence of  $\delta_{D1}^{(0)}$ for $\mu^4{\rm He}^+$ as a function of 
        $K_{\max}$.
        The deviation is the fractional difference from the best estimate, Eq. \eqref{eq:deviation}. Calculations are performed with the nuclear
        force from~\cite{AV18,PuP95}.}
\label{fig:4He_D1_error}
\end{figure}

For $A=3,4$ we have repeated the calculation
with increasing values of $K_{\max}$ until satisfactory
convergence was achieved. However, for these systems the number of HH basis states grows 
rather fast with $K_{\max}$ and
therefore calculations were limited to values of $K_{\max}$ up to about 20.
Nevertheless, even for a hard core nucleon-nucleon potential such as the Argonne $v_{18}$ (AV18)~\cite{AV18} we achieve a sub
percent accuracy in $\delta_{\rm TPE}$.
To demonstrate this point, the convergence of $\delta_{D1}^{(0)}$ in $\mu^4{\rm He}^+$ is presented 
in Fig. \ref{fig:4He_D1_error}. Similarly to Fig. \ref{fig:2H_D1_error}, we plot 
the deviation \eqref{eq:deviation} as a function of $K_{\max}$.
Comparing this figure with Fig. \ref{fig:2H_D1_error} the different convergence patterns
are evident. Due to  the effective interaction we first get a rapid convergence  
to $1\%$ level. Then the results start to oscillate around the asymptotic value and we see a much 
slower rate of convergence. The small decrease seen for the $K_{\max}=20$ point 
might be just coincidental.

%=============================================================================

\section{Uncertainty estimation}
\label{sec:uncert}
The experimental precision in  muonic atom Lamb shift measurements
has achieved such a high level that, currently, the accuracy of the extracted nuclear charge 
radii  is limited by the much larger uncertainties in the theoretical nuclear-structure corrections coming from the two-photon exchange process.
For example, in $\mu^3{\rm He}^+$  theoretical uncertainties in $\delta_{\rm TPE}$
are  five times larger than the experimental uncertainty, see Table~\ref{tab:1}.
Because it is the limiting factor in the analysis of the Lamb shift experiments,
it is of paramount importance to quantify the uncertainties in these
theoretical calculations.
To this end we trace and estimate all possible sources of uncertainty in the presented ab initio calculations.  
We regard each of the uncertainty sources as an independent variable and  present its estimated
standard deviation $\sigma$. Our total uncertainty estimate is computed as 
$\sigma_{\rm total}=\sqrt{\sigma_1^2+\sigma_2^2+\ldots \sigma_n^2}$ where $\sigma_i$ is
the standard deviation of the $i\rm{th}$ uncertainty source.
Below we list all considered error sources and explain their estimation method. Note that this is a global list of uncertainty sources, and each term does not necessarily apply to both $\delta^{A/N}_{\rm Zem}$ and $\delta^{A/N}_{\rm pol}$. The overall uncertainty in $\delta_{\rm TPE}$ is estimated by quadrature sum of the uncertainties in $\delta^{A}_{\rm TPE}$ and $\delta^{N}_{\rm TPE}$. Detailed results of uncertainty evaluation in each muonic atom will be given in Sections~\ref{sec:TPE_N} and \ref{summary}.
\begin{description}
\item [ Numerical accuracy] 
	To estimate the numerical accuracy,
	 calculations are repeated for increasing model spaces until satisfactory 
        convergence is reached. 
        For the HO expansion
        ($A=2$), the model space is controlled by the parameter $N_{\rm max}$. 
        For the EIHH method ($A=3,4$) 
        the model space size is controlled
        by the maximal hyperangular momentum $K_{\rm max}$, as the hyperradial expansion
        converges rapidly. 
	Accordingly, the numerical uncertainty is taken to be 
	the difference between our best value and results obtained with lower 
        $N_{\rm max}$ or $K_{\rm max}$ values.
        
        For $A=3$, calculations with the $\chi$EFT nuclear potentials demonstrated slower convergence than for $A=4$.
        Therefore, in this case additional calculations were performed with the bare interaction, i.e., without applying the
        effective interaction mechanism described in Section~\ref{EIHHsubsection}. These calculations are variational and can be
        readily extrapolated. The final results are weighted averages of the effective interaction and bare results, with their
        respective uncertainty estimates.

\item [Nuclear model]
        The nuclear potentials, which are derived from a phenomenological or effective rather than fundamental 
        theory, introduce another source of uncertainty into the evaluation of the 
        nuclear-structure corrections.
        A simple way to assess this uncertainty is to repeat the calculations with different
        potential models and compare the results. Following this strategy, for $A=3,4$ 
        we employ in the nuclear Hamiltonian either one of the following
        state-of-the-art potentials: 
        (i) the phenomenological AV18/UIX  two-nucleon 
        \cite{AV18} plus three-nucleon \cite{PuP95} force; and (ii) a chiral
        effective field theory $\chi$EFT potential with two-nucleon \cite{Entem03}
        plus three-nucleon \cite{Na07} force.
        Using the difference between these two calculations $\Delta$ to
        evaluate the nuclear-model uncertainty we interpret the value $\pm \Delta/\sqrt{2}$
        as one standard deviation $1 \sigma$.

        For $A=2$, in Refs.~\cite{Hernandez_2014,Hernandez2018} we have performed  a
        more comprehensive study of the nuclear theory 
        uncertainties exploiting the power of the $\chi$EFT formulation. We have studied two sources of error:
        (1) the systematic uncertainty due to the freedom in the specific choice of the functional form of the
        potential~\cite{Hernandez_2014}, and (2) the statistical uncertainty
        due to the scatter in the nuclear input data used to fit nuclear forces~\cite{Hernandez2018}.
        $\chi$EFT, and effective field theories in general, furnish a systematic order-by-order
        description of low-energy processes. As such it can be utilized
        to estimate the systematic uncertainty of $\delta_{\rm TPE}$  by
        truncating at different chiral orders. 
        At any given order, the $\chi$EFT low energy constants (LECs) are fitted to 
        reproduce the appropriate nuclear data. Computational tools
        recently developed by Ekstr\"om et al.~\cite{Ekstrom2016} allow an efficient study relating the scatter in the nuclear data
        to variations of the LECs.
        Analyzing these two effects, it was found that  the
        variation of the LECs has negligible contribution to $\delta_{\rm TPE}$.
        The rigorously estimated systematic uncertainty in $\mu^2{\rm H}$ was found to be about 50$\%$ larger than what we have estimated 
        in our simple approach comparing the AV18 potential and a $\chi$EFT interaction. This finding can hardly be extrapolated to $A=3$ and $4$, since due to the
        presence of three-nucleon forces in these nuclei, nuclear-model uncertainties are larger, as we shall see later.

\item [Isospin symmetry breaking] 
        Isospin symmetry is a useful concept in nuclear physics, however it is an approximate rather 
        than an exact symmetry. In our calculations we have assumed that the total isospin $T$ is a 
        conserved quantity and  that all nucleons have equal mass, taking the average between 
        proton and neutron masses.  
	Ergo, isospin symmetry breaking (ISB) is another source of uncertainty in our
        calculations.
	In the $A=3$ nuclei, for example, 
        most of the ISB effects can be accounted for by allowing the nuclear 
        ground-state wave functions to include both total isospin channels
        $T=1/2,\;T=3/2$, and similarly for the intermediate states spanning the discretized continuum. 
	This, however, increases the number of basis states in each calculation and the associated
        computational cost rises rapidly with $K_{\rm max}$.  
	It was therefore carried out selectively only to estimate the uncertainty associated with
        performing isospin conserving calculations.

\item [Nucleon-size corrections] 
        As we explained in detail, finite nucleon-size effects are included in our calculations by expanding the neutron and
        proton form factors up to first order in $q^2$.
	Additional corrections are expected only for the Zemach and correlation terms that sum to $\delta^{(1)}_{\rm NR}$.
        The Zemach moment is roughly proportional to $r_{\rm nucl}^3\approx (r_A^2 + \tilde{r}_N^2)^{3/2}$, where $r_A$ denotes the point-proton radius and $\tilde{r}_N=(r_p^2 + r_n^2 N/Z)^{1/2}$. One can expand $r_{\rm nucl}^3$ in powers of $\tilde{r}_N/r_A$, 
	\begin{equation} 
	r_{\rm nucl}^3= r_A^3 \left(1 + \frac{3 \tilde{r}_N^2}{2r^2_A} + \frac{3 \tilde{r}_N^4}{8 r_A^4} + \cdots\right),
	\end{equation}
        where the sub-sub-leading term is smaller than the leading one by a factor ${3 \tilde{r}_N^4}/{8 r_A^4}$, and is smaller than the subleading one by ${ \tilde{r}_N^2}/{4 r_A^2}$. To roughly estimate the higher-order nucleon-size correction to $\delta^A_{\rm Zem}$, we assign the error by 
        \begin{equation}
        \label{eq:NS_error}
        \sigma_{\rm NS} [ \delta^A_{\rm Zem}] \approx  {\rm max} \left(\frac{3 \tilde{r}_N^4}{8 r_A^4} |\delta^{(0)}_{Z3}|, \frac{\tilde{r}_N^2}{4 r_A^2} |\delta^{(1)}_{Z1}| \right)~. 
        \end{equation}
        Similarly, the nucleon-size uncertainty in $\delta^A_{\rm TPE}$ is estimated by replacing $\delta^{(1)}_{Z3}$ and $\delta^{(1)}_{Z1}$ in Eq.~\eqref{eq:NS_error} with $\delta^{(1)}_{R3}$ and $\delta^{(1)}_{R1}$ (or $\delta^{(1)}_{Z3}+\delta^{(1)}_{R3}$ and $\delta^{(1)}_{Z1}+\delta^{(1)}_{R1}$ for the case of $\delta^A_{\rm pol}$).

\item [Relativistic corrections] 
	As explained in Section~\ref{sec:relativstic}, electric longitudinal and transverse
	relativistic corrections were included only for the leading non-relativistic term (keeping only the electric dipole contribution). 
	Their sum turned out to be few percent of the non-relativistic value. 
	We therefore estimate the uncertainty, due to the missing relativistic corrections to the higher-order contributions $\delta_{\rm NR}^{(2)}$ of
        the $\eta$-expansion, by assuming that they are of the same relative size with the ratio $| \delta^{(0)}_{L}+\delta^{(0)}_{T} | / |\delta^{(0)}_{D1}|$.

\item [Coulomb corrections]
        Similarly to the relativistic corrections mentioned above, also the effect of Coulomb distortions was calculated only for the leading dipole term ($\delta^{(0)}_{C}$ is the Coulomb correction to $\delta^{(0)}_{D1}$ only). Also here, following Ref.~\cite{Pachucki_2}, we estimate the uncertainty, due to missing Coulomb corrections to the higher-order contributions $\delta_{\rm NR}^{(2)}$, by assuming a similar relative size according to the ratio $|\delta^{(0)}_{C}/\delta^{(0)}_{D1}|$.

\item [$\eta$ expansion] 
        In Section \ref{sec:polar-nonrel} we have argued that the
        dimensionless parameter $\eta$ in the operator expansion is of order $\sqrt{m_r/m_p}$.
	We calculate terms up to second order in this expansion, i.e., the $(0)$, $(1)$ and $(2)$ contributions defined in Eqs.~\eqref{eq:dpol-NR-expand}. 
        This uncertainty, due to the omitted third-order corrections $(3)$ in the $\eta$-expansion, is roughly estimated based on the ratios between the calculated $(0)$, $(1)$ and $(2)$ contributions in each muonic atom.
        We are presently working on improving this uncertainty using other strategies, which give more rigorous estimates of the third-order corrections in the $\eta$-expansion. This will be relevant in particular for the $A=3$ systems, where this uncertainty is larger.

\item [The $\bs {Z\alpha}$ expansion] 
	Except for the logarithmically enhanced Coulomb distortion contribution, 
	we include  all terms of order $(Z\alpha)^5$ in our calculations of $\delta^A_{\rm pol}$. 
	Since $Z\alpha$ is small for light muonic systems, the missing contribution from all the 
        higher-order terms can be approximated by the next order in the series,
        $(Z\alpha)^6$, i.e., a correction to $\delta^A_{\rm pol}$ with a relative size that equals to $Z\alpha$, where $Z\alpha \simeq 0.7\%$ for $Z=1$, and  $Z\alpha \simeq 1.5\%$ for $Z=2$.

\item [Many-body currents]
        In chiral EFT, as the nuclear Hamiltonian admits an expansion in many-body operators, so do the electromagnetic operators. In our calculation we include only one-body operators for the electromagnetic
        charge and current operators, a procedure known as the impulse approximation. The effect of further corrections, i.e., two- and three-body currents,
        are expected to be very small and thus are neglected here. 
        The reason is that the major contribution to $\delta^A_{\rm TPE}$ is due to Coulomb interactions between the muon
        and the nucleus. The latter depend mainly on the nuclear charge density operator. 
        Many-body corrections to this operator appear in 4th order in the chiral expansion,
        and as such are expected to have a negligible effect on  $\delta^A_{\rm TPE}$.

	The magnetic dipole term $\delta^{(0)}_{M}$ comes from the current density operator instead, but it is very small.
        Nevertheless, we do include it and provide an update of its value for $A=2,3$ and new values for $A=4$ in this review. 
        The correction to the magnetic-dipole one-body impulse approximation operator appears at 2nd order in the chiral expansion and
        enhance the strength of $\delta^{(0)}_{M}$ by about 10\% for $A=2$, see, e.g., \cite{Hernandez2017}. 
	As we carried out our calculations of $\delta^{(0)}_{M}$ in the 
        impulse approximation for all light muonic atoms, an uncertainty of 10$\%$ is assigned to $\delta^{(0)}_{M}$, which is negligible
         with respect to other uncertainty sources.

\item[Hadronic corrections]
        For completeness we include in this presentation also the hadronic, i.e., neutron and proton
        TPE contributions to the muonic Lamb shift. As we did not carry these calculations 
        ourselves, we adopt the uncertainties assigned in the respective references, scaled with the number
        of nucleons and the normalization constant $\phi^2(0)$.
\end{description}

\section{Results}
\label{sec:results}

In this Section we will present an overview of results for light muonic systems, from muonic deuterium atoms to  muonic helium ions. While we will primarily focus on our own contributions, we make an effort to put them in the context of other approaches.  Apart from the very early and simplified studies performed in the 60's~\cite{JOACHAIN1961}, we will quote and compare both past and modern results from other groups and review them in the light of the new pressing quests raised after the emergence of the proton-radius puzzle. In these comparisons, special emphasis will be devoted to discussing uncertainties. For clarity, when quoting uncertainty values in percentage we will always specify whether we mean a $1 \sigma$ or a $2 \sigma$ error.

\subsection{$\mu^2{\rm H}$}

In muonic deuterium, the muon orbits the simplest possible compound nucleus, namely the deuteron, a hydrogen isotope made by a bound state of a proton and a neutron. Being  a very simple nucleus, it has been studied extensively in the literature.
The first theoretical studies of polarizability corrections for electronic/muonic deuterium using potential models date back to the 90's.
At that time, the experimental interest was directed towards understanding the isotope shift between ordinary hydrogen and deuterium atoms, where polarizability effects are small but not negligible. Their inclusion was in fact motivated by the striding progress of laser spectroscopy. Pachucki et al.~\cite{PhysRevA.48.R1} first used a square-well potential to approximate the nuclear force and 
shortly after, Lu and Rosenfelder~\cite{Lu:1993nq} analyzed both electronic and muonic deuterium using
simple separable nucleon-nucleon potentials that were lacking the one-pion exchange. 
The first to analyze the nuclear physics uncertainty on electronic and muonic deuterium were Leidemann and Rosenfelder~\cite{Leidemann:1994vq}. They related  $\delta_{\rm TPE}$ to the electromagnetic longitudinal and transverse response functions, following Rosenfelder's original derivation~\cite{Rosenfelder:1983aq}, and implemented a variety of  realistic nucleon-nucleon forces, at that time considered  state-of-the-art. While they did not include Coulomb distortions and other terms, such as the intrinsic proton polarizability term,
they observed that the uncertainty related to the nuclear force should be small and estimated it to be below 2$\%$.

%----------------------------
\begin{center}
 \begin{table}[!tb]
\caption{Comparison of nuclear polarizability contributions (in meV) for muonic deuterium  taken from Refs.~\cite{Pachucki:2011xr,Hernandez_2014,Pachucki_2,Friar:2013rha}. For the first three columns, numbers are obtained with the AV18 potential. Refs.\cite{Pachucki:2011xr,Pachucki_2} do not explicitly quote $\delta^A_{\rm Zem}$, but employ the cancellation of the corresponding elastic and inelastic terms in $\delta^A_{\rm TPE}$. For the purpose of this comparison, we show here the results on $\delta_M^{(0)}$ with the originally used  truncated weight function ${\mathcal F}_M \approx \pi{\sqrt{2\omega/m_r}}$ and not the expression of Eq.~(\ref{eq:mathFM}).}
\label{table:Comparison with Pachucki}
\begin{center}
\footnotesize
\renewcommand{\tabcolsep}{1.0mm}
\begin{tabular}{c c c c c}
\hline\hline
&~~Pachucki~\cite{Pachucki:2011xr}~~ &~~Hernandez et al.~\cite{Hernandez_2014}~~&~~Pachucki and  Wienczek~\cite{Pachucki_2}~~ &~~Friar~\cite{Friar:2013rha}~~ \\
&  (2011) & (2014)& (2015) & (2013) \\ 
\hline
$\delta^{(0)}_{D1}$ &	-1.910	& -1.907 & 	-1.910 & -1.925 \\
$\delta^{(0)}_{L}$ & 0.035 &	0.029 & 0.026 & 0.037 \\
$\delta^{(0)}_{T}$ & $-$  &	-0.012 & $-$ & $-$\\
$\delta_{HO}$      & $-$  &   $-$    & -0.004 & $-$ \\
$\delta^{(0)}_{C}$ & 0.261 & 0.262 & 0.261 & $-$ \\
$\delta^{(0)}_{M}$ & 0.016 & 0.008 & 0.008 & 0.011 \\
$\delta^{(1)}_{Z3}$ &	$-$ & 0.357 & $-$ & $-$ \\
$\delta^{(2)}_{R2}$ & 0.045 &	0.042 &	0.042 & 0.042 \\
$\delta^{(2)}_{Q}$ & 0.066 & 0.061 & 0.061 & 0.061 \\
$\delta^{(2)}_{D1D3}$ &	-0.151 & -0.139 & -0.139 & -0.137 \\
$\delta^{(1)}_{Z1}$ & $-$ & 0.064 & $-$ & $-$ \\
$\delta^{(1)}_{np}$ & $-$  &	0.017 &	0.018 & 0.023 \\
 $\delta^{(2)}_{NS}$	& $-$  &	-0.020 & -0.020 & -0.021 \\                         
\hline
$\delta^A_{\rm pol}$   &    $-$ & -1.240 & $-$ &  $-$\\
$\delta^A_{\rm Zem}$   &    $-$ & -0.421 & $-$ &  $-$\\
$\delta^A_{\rm TPE}$ & -1.638 & -1.661 & -1.657 & -1.909 \\
\hline\hline
\end{tabular}
\end{center}
\end{table}
\end{center}

After the discovery of the proton radius puzzle in 2010~\cite{Pohl:2010zza} and given that the CREMA collaboration planned  to investigate other light muonic atoms, the subject gained a  renewed interest. In 2011 Pachucki~\cite{Pachucki:2011xr} published a thorough calculation of nuclear-structure corrections in muonic deuterium, which included relativistic corrections and Coulomb corrections. He used the modern realistic AV18 nucleon-nucleon potential in his calculations.
In 2013 Friar~\cite{Friar:2013rha}  derived finite nucleon size corrections and analyzed muonic deuterium in zero-range theory,  which allows for an analytical solution. With the correct asymptotic form of the $s$-wave deuteron wave function reproduced, this calculation is similar to pion-less effective field theory at next-to-leading order, whose nuclear-physics uncertainty is expected to be $1\sim2\%$  based on a power counting analysis.
 
In 2014~\cite{Hernandez_2014} we presented our calculation of nuclear-structure corrections using modern nucleon-nucleon potentials derived from $\chi$EFT and pointed out that the effective field theory framework allows for a systematic analysis  of uncertainties related to the non-perturbative nature of nuclear forces, which we estimated to be $0.6\%$.
We also performed calculations with the AV18 potential and compared to Pachucki. 
After mutual verifications, the results  agreed very nicely, as we show in Table~\ref{table:Comparison with Pachucki}.
There, we provide a comparison of all the terms of Refs.~\cite{Pachucki:2011xr,Hernandez_2014,Pachucki_2,Friar:2013rha}, where we neglect the single nucleon polarizability contribution. Despite the additional higher order terms included by Pachucki and Wienczek~\cite{Pachucki_2} (term denoted with $\delta_{HO}$) and the slight difference in the relativistic terms  $\delta^{(0)}_L$ and  $\delta^{(0)}_T$, the final calculations of $\delta^A_{\rm TPE}$ from Ref.~\cite{Hernandez_2014} and \cite{Pachucki_2}, where there is almost a term-by-term correspondence, agree  at the level of 0.25\%. This is very reassuring since the derivation of the formulas and the numerical implementation were done by two independent groups~\footnote{For example, the slight difference in the leading $\delta^{(0)}_{D1}$ is due to the fact that Pachucki uses $m_p \ne m_n$, while we use $m_p = m_n$.}.  

%---------------------------
\begin{figure}[htb]
\centering
 \includegraphics[width=13cm]{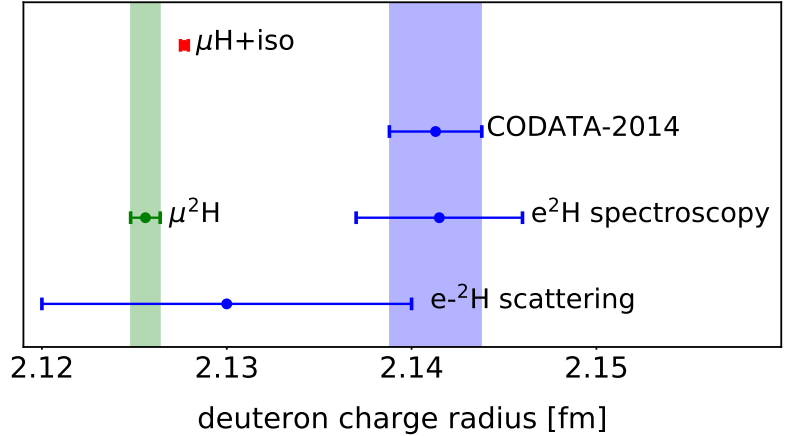}
 \caption{\label{fig_dpuzzle} 
    Deuteron radius puzzle: Recent determination of the deuteron charge radius from $\mu^2{\rm H}$~\cite{Pohl669} in comparison to the CODATA evaluation of 2014~\cite{CODATA_2014} and measurements from ordinary deuterium spectroscopy~\cite{Pohl:2016glp}. Also shown are the results obtained from electron scattering~\cite{Sick98} and the value obtained combining isotope shift~\cite{Parthey10} with muonic hydrogen~\cite{Pohl:2010zza} data, denoted with ``$\mu {\rm H} $+iso''.  Figure adapted from Ref.~\cite{Pohl669}.}
\end{figure}

In 2016, the CREMA collaboration released the  muonic deuterium Lamb shift data~\cite{Pohl669}. In analogy to the proton case, the charge radius $r_d = 2.12562(78)$ fm extracted from muonic deuterium revealed to be smaller, with a  $6.0 \sigma$  (or $7.5 \sigma$)
deviation  from the world averaged  CODATA 2014~\cite{CODATA_2014}
(or CODATA 2010~\cite{Mohr:2012tt})
value. It also deviates  by $3.5 \sigma$ with respect  to spectroscopic extractions from ordinary $e^2{\rm H}$ alone~\cite{Pohl:2016glp}.
Different from the proton case, in the so-called ``deuteron-radius puzzle''  electron scattering data~\cite{Sick98} are not precise enough to discriminate among muonic and electronic deuterium spectroscopy.
By combining the radius squared
difference $r_d^2 - r_p^2$
measured from isotope shift experiments~\cite{Parthey10} with the absolute determinations of the
proton radius from muonic hydrogen~\cite{Pohl:2010zza,Antognini13}, a value $r_d = 2.12771(22)$ fm is obtained. This is denoted with  ``$\mu{\rm H}+$iso'' in Fig.~\ref{fig_dpuzzle},
where it is shown together with the other determinations.
The ``$\mu {\rm H}+$iso'' result is very close to the absolute determination of muonic deuterium, but still differs from it by $2.6 \sigma$.  Such difference can be directly related to the $\delta_{\rm TPE}$ value used in the extraction of $r_d$ from muonic deuterium. 
In fact, from Eq.~(\ref{eq:E2s2p}), one can see that by measuring $\delta_{\rm LS}$ and knowing $\delta_{\rm QED}$,
it is possible to  extract $\delta_{\rm TPE}$ from an experimentally determined
radius. Using the $r_d$ from ``$\mu {\rm H}+$iso'' shown in Fig.~\ref{fig_dpuzzle}  leads to an experimental value  $\delta_{\rm TPE}=-1.7638(68)$ meV~\cite{Pohl669}, which differs from the theoretical summary value of Krauth et al.~\cite{Krauth:2015nja}  by $2.6 \sigma$.
In Ref.~\cite{Pohl669} it was argued that this might be due to missing contributions or underestimated uncertainties in the theoretical derivation of $\delta_{\rm TPE}$.
It is to note that the theoretical summary value of Ref.~\cite{Krauth:2015nja} does not include only results from ab initio calculations, but also results from dispersion relation analyses~\cite{Carlson:2013xea}, which despite suffering from a $35\%$  uncertainty, are compatible with the results obtained from ab initio calculations.

\begin{figure}[htb]
\centering
 \includegraphics[width=13cm]{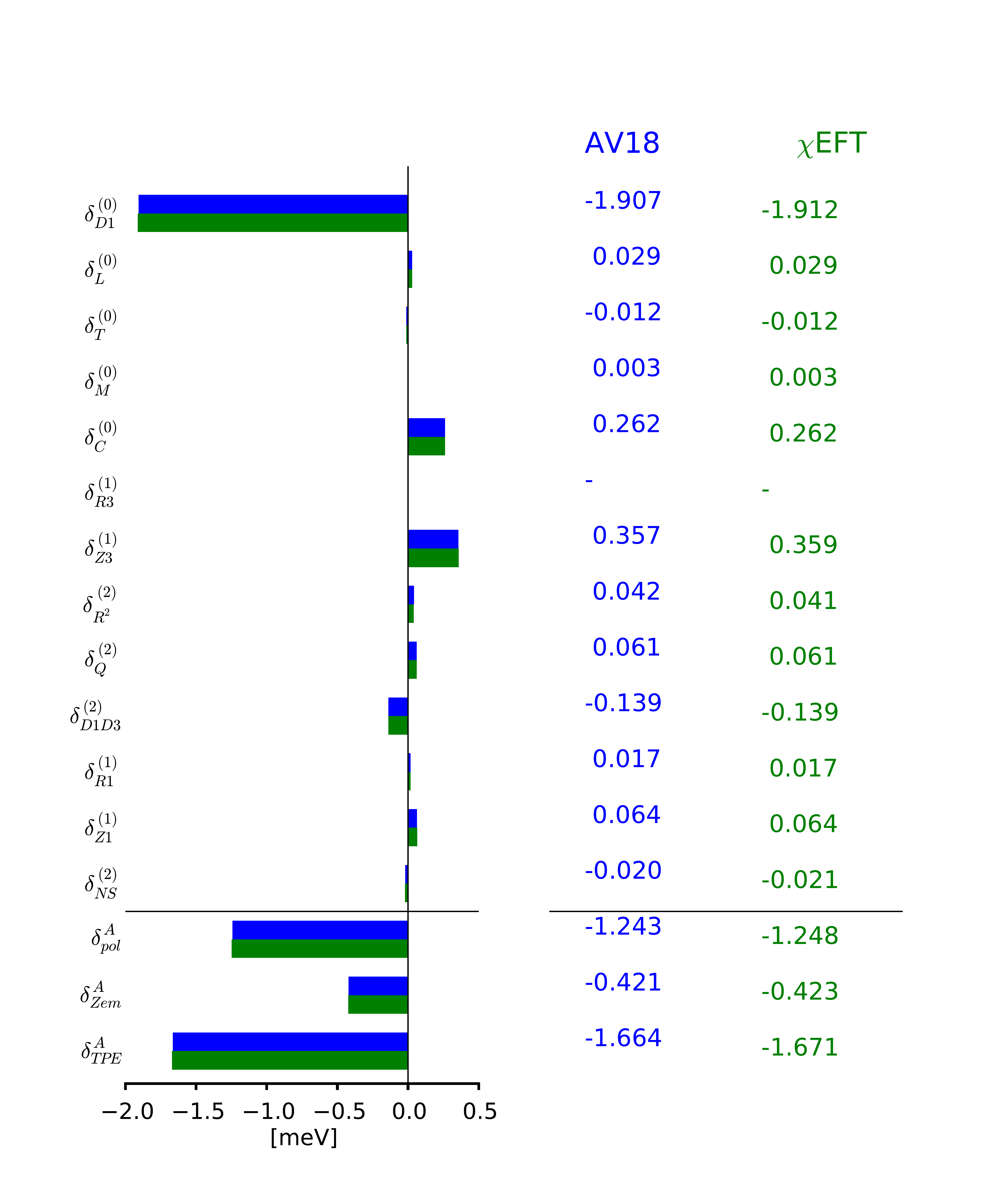}
 \caption{\label{fig:deut} Graphic representation of the various contributions 
		to $\delta^A_{\rm TPE}$ in the  Lamb shift 
		of $\mu^2{\rm H}$ calculated with the AV18~\cite{AV18} and a $\chi$EFT potential~\cite{Entem03}. Numerical values in meV are listed on the right.}
\end{figure}

Very recently, motivated by the above mentioned $2.6 \sigma$ disagreement, we have analyzed statistical and systematic uncertainties in $\chi$EFT~\cite{Hernandez2018} and obtained a result consistent with our previous one, thus still differing from the experimental determination of $\delta_{\rm TPE}$. From an analysis of the various sources of uncertainties, we learned that statistical uncertainties from propagated LECs are small and that systematic uncertainties in nuclear potentials are well under control.
Thus, we deduce that these deuteron discrepancies are unlikely to be explained  by underestimated uncertainties in $\delta_{\rm TPE}$, at least at $(Z\alpha)^5$ presently considered. Recent work by Hill and Paz~\cite{Hill:2016bjv} pointed out that single nucleon contributions might have a larger error bar than previously estimated, thus affecting the deuteron radius puzzle. The actual size of single-nucleon uncertainties still remains debated~\cite{Birse:2017czd,Hill:2017rlj}. In order to further understand the origin of this difference, from the theoretical point of view it might be interesting to investigate higher order terms in the $Z\alpha$ expansion, which require to go beyond second order perturbation theory.
Pachucki et al.~\cite{Pachucki:2018yxe} recently looked into three-photon exchange effects in muonic deuterium and found that their size is small and cannot explain the $2.6 \sigma$ discrepancy.

In Figure~\ref{fig:deut},  we show a graphic representation and the numerical results of various terms composing $\delta_{\rm TPE}$. Presented are values obtained with the AV18 potential and with the chiral nucleon-nucleon force at next-to-next-to-next-to leading order by Entem and Machleidt~\cite{Entem03} denoted by $\chi$EFT.
It is evident that  the $\delta^{(0)}_{D1}$ term dominates and $\delta^{(1)}_{Z3}$ is the second important correction, about a factor of 5 smaller. With respect to $\delta^A_{\rm pol}$, the elastic Zemach term $\delta^A_{\rm Zem}$ is three times smaller.

To evaluate the total nuclear-model uncertainty (amounting to 0.6\% as mentioned before) we did not just take the difference between the AV18 and one $\chi$EFT potential, but rather accounted for the dependence on various cutoffs of the chiral potentials and also estimated the uncertainty related to the chiral order truncation~\cite{Hernandez2018}. Including all the other sources of uncertainty, namely atomic physics, single nucleon contributions etc., as discussed in Section~\ref{sec:uncert},
the overall uncertainty of our muonic deuterium calculation amounts to 1.3$\% (1\sigma)$ on the full $\delta_{\rm TPE}$~\cite{Hernandez2018}. Details on the separate single nucleon contributions will be shown later, in the summary of Tables~\ref{tab_all} and \ref{table_pol_err}, where we will compare muonic deuterium  to  other muonic systems analyzed in this review.

Finally, it is to note that in the various calculations of Refs.~\cite{Pachucki:2011xr,Hernandez_2014,Pachucki_2,Hernandez2018},
an operator expansion was used as described in Section~\ref{sec:polar-nonrel}. Such expansion is truncated at the second order, leading to the $(0)$, $(1)$ and the $(2)$ contributions. Going to third order in this expansion is very complicated and impractical. Thus, in a recent study, we performed a different expansion, 
very similar to the approach adopted by Leidemann and Rosenfelder~\cite{Leidemann:1994vq}, which is more reliable and allows to estimate
the effects of third order corrections in the $\eta$-expansion. The latter are found to be $0.3\%$ of $\delta^A_{\rm TPE}$ for $\mu^2{\rm H}$ in the point-nucleon limit~\cite{Javier_inprep}, see also Section~\ref{summary}.

\subsection{$\mu^3{\rm H}$}

In $\mu^3{\rm H}$ the muon orbits the triton nucleus, an hydrogen isotope made by a bound state of one proton and two neutrons. Triton is a radioactive isotope of hydrogen and as such muonic tritium has not yet been studied in the laboratory. While its radioactivity would not compromise safety when used in very small quantities for  spectroscopic experiments, its low-energy $\beta$-decay emissions would produce a large background in the region where X-rays are measured in muonic atom experiments.
%-----------
\begin{figure}[htb]
\centering
 \includegraphics[width=13cm]{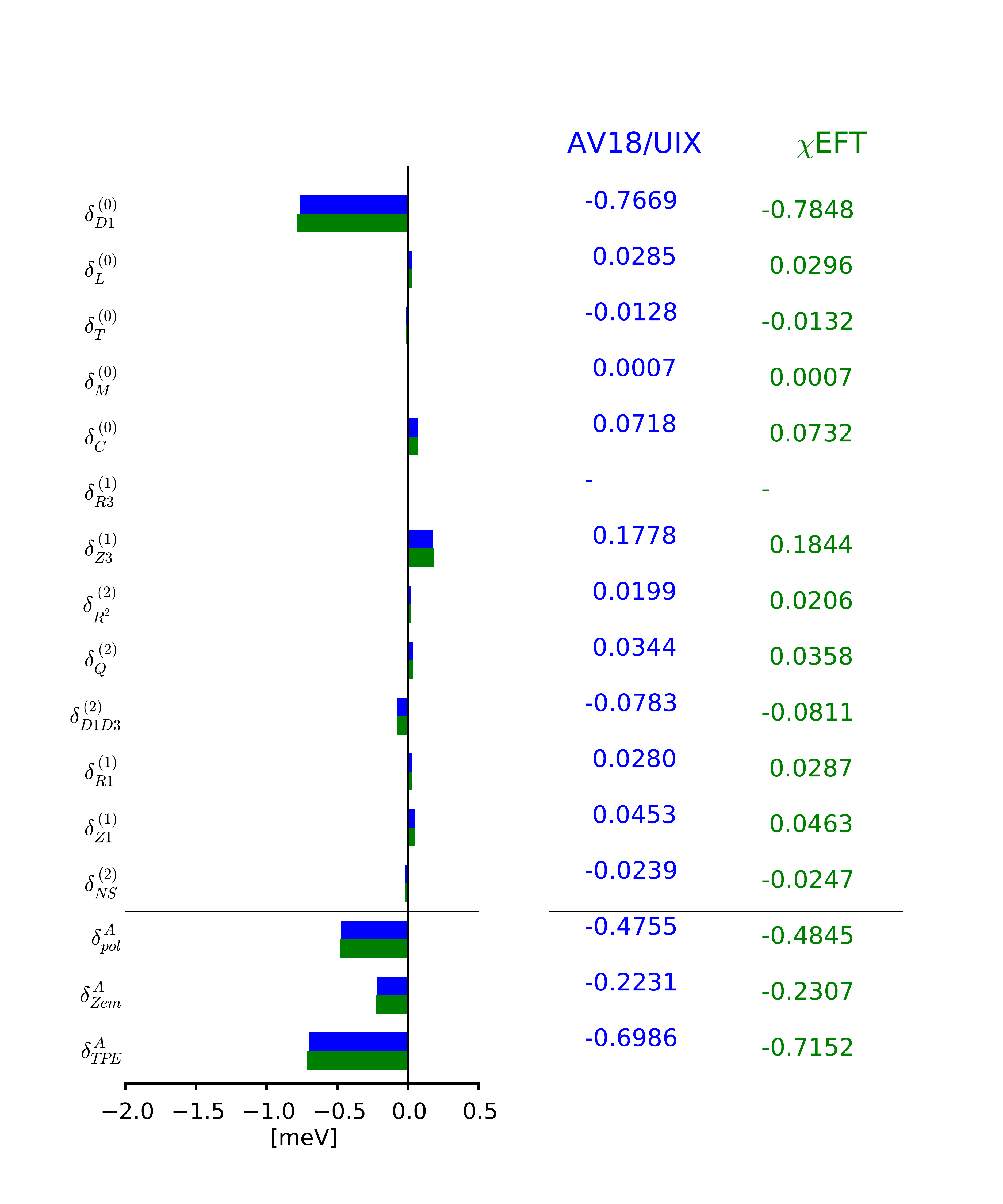}
 \caption{\label{fig:triton}  Graphic representation of the various contributions 
		to $\delta^A_{\rm TPE}$ in the Lamb shift 
		of $\mu^3{\rm H}$, calculated with the AV18+UIX~\cite{AV18,PuP95} and a $\chi$EFT nuclear Hamiltonian~\cite{Entem03,Na07}. Numerical values in meV are listed on the right.}
\end{figure}

From the theoretical point of view, solving a three-nucleon problem  is  more involved than solving the deuteron. The main difference is that one has to supplement the nuclear Hamiltonian with three-nucleon forces, which are less constrained with respect to the two-nucleon forces. Thus, one  expects nuclear physics uncertainties to be larger than in muonic deuterium.

In 2016 we performed the first ab initio calculation of $\delta_{\rm TPE}$ for $\mu^3$H, mostly to investigate structure differences with respect to the muonic deuterium and its nuclear mirror system, namely muonic $^3$He$^+$,  which is discussed in the next subsection.
As in the case of  $\mu^2{\rm H}$, also for $\mu^3{\rm H}$ there is  only one proton, thus all corrections are expected to be of the same order of magnitude due to their $Z\alpha$ dependence. However, the break-up threshold energies are different, with the triton's  being 6.28 MeV with respect to the deuteron's 2.2 MeV. Since $\delta_{\rm TPE}$ is dominated by the dipole electric transitions and $\delta^{(0)}_{D1}$ has an inverse energy weight, one expects the polarizability effects to be smaller in muonic tritium than in muonic deuterium. Indeed, this is what we find.

In Fig.~\ref{fig:triton}, we show our results for all the terms composing $\delta^A_{\rm TPE}$ calculated  using two potential sets:
the phenomenological AV18+UIX~\cite{AV18,PuP95} and one parameterization of the $\chi$EFT Hamiltonian\cite{Entem03,Na07}, both of which consist of two- and three-nucleon potentials.
Similarly to what observed in $\mu^2{\rm H}$,  $\delta^{(0)}_{D1}$ dominates also in $\mu^3{\rm H}$, with $\delta^{(1)}_{Z3}$ being the next important correction. With respect to $\delta^A_{\rm pol}$ the elastic Zemach term $\delta^A_{\rm Zem}$  in $\mu^3{\rm H}$ is only about a factor of two smaller, and not a factor of three as in $\mu^2{\rm H}$. This is mostly due to the smaller polarizability arising from the larger binding energy of $^3$H.
Finally, the potential model dependence in nuclear-structure corrections to $\mu^3$H is of the order of 3$\%$ ($1 \sigma$). The latter is enhanced with respect to $\mu^2{\rm H}$, due to the addition of the less constrained three-nucleon forces.

\subsection{$\mu^3{\rm He}^+$}

$^3$He is an isotope of helium with two protons and one neutron. In muonic $^3$He$^+$ a single muon orbits this nucleus, forming a positively charged ion. $^3$He is the mirror nuclear system with respect to $^3$H, where  protons and neutrons are exchanged. It has a very similar break-up threshold energy, the only difference being the Coulomb interaction between the two protons. Even though the large $\delta^{(0)}_{D1}$ term has an inverse energy weight, the about 1 MeV difference in threshold energy will not lead  to significant difference in nuclear-structure corrections to $\mu^3{\rm He}^+$ versus $\mu^3$H. On the contrary, what will make its TPE larger than in $\mu^3{\rm H}$ and $\mu^2{\rm H}$, is the $Z\alpha$ factor, with $Z$ being 2 for $^3$He, as opposed to 1 in $^3$H and $^2$H.

In recent times, the  interest in nuclear-structure corrections to muonic $^3$He$^+$ has been raised by the activities of the CREMA collaboration. 
Motivated by the above mentioned proton- and deuteron-radius puzzles, the  collaboration has recently measured transitions in the $\mu^3{\rm He}^+$~\cite{Antognini11}, which are presently being analyzed.
The charge radius of $^3$He can be extracted from Lamb shift measurements, provided that their nuclear polarizability is known with sufficient accuracy.

\begin{figure}[htb]
\centering
 \includegraphics[width=13cm]{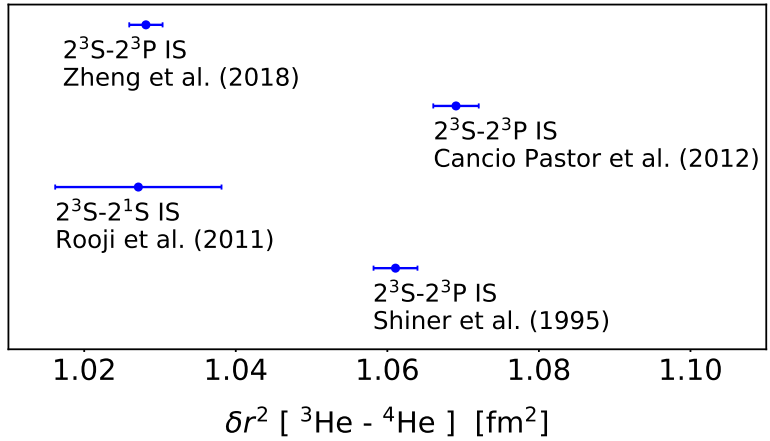}
 \caption{\label{fig_He3puzzle}
   Isotope shift radius puzzle in $^3$He. Figure adapted from Ref.~\cite{Zheng2018}. Experimental data are from Zheng et al.~\cite{Zheng2018}, Cancio Pastor et al~\cite{CancioPastor}, van Rooij et al.~\cite{Rooij}, and Shiner et al.~\cite{Shiner}.}
\end{figure}

The prospects of obtaining an accurate  charge radius for $^3$He is deemed even more interesting given that
it will shed light on yet another radius puzzle pertaining to the isotope shift in ordinary helium atoms.
Indeed, various $2S$--$2P$ atomic transitions in ordinary $^3$He and $^4$He 
have been measured by different groups. Via a combination of spectroscopic isotope shift data and accurate QED calculations,  the difference in squared radii with respect to a reference nucleus, namely $\delta{r^2} = r^2_{^3{\rm He}} - r^2_{^4{\rm He}}$ in this case, can be extracted.   The latest results on this matter were reported by   Zheng et al.~\cite{Zheng2018} and the present situation is displayed in Fig.~\ref{fig_He3puzzle}.  There, one can appreciate the discrepancy among the various measurements amounting, for example, to  $4 \sigma$ between the data from Cancio Pastor et al.~\cite{CancioPastor} and Rooij et al.~\cite{Rooij}.
The atomic physics community is very much looking forward to absolute radii determinations from muonic atoms, since such new data will have the potential to shed light on this isotope shift puzzle.  To exclude either of the determinations presented in Fig.~\ref{fig_He3puzzle} by three respective standard deviations via Lamb shifts measurements, the latter should be determined with an accuracy of 1.5 meV~\cite{Carlson3He}. Since in {$\mu^3{\rm He}^+$ QED contributions are known with an uncertainty of only 0.04 meV and the experimental uncertainty is expected to be of 0.08 meV, the above requirement translates into a direct constraint on the $\delta_{\rm TPE}$, which should be known with an uncertainty of 1.5 meV or better. Given that the size of the total $\delta_{\rm TPE}$ is of about 15 meV, as we shall see below and  in our Summary~\ref{summary}, this corresponds to a 10$\%$ accuracy.

Historically, the  early computation by Rinker in 1976~\cite{Rinker:1976en}, giving a total polarizability of $-4.9$ meV with a roughly estimated uncertainty of $20\%~(1\sigma)$, remained the last word on nuclear-structure corrections on {$\mu^3{\rm He}^+$ for about 40 years. 
Of course, that large of an uncertainty would not allow to shed light on today's isotope shift puzzle.
In 2016, we calculated the full $\delta_{\rm TPE}$ in $\mu^3{\rm H}^+$ for the first time~\cite{Nevo_Dinur_2016}, providing a thorough estimate of the related uncertainties amounting to 2.5\% ($1\sigma$). This was a dramatic reduction  with respect to the previous estimate and is well below the requested 10\% uncertainty to shed light on the isotope shift puzzle. Subsequently, Carlson et al.~\cite{Carlson3He} applied their dispersion relation method to $\mu^3{\rm He}^+$. Thanks to the fact that good experimental data exist for the electromagnetic excitation of $^3$He, the uncertainties in $\delta_{\rm TPE}$ were quite reduced with respect to the deuteron case. While, a comparison of separate contributions may be less accurate due to the fact that ab initio and dispersion relation methods are very different, the total $\delta_{\rm TPE}$ are in nice agreement with each other, with $-15.46(39)$ meV from Ref.~\cite{Nevo_Dinur_2016,Hernandez_2016} and -15.14(49) meV from Ref.~\cite{Carlson3He}.
In the summary \cite{Franke:2017tpc} prepared in anticipation of the future analysis of $\mu^3{\rm He}^+$, the average value of the previous two has been chosen.

\begin{figure}[htb]
\centering
 \includegraphics*[width=12.cm]{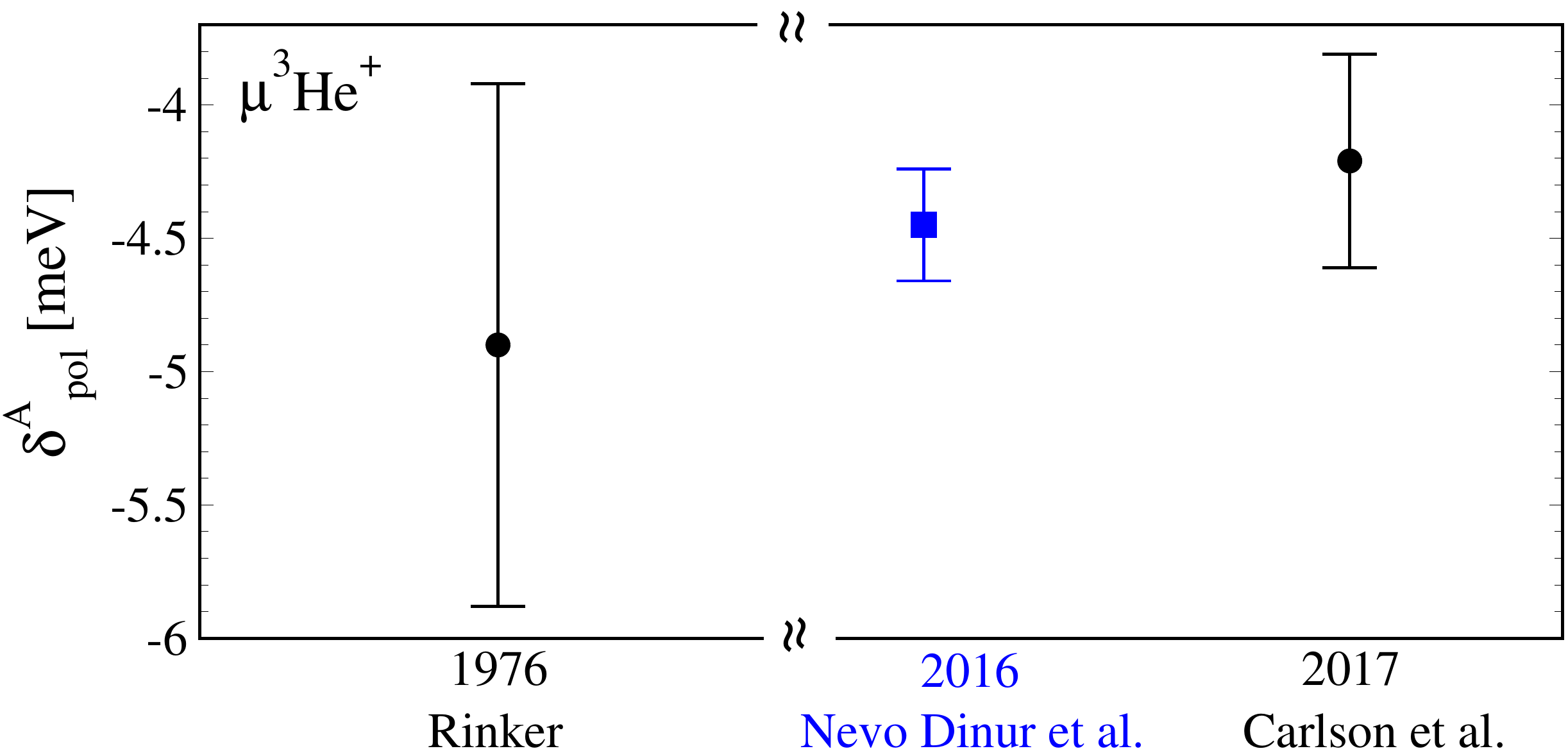}
\caption{\label{time_3He} Time evolution of  
		 $\delta^A_{\rm pol}$ estimates for $\mu^3{\rm He}^+$ with corresponding $\pm 1\sigma$ uncertainties. The ab initio computation is denoted by the square. See text for details.}
\end{figure}

In Fig.~\ref{time_3He}, we present a historical overview of the calculations of $\delta^A_{\rm pol}$ in muonic $^3$He$^+$. It is evident that ab initio computations provided  so far the most precise determination. They are consistent with dispersion relation results which have only a slightly larger uncertainty. The precision by which we know $\delta_{\rm TPE}$ today will allow the CREMA collaboration to obtain a competitive measurement of the helium radii and discern among isotopic shifts data, which would not have been possible with the $20\%$ uncertainties of the older TPE calculation.

\begin{figure}[htb]
\centering
 \includegraphics*[width=13.cm]{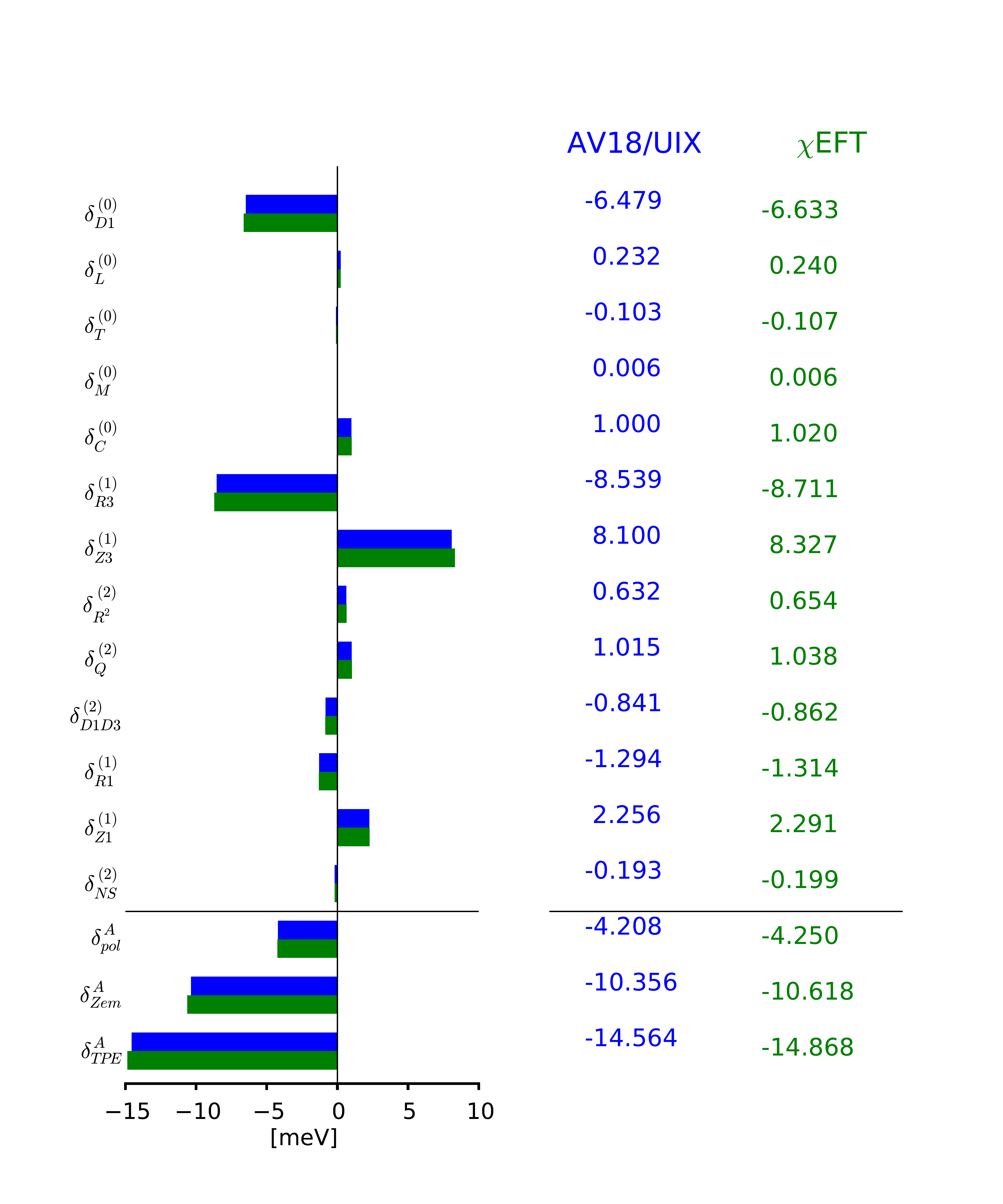}
 \caption{\label{fig_comp_3He} Same as Fig.~\ref{fig:triton} but for $\mu^3{\rm He}^+$.}
\end{figure}

In Fig.~\ref{fig_comp_3He} we finally show our results for all the terms composing $\delta^A_{\rm TPE}$  using two potential sets:
the phenomenological AV18+UIX~\cite{AV18,PuP95} and one parameterization of $\chi$EFT~\cite{Entem03,Na07}.
As previously anticipated, nuclear-structure corrections are larger in $\mu^3{\rm He}^+$ than in $\mu^3$H due to the $Z$ dependence.
Also in the $\mu^3{\rm He}^+$ case, $\delta^{(0)}_{D1}$ is the most important correction to $\delta_{\rm pol}^A$, since the two large terms appearing in $\delta^{(1)}_{\rm NR}$, namely $\delta^{(1)}_{R3}$ and $\delta^{(1)}_{Z3}$ are opposite in sign and mostly cancel out. As already mentioned in Section~\ref{sec:polar-nonrel},  $\delta^{(1)}_{R3}$ is different from zero only for $Z\ne 1$ nuclei, where  more than one proton exists. The hierarchy in the various terms of the $\eta$-expansion is also preserved. For $\mu^3{\rm He}^+$, we note that $\delta^A_{\rm Zem}$ is larger than that of $\mu^3$H, mainly due to the $(Z\alpha)^5$ scaling and it is even larger than $\delta^A_{\rm pol}$.

\subsection{$\mu^4{\rm He}^+$}

In muonic $^4$He$^+$, a single muon orbits a nucleus of $^4$He, also called alpha-particle, which is a bound state of two protons and two neutrons. Being a $Z=2$ nucleus, muonic $^4$He$^+$ is  a positively charged ion. Moreover since the nuclear spin is zero,  the structure of its atomic spectrum is simplified, in that there is no hyperfine splitting. The QED theory of the Lamb shift and of the fine structure of muonic atoms with spinless nuclei, including   $\mu^4{\rm He}^+$,  was recently revisited in Ref.~\cite{Karsheboim2018}, to which we refer the interested reader.

The first studies of nuclear-structure corrections in $\mu^4{\rm He}^+$ date back to the 70's, when the precision of contemporary experiments on muonic atoms performed at CERN~\cite{BERTIN1975411} reached a level where it became necessary to take into account the effects of the internal excitation of the nucleus.

Various theorists tackled this issue, at first aiming to estimate these effects and only later at understanding the
associated uncertainties. Bernabeau and Karlskog~\cite{Bernabeu:1973uf} used dispersion relations to connect the polarizability to electron scattering data even before the CERN experiment was performed and obtained a value of $-3.1$ meV, with no uncertainty estimates. Later Henley, Kriejs and Wilets~\cite{HENLEY1976349} used a harmonic oscillator model to microscopically describe polarizability effects and obtained a value three times larger, namely $ -13.1 ~{\rm meV} \le \delta^A_{\rm pol} \le -12.2$ meV. In 1976 Rinker~\cite{Rinker:1976en} related
polarizability corrections in muonic atoms to various measured electromagnetic transitions  and obtained a value of $-3.1$ meV, compatible with Bernabeau and Karlskog, indicating that the value from Ref.~\cite{HENLEY1976349} was wrong. His very rough uncertainty estimate was of about 20$\%$ ($1 \sigma$). Finally Friar settled the situation by confirming that a reasonable estimate of nuclear polarizability and its uncertainty should be $-3.1\pm 20\%$~\cite{Friar:1977cf}.
It should be noted that the fact that three of the early calculations obtained a value of 3.1 meV is partially a coincidence, since for example two inaccurate treatments were made by Bernabeau and Karlskog which canceled out, as pointed out by Friar.

\begin{figure}[htb]
\centering
 \includegraphics[width=10cm]{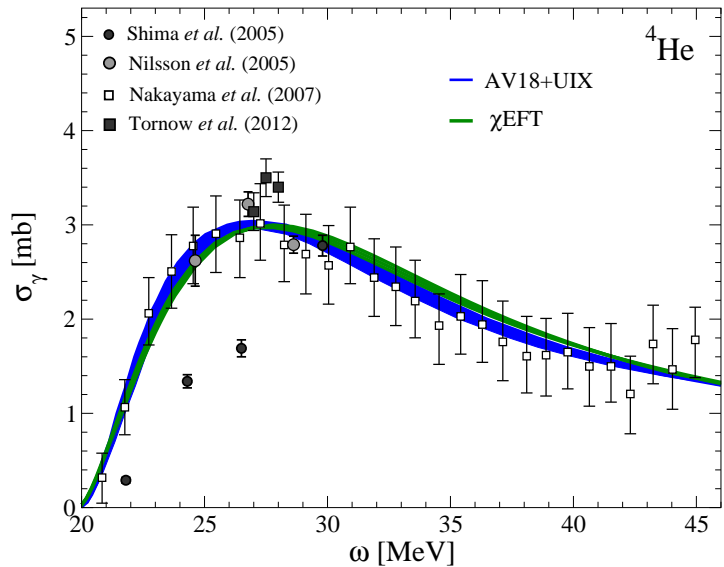}
 \caption{\label{fig_photoabs} Photodisintegration cross section $\sigma_{\gamma}$ as a function of the energy transfer $\omega$: recent experimental data (dark circles from Ref.~\cite{Shima:2005ix}, light circles from Ref.~\cite{Nilsson}, light squares from Refs.~\cite{Nakayama}, and dark squares from Ref.~\cite{Tornow:2012zz,Raut:2012zz}) compared with theoretical calculations and corresponding uncertainty bands (dark band from Ref.~\cite{Doron2006} and light band from Ref.~\cite{Qua07}). The relation of this cross section to the dipole response function is simply $\sigma_{\gamma}(\omega)=4\pi^2\alpha \omega S_{D1} (\omega)$. Figure adapted from Ref.~\cite{BaccaPastore}.}
\end{figure}
In his work, Friar derived a connection between $\delta_{\rm TPE}$ to sum rules of the photoabsoprtion cross section $\sigma_{\gamma}$, which was known experimentally. He was the first to seriously discuss uncertainties,   primarily related to the experimental data on photonuclear cross sections. Even today, if one tried to estimate nuclear-structure corrections from modern data on $^4$He, the situation would not be much different, as explained in Fig.~\ref{fig_photoabs}. An up-to-date picture of recent experimental data is presented  in comparison to theoretical calculations obtained with the Lorentz integral transform method using state-of-the-art realistic two- and three-nucleon potentials as input. One can readily see that experimental data differ from each other or suffer from large error bars. As a consequence the accuracy of the extracted $\delta_{\rm TPE}$, or better of the extracted dominant dipole contribution $\delta^{(0)}_{D1}$, would not be satisfactory. In Ref.~\cite{Antognini11}, it was pointed out that in
muonic helium, to determine the nuclear radii with a relative accuracy of $3 \times 10^{-4}$,  $\delta_{\rm TPE}$ 
needs to be known at the $\sim 5\%$ ($1 \sigma$) level. 
It is evident from Fig.~\ref{fig_photoabs}  that ab initio theory based on modern potentials has a real chance to provide us with the necessary accuracy.

In fact, in 2013 we provided the first ab initio computation of the full $\delta_{\rm TPE}$~\cite{Ji13} using the realistic nuclear Hamiltonian, AV18+UIX~\cite{AV18,PuP95} and one parameterization of $\chi$EFT~\cite{Entem03,Na07}.
 We have used the difference of the two potentials as a way to probe 
nuclear-model uncertainties, obtaining a 4$\%$ ($1 \sigma$) effect. This indetermination comes mostly from what we call nuclear physics uncertainty and is very much connected to the fact that three-nucleon forces are less constrained than two-nucleon forces. The overall uncertainty budget, including atomic physics errors and other sources, is of $6\%$ ($1 \sigma$).

In Fig.~\ref{time_4He}, the time evolution of $\delta^A_{\rm pol}$ estimates is shown, from the first calculations in the 70' until now, leaving out the most likely wrong result of Ref.~\cite{HENLEY1976349}. 
When the uncertainty was not estimated, it might be considerably large, and we indicate it by the dashed line. 
 One can readily see that while the number has been coincidentally stable in the early calculations, our ab initio descriptions  were the first to actually reduce the $1 \sigma$ error bar from a $20\%$ to a $6\%$. This will allow for a sensible extraction of the charge radius from $\mu^4{\rm He}^+$ spectroscopic measurements by the CREMA collaboration.

\begin{figure}[htb]
\centering
 \includegraphics*[width=12.cm]{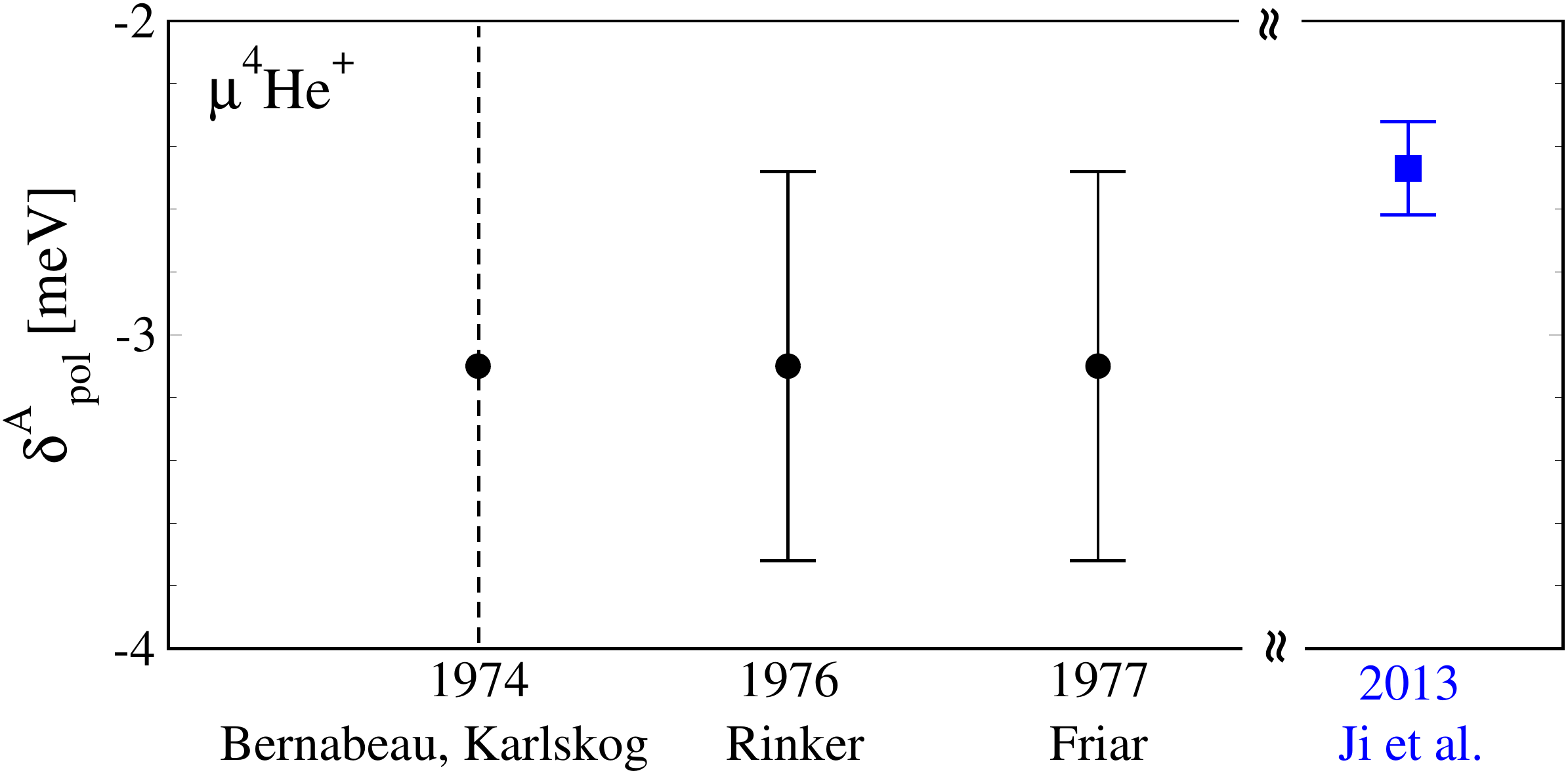}
\caption{\label{time_4He} Time evolution of  
		 $\delta^A_{\rm pol}$ estimates for $\mu^4{\rm He}^+$  with corresponding $2\sigma$ uncertainties. In the earliest calculations uncertainties were not estimated and are potentially large as indicated by the dashed line. The ab initio result is indicated by the square symbol. See text for details.}
\end{figure}

\begin{figure}[htb]
\centering
 \includegraphics[width=13cm]{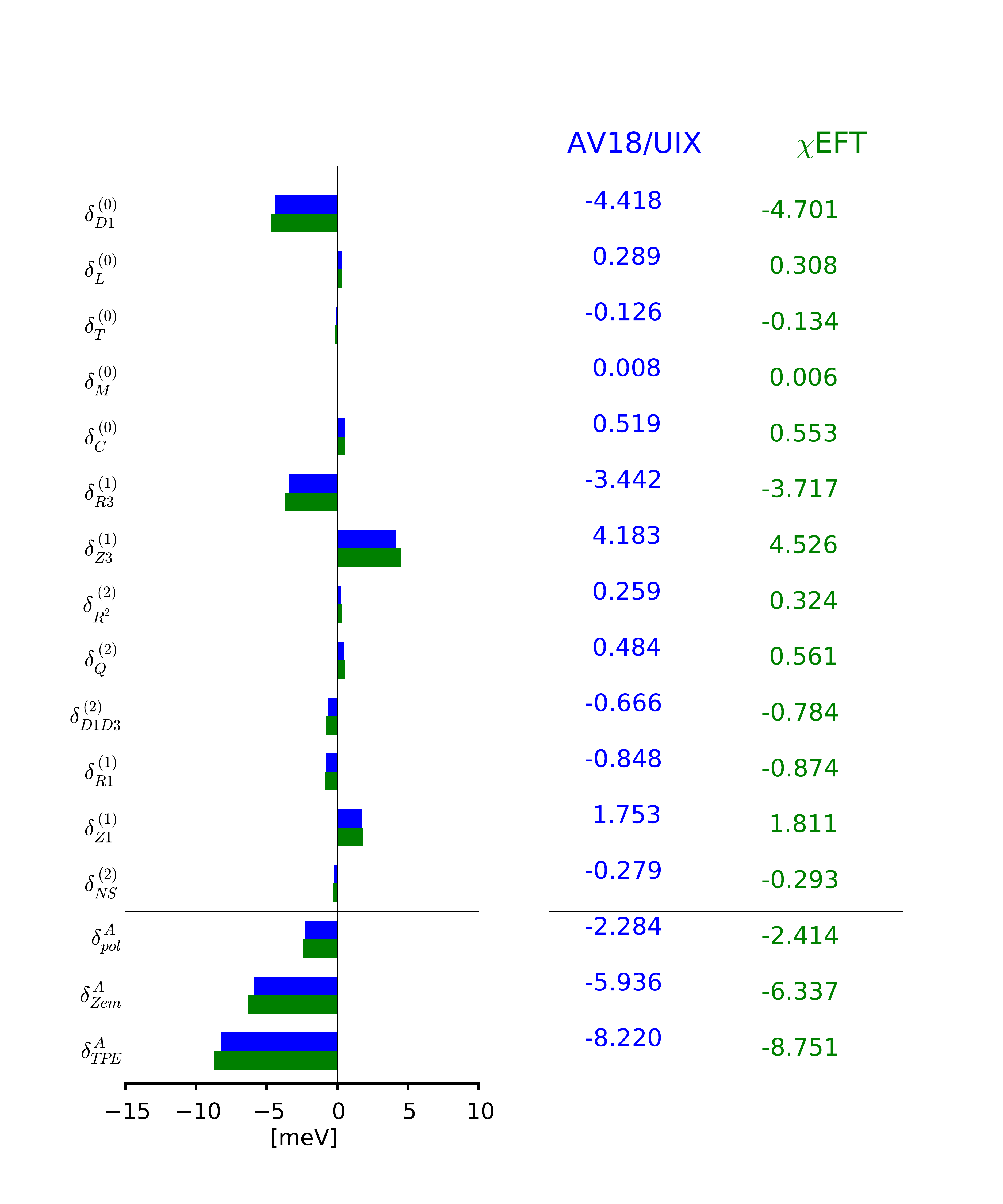}
 \caption{\label{fig:4He}
    Same as Fig.~\ref{fig:triton} but for $\mu^4{\rm He}^+$. }
\end{figure}
Finally, our broken down $\delta^A_{\rm TPE}$  results with a few updates  are shown in Fig.~\ref{fig:4He}. We observe that $\delta^{(0)}_{D1}$ is again the dominant piece, with $\delta_{Z3}^{(1)}$ and $\delta^{(1)}_{R3}$ being opposite in sign and thus largely canceling out. The overall strengths of the various terms is very similar to the case of $\mu^3{\rm He}^+$.

We would like to emphasize that in this review we present the first results for the $\delta^{(0)}_{M}$ term in $\mu^4{\rm He}^+$, which were  neglected in Ref.~\cite{Ji13}. In this respect we provide here a more accurate  calculations of the   $\delta^A_{\rm TPE}$ and thus of the total $\delta_{\rm TPE}$.

%=============================================================
\subsection{Intrinsic nucleonic two-photon exchange} 
\label{sec:TPE_N}

The evaluation of the hadronic part of the two-photon exchange contribution in muonic atoms is beyond the scope of nuclear ab initio calculations, in which the internal nucleonic degrees of freedom are not explicitly included.

Using dispersion relation analyses on electron-proton scattering data, $\delta_{\rm TPE}^N$ in $\mu {\rm H}$ is obtained as a combination of four components: elastic, non-pole Born, inelastic, and subtraction terms. For $\mu$H, Carlson and Vanderhaeghen obtained the elastic, non-pole Born, and inelastic terms to be respectively $-0.0295(13)$ meV, $0.0048$ meV, and $-0.0127(05)$ meV~\cite{Carlson:2011zd}. Using chiral perturbation theory, the subtraction term was calculated by Birse and McGovern to be $0.0042(10)$ meV~\cite{Birse:2012eb}. The combination of elastic and non-pole Born contributions yields $\delta_{\rm Zem}^N$, and the sum of inelastic and subtraction terms gives $\delta_{\rm pol}^{N}$. The uncertainties are obtained in quadrature sum. Each contributions are shown in Table~\ref{tab_tpeN_dispersion}.

The dispersion relation was also applied to evaluate two-photon exchange effects in $\mu^2{\rm H}$~\cite{Carlson:2013xea} and $\mu^3{\rm He}^+$~\cite{Carlson3He}. $\delta_{\rm TPE}$ in these muonic atoms are also separated into elastic, non-pole Born, inelastic, and subtraction terms under dispersion relation analysis. However, these contributions were given in Refs.~\cite{Carlson:2013xea,Carlson3He} with the nuclear and hadronic parts combined, with only a few exceptions. For comparison, we list in Table~\ref{tab_tpeN_dispersion} only the  explicit hadronic terms in $\mu^2{\rm H}$ and $\mu^3{\rm He}^+$ calculated by dispersion relation analyses.

\begin{table}[htb]
\centering
\caption{$\delta_{\rm TPE}^{N}$ (in meV) evaluated in dispersion relation analyses.}
\label{tab_tpeN_dispersion}
\begin{center}
\footnotesize
\renewcommand{\tabcolsep}{1.0mm}
\begin{tabular}{l cr cr cr}
\hline\noalign{\smallskip}
&&  $\mu{\rm H}$ && $\mu^2{\rm H}$ && $\mu^{3}{\rm He}^{+}$ \\
\hline\noalign{\smallskip}
$\delta^{N}_{\rm Zem}$        && $-$0.0247(13) &&                &&            \\
$\delta^{N}_{\rm pol}$        && $-$0.0085(11) &&                && $-$0.10(4) \\
\quad inelastic               && $-$0.0127(05) &&  $-$0.028(2)   && $-$0.31(2) \\
\quad subtraction             &&    0.0042(10) &&                &&    0.21(3) \\
$\delta^{N}_{\rm TPE}$        && $-$0.0332(17) &&                &&            \\
\noalign{\smallskip}\hline
\end{tabular}
\end{center}
\end{table}

Using the scaling relations~(\ref{eq:Zem_N_scal},\ref{eq:pol_N_scal}), we can relate $\delta_{\rm Zem}^N$ and $\delta_{\rm pol}^{N}$ in $\mu {\rm H}$, evaluated in dispersion relation analyses, to the corresponding contributions in other light muonic atoms. The results are shown in Table \ref{tab_tpeN_scaling}. Regarding  $\delta^{N}_{\rm Zem}$, the main difference across the various systems is given by the $Z^4$ scaling, slightly adjusted by an additional scaling from the muon reduced mass. The uncertainty of $\delta_{\rm Zem}^N$ is estimated based on that in $\mu{\rm H}$, multiplied with the scaling coefficient in Eq.~\eqref{eq:Zem_N_scal}. 

\begin{table}[htb]
\centering
\caption{$\delta_{\rm TPE}^{N}$ contributions (in meV) to muonic atoms are calculated using scaling relation with $\delta_{\rm TPE}^{N}$ in $\mu{\rm H}$.}
\label{tab_tpeN_scaling}
\begin{center}
\footnotesize
\renewcommand{\tabcolsep}{1.0mm}
\begin{tabular}{l cr cr cr cr}
\hline\noalign{\smallskip}
&& $\mu^2{\rm H}$ && $\mu^{3}{\rm H}$ && $\mu^{3}{\rm He}^{+}$ && $\mu^{4}{\rm He}^{+}$  \\
\hline\noalign{\smallskip}
$\delta^{N}_{\rm Zem}$        &&  $-$0.030(02)   && $-$0.033(02)  && $-$0.52(03)  && $-$0.54(03) \\
$\delta^{N}_{\rm pol}$        &&  $-$0.020(10)   && $-$0.031(17)  && $-$0.25(13)  && $-$0.34(20) \\
\quad inelastic               &&  $-$0.030(02)   && $-$0.047(06)  && $-$0.38(05)  && $-$0.52(10) \\
\quad subtraction             &&     0.010(10)   &&    0.016(16)  &&    0.12(12)  &&    0.17(17) \\
$\delta^{N}_{\rm TPE}$        &&  $-$0.050(10)   && $-$0.064(17)  && $-$0.77(14)  && $-$0.89(20) \\
\noalign{\smallskip}\hline
\end{tabular}
\end{center}
\end{table}

To compare with the dispersion relation analyses, we separate $\delta^{N}_{\rm pol}$ into inelastic and subtraction terms, which are both obtained by the scaling relation in Eq.~\eqref{eq:pol_N_scal}. The inelastic part of $\delta^{N}_{\rm pol}$ in $\mu^2{\rm H}$ is $-0.030$ meV, compared to $-0.028$ meV from dispersion relation analyses. We consider this difference, $\sigma_{\rm med}^N(\mu^2{\rm H})\equiv 0.002$ meV, as a result of nuclear medium effects and nucleon-nucleon interferences, neglected in Eq.~\eqref{eq:pol_N_scal}. We assume the medium effect in another muonic atom/ion $\mu{\rm X}$ is amplified by a factor of $A(A-1)/2$, considering the number of nucleon-nucleon pairs inside nucleus $X$. Therefore, the uncertainty of the inelastic term is estimated by
\begin{eqnarray}
\fl
\sigma^{N}_{\rm inel}(\mu{\rm X})
&=&  \left\{ 
\left[A \frac{\phi_{\mu{}{\rm X}}^2(0)}{\phi^2_{\mu {\rm H}}(0)} \sigma^{N}_{\rm inel}(\mu{\rm H}) \right]^2 
+\left[ \frac{A(A-1)}{2} \frac{\phi^2_{\mu{}{\rm X}}(0)}{\phi^2_{\mu^2 {\rm H}}(0)} \sigma^N_{\rm med}(\mu^2{\rm H}) \right]^2
\right\}^{1/2}
\cr
\fl
&=&
 A Z^3 \left\{ \frac{m_r^6(\mu{\rm X})}{m_r^6(\mu {\rm H})}
\left[ \sigma^{N}_{\rm inel}(\mu{\rm H}) \right]^2 
+ \frac{(A-1)^2 m_r^6(\mu{\rm X})}{4\, m_r^6(\mu^2 {\rm H})} 
\left[\sigma^N_{\rm med}(\mu^2{\rm H}) \right]^2
\right\}^{1/2}
{.}
\end{eqnarray}
The evaluation of the subtraction term is generally model dependent. Similar values of the subtraction term in $\mu{\rm H}$ were obtained in Refs.~\cite{Carlson:2011zd,Miller:2012ne,Hill:2016bjv,Birse:2017czd,Hill:2017rlj}, only its uncertainty being disputed. We follow the strategy of Ref.~\cite{Krauth:2015nja} by assigning a $\pm 100\%$ error to the subtraction term in muonic atoms. The uncertainties in $\delta^{N}_{\rm pol}$ and $\delta^{N}_{\rm TPE}$ are then obtained by using relevant quadrature sums. Our results of hadronic terms in $\mu^3{\rm He}^+$ are consistent with those obtained in dispersion relation analysis~\cite{Carlson3He}.

%==========================================

\subsection{Summary}
\label{summary}

In this summary we will compare all light muonic atoms discussed above against each other and include single nucleon contributions to provide values for the total $\delta_{\rm TPE}$.

\begin{table}[!tb]
\caption{Nuclear-structure corrections to the  Lamb shift (in meV), broken down into all terms composing $\delta^A_{\rm TPE}$.  All light muonic atoms are computed here with AV18/UIX nuclear potential (in case of the deuteron only AV18). }
\label{table_pol_eftem}
\begin{center}
\footnotesize
\renewcommand{\tabcolsep}{1.0mm}
\begin{tabular}{cl cr cr cr cr}
\hline\hline  &        &&$\mu^2{\rm H}$ &&$\mu^3{\rm H}$  &&$\mu^3{\rm He}^+$ && $\mu^4{\rm He}^+$\\ 
\hline    
&$\delta^{(0)}_{D1}$   &&$-$1.907   &&$-$0.7669 && $-$6.479 &&$-$4.418 \\
&$\delta^{(0)}_{L}$    &&   0.029   &&   0.0285 &&    0.232 &&   0.289 \\
&$\delta^{(0)}_{T}$    &&$-$0.012   &&$-$0.0128 && $-$0.103 &&$-$0.126 \\
&$\delta^{(0)}_{M}$    &&   0.003   &&   0.0007 &&    0.006 &&   0.008\\
&$\delta^{(0)}_{C}$    &&   0.262   &&   0.0718 &&    1.000 &&   0.519 \\
&$\delta^{(1)}_{R3}$   &&   0       &&   0      && $-$8.539 &&$-$3.442 \\
&$\delta^{(1)}_{Z3}$   &&   0.357   &&   0.1778 &&    8.100 &&   4.183 \\
&$\delta^{(2)}_{R^2}$  &&   0.042   &&   0.0199 &&    0.632 &&   0.259 \\
&$\delta^{(2)}_{Q}$    &&   0.061   &&   0.0344 &&    1.015 &&   0.484 \\
&$\delta^{(2)}_{D1D3}$ &&$-$0.139   &&$-$0.0783 && $-$0.841 &&$-$0.666 \\
&$\delta^{(1)}_{R1}$   &&   0.017   &&   0.0280 && $-$1.294 &&$-$0.848 \\
&$\delta^{(1)}_{Z1}$   &&   0.064   &&   0.0453 &&    2.256 &&   1.753 \\
&$\delta^{(2)}_{NS}$   &&$-$0.020   &&$-$0.0239 && $-$0.193 &&$-$0.279 \\
\hline
&$\delta^A_{\rm pol}$                  &&$-$1.243   &&$-$0.4755 && $-$4.208 &&$-$2.284 \\
&$\delta_{\rm Zem}^{A}$                &&$-$0.421   &&$-$0.2231 &&$-$10.356 &&$-$5.936 \\
&$\delta^A_{\rm TPE}$                  &&$-$1.664   &&$-$0.6986 &&$-$14.564 &&$-$8.220 \\
\hline\hline
\end{tabular}
\end{center}
\end{table}

\begin{table}[!tb]
  \caption{Same as in Table~\ref{table_pol_eftem}, but for one parameterization of the $\chi$EFT nuclear potential (in case of the deuteron only nucleon-nucleon interaction). Numbers are given in meV.}
\label{table_pol_av18u9}
\begin{center}
\footnotesize
\renewcommand{\tabcolsep}{1.0mm}
\begin{tabular}{cl cr cr cr cr}
\hline\hline               &                       &&$\mu^2{\rm H}$ &&$\mu^3{\rm H}$  &&$\mu^3{\rm He}^+$ && $\mu^4{\rm He}^+$\\ 
\hline    
 & $\delta^{(0)}_{D1}$   &&$-$1.912 &&$-$0.7848 && $-$6.633 &&$-$4.701 \\                                       
               & $\delta^{(0)}_{L}$    &&   0.029 &&   0.0296 &&    0.240 &&   0.308 \\ 
               & $\delta^{(0)}_{T}$    &&$-$0.012 &&$-$0.0132 && $-$0.107 &&$-$0.134 \\ 
               & $\delta^{(0)}_{M}$    &&   0.003 &&   0.0007 &&    0.006 &&   0.006\\ 
               & $\delta^{(0)}_{C}$    &&   0.262 &&   0.0732 &&    1.020 &&   0.553 \\ 
 & $\delta^{(1)}_{R3}$   &&   0     &&   0      && $-$8.711 &&$-$3.717 \\ 
               & $\delta^{(1)}_{Z3}$   &&   0.359 &&   0.1844 &&    8.327 &&   4.526 \\ 
 & $\delta^{(2)}_{R^2}$  &&   0.041 &&   0.0206 &&    0.654 &&   0.324 \\ 
               & $\delta^{(2)}_{Q}$    &&   0.061 &&   0.0358 &&    1.038 &&   0.561 \\ 
               & $\delta^{(2)}_{D1D3}$ &&$-$0.139 &&$-$0.0811 && $-$0.862 &&$-$0.784 \\ 
 & $\delta^{(1)}_{R1}$   &&   0.017 &&   0.0287 && $-$1.314 &&$-$0.874 \\ 
               & $\delta^{(1)}_{Z1}$   &&   0.064 &&   0.0463 &&    2.291 &&   1.811 \\ 
               & $\delta^{(2)}_{NS}$   &&$-$0.021 &&$-$0.0247 && $-$0.199 &&$-$0.293 \\ 
\hline
&$\delta^A_{\rm pol}$                  &&$-$1.248 &&$-$0.4845 && $-$4.250 &&$-$2.414\\
&$\delta_{\rm Zem}^{A}$                &&$-$0.423 &&$-$0.2307 &&$-$10.618 &&$-$6.337 \\
&$\delta^A_{\rm TPE}$                 &&$-$1.671 &&$-$0.7152 &&$-$14.868 &&$-$8.751 \\
\hline\hline
\end{tabular}
\end{center}
\end{table}

First, in Tables~\ref{table_pol_eftem} and \ref{table_pol_av18u9} we detail all the terms composing  $\delta^A_{\rm TPE}$ computed with the
AV18/UIX and $\chi$EFT potentials, respectively.
While these numerical values have already been shown in Figs.~\ref{fig:deut}, \ref{fig:triton}, \ref{fig_comp_3He} and \ref{fig:4He},
these new tables allow for a comparison of the different light muonic systems. Indeed, one primarily appreciates the $Z$ dependence, which is making each term of $\mu^3{\rm He}^+$ and  $\mu^4{\rm He}^+$ larger than the corresponding one in  $\mu^2{\rm H}$ and   $\mu^3{\rm H}$. Looking at $\delta_{\rm TPE}^A$, it is also interesting to note that overall nuclear-structure corrections due to the few-nucleon dynamics are the largest in $\mu^3{\rm He}^+$ and the smallest in $\mu^3{\rm H}$. This fact is simply explained by a combination of the $Z$ dependence and differences in nuclear threshold effects, as we  have argued above.

\begin{table}[!tb]
\caption{Contributions to $\delta_{\rm TPE}$ of the Lamb shift in light muonic atoms (in meV), with separated contributions from the few-nucleon dynamics and from the individual nucleons. The uncertainty associated with each value is given in brackets. Due to cancellation of the elastic and inelastic Zemach term, the uncertainty in $\delta_{\rm TPE}$ can differ from the quadrature sum of the corresponding terms in the table.
}
\label{tab_all} 
\begin{center}
\footnotesize
\renewcommand{\tabcolsep}{1.0mm}
\begin{tabular}{l|crcrcrcr|cr}
\hline\hline \noalign{\smallskip}
 && $\delta^{A}_{\rm Zem}$ && $\delta^{A}_{\rm pol}$ && $\delta^{N}_{\rm Zem}$ && $\delta^{N}_{\rm pol}$ && $\delta_{\rm TPE}$  \\
\hline
$\mu^2{\rm H}$      && $-$0.423(04) && $-$1.245(13)  && $-$0.030(02) && $-$0.020(10) && $-$1.718(17) \\
$\mu^3{\rm H}$      && $-$0.227(06) && $-$0.480(11)  && $-$0.033(02) && $-$0.031(17) && $-$0.771(22)  \\
$\mu^3{\rm He}^{+}$ && $-$10.49(23) && $-$4.23(18)   && $-$0.52(03)  && $-$0.25(13)  && $-$15.49(33) \\
$\mu^4{\rm He}^{+}$ && $-$6.14(31)  && $-$2.35(13)   && $-$0.54(03)  && $-$0.34(20)  && $-$9.37(44) \\
\hline
\end{tabular}
\end{center}
\end{table}

In Table~\ref{tab_all} we include the intrinsic single nucleon contributions to each muonic system by adding $\delta^{N}_{\rm Zem}$ and $\delta^{N}_{\rm pol}$, so as to compose the total $\delta_{\rm TPE}$. Values and uncertainties of the single nucleon terms are given in subsection~\ref{sec:TPE_N}. They do not depend on the $A$-nucleon dynamics, nor on the potential. 
One can observe that generally single nucleon contributions to $\delta_{\rm TPE}$ are smaller than the parts determined by the few-nucleon dynamics, thus making it crucial to reliably determine the latter and the related uncertainties. As demonstrated above for each light muonic atom, ab initio computations of the few-nucleon dynamics have enabled to reach a precision level which was not available before.  The uncertainty associated with each value of Table~\ref{tab_all} for $\delta^A_{\rm pol}, \delta^A_{\rm Zem}$ and $\delta^A_{\rm TPE}$ is given in the brackets and in Table ~\ref{table_pol_err} is broken down into all the uncertainties sources explained in Section~\ref{sec:uncert}.  In particular, it is to note that while for $\mu^2{\rm H}$ the largest source of uncertainty comes from atomic physics, for $\mu^4{\rm He}^+$ it comes from nuclear modeling.

\begin{table}[!tb]
  \caption{Relative uncertainties (in $\%$) for $\delta^A_{\rm pol}$, $\delta^A_{\rm Zem}$, and $\delta^A_{\rm TPE}$. The estimation is based on sources of uncertainty listed in Section~\ref{sec:uncert}. The total uncertainties are obtained from a quadrature sum.}
\label{table_pol_err}
\begin{center}
\footnotesize
\renewcommand{\tabcolsep}{1.2mm}
\begin{tabular}{l |l c c c |l c c c |l c c c |l c c c}
\hline\hline
               && \multicolumn{3}{c|}{$\mu^2{\rm H}$}    && \multicolumn{3}{c|}{$\mu^3{\rm H}$}  
               && \multicolumn{3}{c|}{$\mu^3{\rm He}^+$} && \multicolumn{3}{c}{$\mu^4{\rm He}^+$}\\ 
               && $\delta_{\rm pol}^A$ & $\delta_{\rm Zem}^A$ & $\delta_{\rm TPE}^A$
               && $\delta_{\rm pol}^A$ & $\delta_{\rm Zem}^A$ & $\delta_{\rm TPE}^A$
               && $\delta_{\rm pol}^A$ & $\delta_{\rm Zem}^A$ & $\delta_{\rm TPE}^A$
               && $\delta_{\rm pol}^A$ & $\delta_{\rm Zem}^A$ & $\delta_{\rm TPE}^A$\\
\hline    
Numerical        && 0.0 & 0.0 & 0.0 && 0.1 & 0.0 & 0.1 && 0.4 & 0.1 & 0.1 && 0.4 & 0.3 & 0.4\\
Nuclear model    && 0.3 & 0.5 & 0.4 && 1.3 & 2.4 & 1.7 && 0.7 & 1.8 & 1.5 && 3.9 & 4.6 & 4.4\\
ISB              && 0.2 & 0.2 & 0.2 && 0.7 & 0.2 & 0.5 && 1.8 & 0.2 & 0.5 && 2.2 & 0.5 & 0.5\\
Nucleon size     && 0.3 & 0.8 & 0.0 && 0.6 & 0.9 & 0.2 && 1.2 & 1.3 & 0.9 && 2.7 & 2.0 & 1.2\\
Relativistic     && 0.0 &  -  & 0.0 && 0.1 &  -  & 0.1 && 0.4 &  -  & 0.1 && 0.1 &  -  & 0.0\\
Coulomb          && 0.4 &  -  & 0.3 && 0.5 &  -  & 0.3 && 3.0 &  -  & 0.9 && 0.4 &  -  & 0.1\\
$\eta$-expansion && 0.4 &  -  & 0.3 && 1.3 &  -  & 0.9 && 1.1 &  -  & 0.3 && 0.8 &  -  & 0.2\\
Higher $Z\alpha$ && 0.7 &  -  & 0.5 && 0.7 &  -  & 0.5 && 1.5 &  -  & 0.4 && 1.5 &  -  & 0.4\\
\hline
Total            && 1.0 & 0.9 & 0.8 && 2.3 & 2.2 & 2.0 && 4.2 & 2.2 & 2.1 && 5.5 & 5.1 & 4.6\\
\hline\hline
\end{tabular}
\end{center}
\end{table}

\section{Conclusions}
\label{sec:conclude}

In this work we reviewed our recent activities devoted to the calculation of nuclear-structure corrections in light muonic atoms.
This subject has gained a renewed interest following the emergence of the proton-radius puzzle and the subsequent experimental campaign of 
the CREMA collaboration directed towards Lamb shift measurements in light muonic systems, with the goal of extracting their nuclear charge radii and compare them to ordinary atom spectroscopy or electron scattering data.

While $\delta_{\rm TPE}$ is not a ``traditional'' observable for ab initio nuclear theory studies, we showed that, taking advantage of methodologies and few-body techniques developed for nuclear-structure physics, it could be computed with unprecedented precision.
In particular, for the $\mu^3{\rm He}^+$ and  $\mu^4{\rm He}^+$ we have reduced uncertainties by a factor of $5$, when compared to the previously available estimates of $\delta^A_{\rm pol}$. For  $\mu^2{\rm H}$, instead, we have presented the most thorough  estimates of the uncertainty related to the non-perturbative nature of nuclear forces. 
These achievements are crucial for the muonic atom measurements, where  $\delta_{\rm TPE}$ is needed for the extraction of nuclear charge radii and is presently the bottle-neck to exploit the experimental precision.

State-of-the-art calculations of $\delta_{\text{TPE}}$ presented in this work
employ nuclear potentials derived either phenomenologically or from a low-energy
expansion of quantum chromodynamics, namely $\chi$EFT.
Effective field theories in general, unlike
phenomenological models, furnish a systematic, i.e, order-by-order,
description of low-energy processes at a chosen level of
resolution.
$\chi$EFT presently constitutes the
modern paradigm of analyzing nuclear forces, which are built from a sum of  pion-exchange contributions 
and nucleon contact terms, see,
e.g., Refs.~\cite{Epelbaum05,Machleidt:2011zz}.
Power counting enables
to determine the importance of individual terms in the low-energy
expansion and thereby also facilitates a meaningful truncation of higher-order diagrams that build the potential.
 Given the chiral expansion, contributions with a low power
 are more important than terms at higher powers. Starting from
the leading order (LO),  higher orders are  denoted
as next-to-leading order (NLO), next-to-next-to-leading (N$^2$LO), etc.
 At each order  of the chiral EFT potential, there
is a finite set of parameters,
the LECs, that determine the strength of various pion-nucleon and
multi-nucleon operators. The LECs are not provided by the theory
itself but can be obtained from fitting to selected experimental data,
such as NN and $\pi$N scattering cross sections, and other few-body
ground state observables, such as binding
energies. Different fitting procedures exist, which in turn lead to different potentials.
The optimal way to exploit such an approach is to
 explore an ensemble of parameterization of the potentials to probe both statistical and systematic uncertainties, order by order.
 We have performed such a study so far only for $\mu^2{\rm H}$~\cite{Hernandez2018}. Remarkably, we find that statistical errors are small
 and that systematic errors are underestimated by maximally 50\% when comparing only one choice of $\chi$EFT potential with a phenomenological interaction. This result cannot be easily extrapolated to the three and four-body systems. There, in fact three-body forces need to be included, which are much less constrained than the nucleon-nucleon force. Nevertheless, we expect that our calculations based on two substantially different interactions should capture the bulk of the uncertainty. A more thorough systematic and statistical analysis is useful and is presently called for.

 It is to be noted that the nuclear theory community is at the moment devoting a fair amount of resources towards the development of nucleon-nucleon forces at N$^5$LO and three-body forces at N$^{4}$LO. In the future, these will potentially provide the possibility to reduce uncertainties in $\delta_{\rm TPE}$. A  reduction of the uncertainties may be also obtained by fitting the LECs to radii extracted from muonic atoms with high precision. This strategy is presently under discussion.
 
 One has to keep in mind that uncertainties on $\delta_{\rm TPE}$ are  much larger than the experimental precision, as shown in Table~\ref{tab:1} in the introduction to this review. As opposed to what is possible today for  $\delta_{\rm QED}$, reducing nuclear-model uncertainties in $\delta_{\rm TPE}$ at the level of being able to match or even supersede the experimental precision is currently unrealistic and infeasible. The results presented in this review represent the state-of-the-art, some improvement might be achieved by exploring the pathways alluded to above, or by systematically adding the effect of meson exchange currents. 

 On the other hand, the CREMA collaboration aims at measuring the Lamb shift also in muonic lithium and beryllium ions. In preparation for that, the QED community has already started to calculate the corresponding QED corrections, see, e.g.,~\cite{Krutov}. The TPE contributions are presently based on rough estimates from the Rinker semi-empirical formula and/or from experimental data, as explained in Ref.~\cite{Drake}. They do not contain all the terms we derived in this review and are plagued by large uncertainties.
 For example, for $\mu^6{\rm Li}^{2+}$ the uncertainty is of  27\% ($1 \sigma$).
Computing $\delta_{\rm TPE}$ in such systems with ab initio methods should be possible, and this would  almost certainly lead to a reduction in the related uncertainty. Obviously, this does not come without challenges since $^6$Li has a pronounced $^4$He-$^2$H cluster structure, so that very large model spaces are needed to capture these features when using basis function expansions. First steps towards developing the necessary machinery to achieve these computations are presently underway.

Another natural extension of this work is to explore nuclear-structure corrections to other transitions relevant in muonic atom spectroscopy, e.g., the hyperfine splitting. While $\mu{\rm H}$ is the first system that will be measured by several groups around the world (Switzerland, England and Japan), eventually also the hyperfine splitting in
$\mu^3{\rm He}^+$ and $\mu^2{\rm H}$ will be investigated by the CREMA collaboration. We are presently in the process of deriving the necessary formalism to achieve this goal.

Finally, we would like to remark that in precision physics it is of paramount importance that different methods are employed to compute the $\delta_{\rm TPE}$ corrections and compared to the approach presented in this review. For example, the work based on dispersion relation is very relevant and, depending on the available experimental data, can reach comparable precision to the ab initio approach. Also, pion-less effective field theories may be applied to this problem, as well as other few-body methods that can deal with precise calculation of the response function. 

\ack{ We would like to acknowledge very fruitful discussions with Randolf Pohl, Beatrice Franke, Julian J. Krauth, Marc Vanderhaeghen, Carl E. Carlson, Krzysztof Pachucki, and Savely G. Karshenboim.
  This work was supported in parts by the Natural Sciences and
  Engineering Research Council (NSERC), the National Research Council
  of Canada, by the Deutsche Forschungsgemeinschaft DFG through the
  Collaborative Research Center [The Low-Energy Frontier of the
    Standard Model (SFB 1044)], and through the Cluster of Excellence
  [Precision Physics, Fundamental Interactions and Structure of Matter
    (PRISMA)], and the Pazy Foundation.}

\newpage
\appendix
\section{Wigner-Eckart Theorem}
\label{app:wiec}
%=============================================================================

The reduced matrix element is defined by Wigner-Eckart Theorem~\cite{Edmonds:1996} as
\begin{equation}
\label{eq:WigEck}
\bra N_0 J_0 M_0 |T^{(k)}_\nu|N J M \ket = (-1)^{k-J+J_0} \frac{ \bra k \nu J M |J_0 M_0\ket}{\sqrt{2J_0+1}} \bra N_0 J_0 || T^{(k)} || N J\ket {,}
\end{equation}
\begin{equation}
\label{eq:App-I001}
\bra NJ M |T^{(k)\dagger}_\nu|N_0 J_0 M_0 \ket = (-1)^{k} \frac{\bra k \nu JM| J_0 M_0\ket}{\sqrt{2J_0+1}}\bra NJ || T^{(k)} ||N_0 J_0\ket {.}
\end{equation}
\begin{equation}
\label{eq:App-I003}
\bra NJ || T^{(k)} ||N_0 J_0\ket = (-1)^{J-J_0}\bra N_0 J_0 || T^{(k)} ||NJ\ket^* {.}
\end{equation}

Using the relations in Eqs.~(\ref{eq:WigEck}, \ref{eq:App-I001}, \ref{eq:App-I003}), we have
\begin{eqnarray}
\label{eq:App-I0}
&&\sum\limits_M \bra N_0 J_0 M_0| A^{(k)} | NJM\ket \cdot \bra  NJ M| B^{(k)} |N_0 J_0 M_0\ket
\nn
&=& \frac{(-1)^{J_0-J}}{2J_0+1} \bra N_0 J_0 || A^{(k)} || N J\ket
\bra  NJ || B^{(k)} ||N_0 J_0 \ket
\nn
&=&\frac{1}{2J_0+1} \bra N_0 J_0 || A^{(k)} || N J\ket\,\bra  N J || B^{(k)\dagger} ||N_0 J_0 \ket {,}
\end{eqnarray}
\begin{eqnarray}
\label{eq:App-I2}
\bra N_0 J_0 | A^{(k)} | NJ\ket \cdot \bra  NJ | A^{(k)} |N_0 J_0 \ket 
&=&\frac{1}{2J_0+1} |\bra N_0 J_0 || A^{(k)} || N J\ket|^2 {.}
\end{eqnarray}

The matrix element of the scalar product of two tensor operators obeys Eq. (7.1.6) in Ref.~\cite{Edmonds:1996}, which yields
\begin{eqnarray}
\label{eq:App-II}
\fl
\sum_{J' M'} |\bra j_1' j_2' J' M' | T^{(k)}\cdot U^{(k)} | j_1 j_2 J M\ket|^2
= \sixj{j_1'}{k}{j_1}{j_2}{J}{j_2'}^2\, |\bra j_1' || T^{(k)} || j_1\ket|^2\, |\bra j_2' || U^{(k)} || j_2\ket|^2 {.}
\nn
\end{eqnarray}

A single operator in coupled scheme obeys Eq. (7.1.7) in Ref.~\cite{Edmonds:1996}, which leads to
\begin{equation}
\label{eq:App-Tk}
\fl
|\bra j_1' j_2 J' || T^{(k)} || j_1 j_2 J\ket|^2
=(2J+1)(2J'+1) \sixj{j_1'}{J'}{j_2}{J}{j_1}{k}^2 |\bra j_1' || T^{(k)} || j_1\ket|^2 {,}
\end{equation}
\begin{equation}
\label{eq:App-Tk2}
\bra \ell_0 || Y_k ||\ell\ket = (-1)^{\ell_0} \sqrt{\frac{(2\ell_0+1)(2k+1)(2\ell+1)}{4\pi}} \threej{\ell_0}{k}{\ell}{0}{0}{0} {,}
\end{equation}
with $\left(:::\right)$ indicates the 3j-symbol. For $k=1$, we have
\begin{equation}
\label{eq:app-Y1}
|\bra \ell_0 || Y_1 ||\ell\ket|^2 = \frac{3}{4\pi}\left[(\ell_0+1)\delta_{\ell,\ell_0+1} + \ell_0 \delta_{\ell,\ell_0-1}\right] {.}
\end{equation}

\section{Coulomb integrals}
\label{app:coulomb}

The radial Coulomb Green's function $g_\ell$ can be expressed in the form of Whittaker functions~\cite{Friar:1978wv} as
\begin{equation}
g_\ell(-\omega_N;r,r') = -2m_r \xi \frac{\Gamma(\ell+1-2\kappa)}{\Gamma(2\ell+2)}\, \mathcal{M}_{2\kappa,\ell+\frac{1}{2}}\left(\frac{r_{<}}{\xi}\right)\, \mathcal{W}_{2\kappa,\ell+\frac{1}{2}}\left(\frac{r_{>}}{\xi}\right) {,}
\end{equation}
where $\kappa\equiv \frac{Z\alpha}{4}\sqrt{\frac{2m_r}{\omega_N}}$ and $\xi = 1/\sqrt{8m_r \omega_N}$ are defined. The function $\mathcal{M}$ and $\mathcal{W}$ are two Whittaker functions, which are regular respectively at $r\rightarrow0$ and $r\rightarrow\infty$. $g_\ell$ is therefore regular at both origin and infinity. The Whittaker functions satisfy the Wronskian relation that~\cite{Friar:1978wv}
\begin{equation}
\mathcal{M} \frac{d \mathcal{W}}{dr} - \mathcal{W} \frac{d \mathcal{M}}{dr} = -\frac{\Gamma(2\ell+2)}{\xi\Gamma(\ell+1-2\kappa)} {.}
\end{equation}

In Section~\ref{sec:Coul-corr}, we only need to evaluate the Coulomb integral $\mathcal{F}_{01}$, which can be analytically solved by using generating functions based on the double-Laplace transform of the Green's function~\cite{Bodine1982,Swainson1991}. This solution is related to the hypergeometric function ${}_2F_1$ by
\begin{eqnarray}
\label{eq:c-int1}
\fl
\mathcal{F}_{01}(\omega_N) &=& \int_0^\infty dr\, \int_0^\infty dr' R_{20}\left(\frac{m_r Z\alpha r}{2}\right)\; R_{20}\left(\frac{m_r Z\alpha r'}{2}\right)\; \frac{g_\ell(-\omega_N;r,r')}{rr'}
\nn
\fl
&=& -\frac{1}{Z\alpha} \left[\frac{3}{4}-\frac{1}{4\kappa^2}+\frac{24\kappa(1-\kappa)}{(1+\kappa)^6} \,
{}_2F_1\left(4,2-2\kappa,3-2\kappa,\frac{(1-\kappa)^2}{(1+\kappa)^2}\right)
\right]  {.}
\end{eqnarray}
Expanding $\mathcal{F}_{01}$ in powers of $\kappa$ yields, 
\begin{eqnarray}
\label{eq:F201}
\mathcal{F}_{01} 
&=& -\frac{1}{Z\alpha} \left[4\kappa+16\kappa^2\ln(4\kappa)+2\kappa^2+\mathcal{O}(\kappa^3)\right]
\nn
&=& -\sqrt{\frac{2m_r}{\omega_N}}  - \frac{Z\alpha m_r}{\omega} \ln\frac{2(Z\alpha)^2 m_r}{\omega}
 - \frac{Z\alpha m_r}{4\omega}+\cdots {.}
\end{eqnarray}

%=============================================================================

\section*{References}
\bibliographystyle{iopart-num}
\providecommand{\newblock}{}

%\bibliographystyle{unsrt}
%\bibliography{muA.bib}

\end{document}